\documentclass[aps,rmp,preprint]{revtex4-1}
\usepackage[USenglish]{babel} 
\usepackage[T1]{fontenc}
\usepackage[ansinew]{inputenc}
\usepackage{lmodern} 
\usepackage{graphicx} 
\usepackage{amsmath}
\usepackage{amsthm}
\usepackage{amsfonts}
\usepackage{multirow}
\usepackage[caption=false]{subfig}
\usepackage{slashed}
\usepackage{natbib}
\usepackage{array}
\usepackage{setspace}

\newcommand{\se}{\hat{s}_e}

\begin{document}
\newcommand{\HRule}{\rule{\linewidth}{0.5mm}}
\newcommand{\cev}[1]{\reflectbox{\ensuremath{\vec{\reflectbox{\ensuremath{#1}}}}}}

\title{Symmetry violations in nuclear and neutron $\beta$ decay}
\author{K.K. Vos}
\affiliation{Van Swinderen Institute for Particle Physics and Gravity, University of Groningen, Nijenborgh 4, 9747 AG Groningen, The Netherlands}
\author{H.W. Wilschut}
\affiliation{Van Swinderen Institute for Particle Physics and Gravity, University of Groningen, Nijenborgh 4, 9747 AG Groningen, The Netherlands}

\author{R.G.E. Timmermans}
\affiliation{Van Swinderen Institute for Particle Physics and Gravity, University of Groningen, Nijenborgh 4, 9747 AG Groningen, The Netherlands}


\begin{abstract}
The role of $\beta$ decay as a low-energy probe of physics beyond the Standard Model is reviewed. Traditional searches for deviations from the Standard Model structure of the weak interaction in $\beta$ decay are discussed in the light of constraints from the LHC and the neutrino mass. Limits on the violation of time-reversal symmetry in $\beta$ decay are compared to the strong constraints from electric dipole moments. Novel searches for Lorentz symmetry breaking in the weak interaction in $\beta$ decay are also included, where we discuss the unique sensitivity of $\beta$ decay to test Lorentz invariance. We end with a roadmap for future $\beta$-decay experiments. 

\end{abstract}
\maketitle

\tableofcontents

\newpage
\section{Introduction}
\noindent
The study of nuclear and neutron $\beta$ decay has played a major role in uncovering the structure of the weak interaction, and therefore in the development of the electroweak sector of the Standard Model (SM) of particle physics. The intensity and the variety of $\beta$ emitters, combined with the high precision with which $\beta$-decay parameters can be measured, ensured that $\beta$ decay remained important in searches for new physics beyond the SM (BSM). Novel techniques of laser cooling and atom trapping \cite{Beh09,Spr97} made it possible to detect the momentum of the recoiling nucleus, allowing for further searches in unexplored observables that became available. New sources for slow neutrons enabled further progress in the study of neutron $\beta$-decay observables \cite{Abe08, Nic09, Dub11}. The motivation for these modern experiments is on the one hand to improve the accuracy of SM parameters, and on the other hand to search for physics BSM. 

Searches for BSM physics in $\beta$ decay look for deviations from the left-handed vector-axial-vector (``$V-A$'') space-time structure of the weak interaction, see \textcite{Severijnstest, Hol14} and references therein. High-precision $\beta$-decay experiments are sensitive to possible contributions of non-SM (or exotic) currents, in particular right-handed vector, scalar, and tensor currents, that couple to hypothetical new, heavy particles. These exotic currents can also give additional violations of the discrete symmetries parity (P), charge conjugation (C), and time-reversal invariance (T). 

Traditionally, $\beta$ decay has been viewed as complementary to the direct searches for new, heavy particles at high-energy colliders. However, with the availability of meson factories the emphasis of searching for new physics in precise measurements of semileptonic decay parameters has shifted from $\beta$ decay. New physics has also been severely constrained by the emergence of the new field of neutrino oscillations and by the ultra-precise measurements of static observables such as the weak charges of quarks and electrons and the P- and T-odd electric dipole moments (EDM) of particles, atoms, or molecules. Moreover, theoretical developments made it clear how various observables are interconnected, and therefore how the discovery potential of $\beta$-decay experiments compares to that of other fields. 

Recently, another twist has been added to $\beta$ decay as a promising precision laboratory to test the invariance of the weak interaction under Lorentz transformations, that is, boosts and rotations. The available evidence for the Lorentz invariance of the weak interaction is, in fact, surprisingly poor. The possibility to break Lorentz and the closely related CPT invariance \cite{Gre02} occurs in many proposals that attempt to unify the SM with general relativity, one of the central open issues in theoretical high-energy physics. During the last decade, the phenomenological consequences of such a breakdown of Lorentz symmetry have been charted \cite{kossme}, and recently such theoretical studies have been extended to $\beta$ decay \cite{Noo13a}.

This review gives a broad overview of the searches for symmetry violations in nuclear and neutron $\beta$ decay and discusses their significance compared to various other observables, both in precision measurements and in collider searches. In this way, it attempts to identify which $\beta$-decay studies are the most relevant to pursue. In Sec.~\ref{sec:formalism} we first introduce the effective field theory (EFT) framework, which enables us to compare various experiments in a model-independent approach. We define the $\beta$-decay observables in Sec.~\ref{sec:obs}.  

In Sec.~\ref{sec:constraints} we review the best bounds on exotic right-handed vector, scalar, and tensor couplings. We first address the most sensitive $\beta$-decay experiments, in which we also include limits from pion-decay experiments. 

Second, we discuss how the neutrino mass and data from the LHC experiments constrain BSM physics. We compare the bounds from these two sectors with the bounds from $\beta$-decay experiments.    
The violation of time-reversal invariance is discussed in Sec.~\ref{sec:time}. In $\beta$ decay, T-violation manifests itself in nonzero imaginary parts of the couplings, which are probed by triple-correlation observables in $\beta$ decay. We discuss how these bounds compare 
to those derived from the stringent upper limits on the values of EDMs. 

In Sec.~\ref{sec:LIV}, we address the possibility that the weak interaction violates Lorentz symmetry, and in particular rotational invariance, in nuclear and neutron $\beta$ decay. Such Lorentz violation (LV) would give rise to unique signals with no SM ``background,'' which, even when tiny, could be experimentally detectable. Nuclear and neutron $\beta$ decay offer a unique sensitivity to some Lorentz-violating parameters, especially in the gauge and neutrino sector, which we discuss separately. 

We conclude with a roadmap for the opportunities in future $\beta$-decay studies, in the light of the obtained and foreseen bounds from other frontiers.

\newpage\section{Formalism}\label{sec:formalism}

\begin{table}
\begin{tabular}{|c|c|c|c|c|c|}
  \hline
 
 & F & GT & mixed & $1^{st}$ unique forbidden& Section \\ 
  &$\Delta J=0$ & $\Delta J = 0,\pm 1 $& $\Delta J=0$ & $\Delta J=\pm 2$& \\\cline{2-5}
  &\multicolumn{3}{c|}{$\pi_i\pi_f =+1$} &  {$\pi_i\pi_f =-1$}& \\ \hline\hline
SM&$V_{ud}$ & $\rho$ & $\rho, V_{ud},\lambda$ & & \ref{sec:obs} \\ 
parameter&&&&& \\ \hline
BSM & $A_{L,R}$ & $\alpha_{L,R}$ &  $\alpha_{L,R}$ & & \ref{sec:betabounds} \\ 
T-even& &&&&\\ \hline
BSM&  - & Im~$\alpha_L$ & Im~$A_L$ \textrm{and} Im~$\alpha_L$  & & \ref{sec:time} \\
T-odd& && Im~$a_{LR}$&&\\ \hline 
LV & $\chi_{r,s}^{\mu \nu}$ & $\chi_{i,a}^{\mu \nu}$  & & $\chi_{\mu \nu}$&\ref{sec:LIV} \\
& - & - & $a_{\textrm{LV}}$ & - &  \\

  \hline\hline
\end{tabular}
\caption{Classification of nuclear $\beta$ decays and their characteristic use in the SM and in the search for BSM physics. }
\label{tab:class}
\end{table}
Nuclear and neutron $\beta$ decay are semileptonic processes, mediated by the $W$ gauge boson of the electroweak interaction. This interaction is described by a spontaneously broken $SU(2)_L\times U(1)_Y$ gauge symmetry. Under $SU(2)_L$ symmetry, left-handed leptons transform as a doublet, while right-handed particles are $SU(2)_L$ singlets. This is denoted by
\begin{equation}\label{eq:LandR}
L_A = \left(\nu_A, l_A\right)_L \;, \; \; R_A=(l_A)_R \; ,
\end{equation}
where $A$ is the flavor index and the left- and right-handed fields are  
\begin{equation}
\psi_L \equiv \frac{1}{2}(1-\gamma_5)\psi \;, \; \; \psi_R \equiv \frac{1}{2}(1+\gamma_5)\psi\;.
\end{equation}
The $W$ boson only interacts with left-handed fermions, which reflects the maximal violation of parity (P) symmetry in the weak interaction. In the minimal SM neutrinos are assumed to be massless, and right-handed neutrinos are absent. The role of the neutrino mass is discussed in Sec.~\ref{sec:neutrino}.  

 The $\beta^- (\beta^+)$ decay transition $d\rightarrow u e^- \bar{\nu}_e \;(u\rightarrow d e^+ \nu_e) $ is, in the limit of infinite $W$-boson mass, described by the effective Lagrange density
\begin{equation}
\mathcal{L}_{\textrm{SM}} = \frac{G_F V_{ud}}{\sqrt{2}} \bar{e} \gamma_{\mu} (1-\gamma_5)\nu_e \;\bar{u}\gamma^\mu (1-\gamma_5)d + \textrm{h.c.}  \; ,
\end{equation}
where $G_{F}$ is the Fermi coupling constant, $V_{ud}$ is the $ud$ entry of the Cabibbo-Kobayashi-Maskawa (CKM) mixing matrix, and h.c. denotes the hermitian conjugate. We work in natural units, $\hbar=c=1$, and use $\gamma^5\equiv i \gamma^0\gamma^1\gamma^2\gamma^3$ and $\epsilon^{0123}=-\epsilon_{0123}=1$. 

At the nucleon level, all possible quark bilinears and their associated form factors need to be inserted \cite{Wei58}, such that
\begin{align}\label{eq:compva}
\langle p | \bar{u} \gamma_\mu d | n \rangle {}& =  \bar{p} \left[g_V ( q^2 ) \gamma_\mu+ \frac{g_M(q^2)}{M} \sigma_{\mu\nu} q^\nu + \frac{\tilde{g}_S(q^2)}{2M}q_\mu \right]n \ , \nonumber
\\ 
\langle p | \bar{u} \gamma_\mu \gamma_5 d | n \rangle {}& =  \bar{p} \left[g_A ( q^2 ) \gamma_\mu \gamma_5 + \frac{\tilde{g}_T(q^2)}{2M} \sigma_{\mu\nu}q^\nu\gamma_5 + \frac{\tilde{g}_P(q^2)}{2M}q^\mu\gamma_5\right] n  \ ,
\end{align}
where $q=p_n -p_p$ is the momentum transfer and $M$ is the nucleon mass. The vector form factor $g_V$ and the axial-vector form factor $g_A$ give the leading contributions to $\beta$ decay, because the nuclei can be treated nonrelativistically. In the isospin limit, the induced form factor $g_M$, called weak magnetism, is given by $(\mu_p-\mu_n)/2$, i.e. the difference between the magnetic moments of the proton and the neutron. Given the current experimental precision, this form factor can be neglected, but future experiments might reach a level of precision for which weak magnetism has to be taken into account, see Sec.~\ref{sec:efforts}. In the isospin limit the induced scalar form factor $\tilde{g}_S$ and tensor form factor $\tilde{g}_T$ vanish \cite{Wei58}, and we can neglect them at present. The induced pseudoscalar form factor $\tilde{g}_P$ gets an additional suppression of $q/M$, because of the $\bar{p}\gamma_5n$ structure. We comment on pseudoscalar couplings in Sec.~\ref{sec:pseudo}. 

The leading-order SM expression for neutron decay is  
\begin{equation}
\mathcal{L}_{\textrm{SM}} = \frac{G_F V_{ud}}{\sqrt{2}}g_V(q^2) \bar{e} \gamma_{\mu} (1-\gamma_5)\nu_e \bar{p}\gamma^\mu (1-\frac{|g_A(q^2)|}{g_V}\gamma_5)n + \textrm{h.c.}  \;
\end{equation}
In the limit of $q^2 \rightarrow 0$, the vector charge is $g_V(0)=1$, up to small corrections. This is dictated by the hypothesis of the conserved vector current (CVC). The axial-vector charge $g_A$ is only partially conserved (PCAC). The best current value is derived from neutron $\beta$-decay experiments, $|g_A|=1.2723(23)$ \cite{pdg2014}. 
 
In nuclear $\beta$ decay one can exploit the properties of the parent and daughter nucleus to select particular parts of the interaction. Pure Fermi (F) transitions probe the vector currents ($\gamma^\mu$), while pure Gamow-Teller (GT) transitions probe the axial-vector currents ($\gamma_5\gamma^\mu$). Mixed transitions always require knowledge of the Fermi and Gamow-Teller transition matrix elements, $M_{F}\equiv \left\langle f | 1 | i\right\rangle$ and $M_{GT}\equiv\left\langle f | \vec{\sigma} | i\right\rangle$, respectively. The conditions for spin change ($\Delta J$) and parity change ($\pi_i\pi_f$) for Fermi and Gamow-Teller transitions are given in Table \ref{tab:class}. This Table also lists for which aspect in SM and BSM research these transitions are used. We have defined the Fermi-Gamow-Teller mixing ration
\begin{equation}\label{eq:rho}
\rho\equiv g_A M_{GT}/g_V M_F \ ,
\end{equation} 
and
\begin{equation}\label{eq:lamb}
\lambda \equiv |g_A|/g_V \ .
\end{equation}
It is desirable to reduce the uncertainties of nuclear structure and select the simplest isotopes. For Fermi transitions the superallowed $0^+\rightarrow 0^+$ transitions are of most interest. For mixed transitions, mirror nuclei are preferred. For general mirror nuclei $\rho$ has to be measured, while neutron decay ($J^\pi=1/2^+ \rightarrow J^\pi=1/2^+$, $|M_F|^2=1$ and $|M_{GT}|^2=3$) allows for the determination of the value of $\lambda$ \cite{Abe08, Nic09, Dub11}. An elaborate compilation of neutron-decay amplitudes is given in \textcite{Iva13}. 

When searching for physics BSM, nuclei serve as ``micro-laboratories'' that can be judiciously chosen to look for certain manifestations of new physics. In this review, we address both the traditional searches for exotic couplings and the novel searches for Lorentz violation. In the latter, the possibility of angular-momentum violation needs to be considered, where the simplest of the forbidden decays, first-forbidden unique transitions, become relevant \cite{Noo13b}. Both fields search for BSM physics generated by an unknown fundamental theory at a high-energy scale. To study the effect of new physics at low energies, we work in an EFT approach. Within this framework the effects of new physics at low energies are described in a model-independent way with an effective Lagrangian of the form
\begin{equation}\label{eq:Lagrangian}
\mathcal{L}^{(\textrm{eff})} =\mathcal{L}_{\textrm{SM}} + \mathcal{L}_{\textrm{BSM}} \ .
\end{equation}
The search for exotic couplings focuses on right-handed vector, scalar, and tensor couplings. These non-SM interactions can be included in the Lagrangian by adding higher-dimensional operators to $\mathcal{L}_{\textrm{BSM}}$. The effects of Lorentz violation can also be described in an EFT framework \cite{kossme, Noo13a}. We discuss both frameworks separately.

\subsection{Exotic couplings}
In EFT, deviations from the $V-A$ structure due to exotic couplings are generated by higher-dimensional operators, which are suppressed by the high-energy scale $\Lambda$. 
The effective Lagrangian is parametrized as
\begin{equation}\label{eq:efflag}
\mathcal{L}^{(\textrm{eff})} =\mathcal{L}_{\textrm{SM}} + \frac{1}{\Lambda^k} \mathcal{L}^{(4+k)}  \ ,
\end{equation}
where
\begin{equation}\label{eq:lag6}
\mathcal{L}^{(4+k)} = \sum_i c_i \mathcal{O}_i^{(4+k)} \ ,
\end{equation}
and where the $c_i$ are dimensionless constants and $\mathcal{O}_i^{(4+k)}$ are dimension-$(4+k)$ operators. The SM only contains operators with mass dimension 3 or 4. For Lorentz-symmetric BSM physics, the lowest term we could add is $\mathcal{L}^{(5)}$. There is, however, only one dimension-5 operator, namely the operator that generates Majorana neutrino masses \cite{Wei79}. In searches for exotic couplings we assume the neutrino mass to be small, and therefore we neglect this operator. We focus only on $\mathcal{L}^{(6)}$, as even higher-dimensional terms are suppressed by additional powers of the large scale $\Lambda$. 

The $\mathcal{O}_i^{(6)}$ that contribute to semileptonic charged decays are listed in \textcite{Cir10, Cir13}. At low energies these dimension-6 operators generate the original vector $(C_V)$, axial-vector $(C_A)$, scalar $(C_S)$, pseudoscalar $(C_P)$, and tensor $(C_T)$ couplings of \textcite{Lee56}. At the quark level, the effective Lagrangian for $\beta$ decay, with non-derivative four-fermion couplings, is\footnote{We follow \textcite{Her01}, except for a factor $G_F V_{ud}/\sqrt{2}$ that we have extracted.}
\begin{eqnarray}\label{eq:lageff}
\nonumber\mathcal{L}^{(\textrm{eff})}  & = &\frac{4G_{F}V_{ud}}{\sqrt{2}} \sum_{\epsilon,\delta = L, R} \left\{\frac{}{}a_{\epsilon\delta}\:\bar{e} \gamma^{\mu}\nu^{\epsilon}_e\cdot\bar{u}\gamma_\mu d_{\delta}\right. \\
&& \left. + A_{\epsilon\delta}\; \bar{e} \nu^{\epsilon}_e\cdot\bar{u}d_{\delta} + \alpha_{\epsilon}\: \bar{e}\frac{\sigma^{\mu\nu}}{\sqrt{2}}\nu^{\epsilon}_e\cdot \bar{u} \frac{\sigma_{\mu\nu}}{\sqrt{2}} d_{\epsilon} \right\}\ ,
\end{eqnarray}
where we sum over the chirality ($L$, $R$) of the final states. 

The coefficients represent
\begin{itemize}
	\item $a_{\epsilon\delta}$: all possible $V$ and $A$ couplings,
	\item $A_{\epsilon\delta}$: exotic scalar/pseudoscalar couplings (where $\epsilon$ denotes the chirality of the neutrino and $\delta$ the chirality of the $d$ quark),
	\item $\alpha_{\epsilon}$: exotic tensor couplings (where $\epsilon$ denotes the chirality of both the neutrino and the $d$ quark).
\end{itemize}
These coefficients are related to the couplings $C_i$ and $C'_i$ $(i = S, V, A, T, P)$ of \textcite{Lee56} by Eqs.~\eqref{eq:Clist} and \eqref{eq:list} of Appendix \ref{sec:app}. 
In the SM all couplings except $a_{LL}=1$ are zero. For tensor couplings, only $\alpha_{L}$ and $\alpha_{R}$ occur, since $\sigma_{\mu\nu}\gamma_5 = \frac{i}{2}\epsilon_{\mu\nu\alpha\beta}\sigma^{\alpha\beta}$. The constants $a_{\epsilon\delta}$, $A_{\epsilon\delta}$, and $\alpha_{\epsilon}$ can be related to $c_i$, by matching their values at the low-energy scale with standard EFT techniques. 
The chiral structure of the coefficients is expressed by the first and second index, which denote the chirality of the neutrino and the $d$-quark, respectively. All couplings with first index $R$ involve a right-handed neutrino. In the SM, right-handed neutrinos are absent, but they are present in many new-physics models. The role of the right-handed neutrino is discussed in Sec.~\ref{sec:neutrino}. The new exotic couplings can be complex, representing the possibility of time-reversal (T) violation (Sec.~\ref{sec:time}). The introduction of left-handed and right-handed couplings leads to parity violation when the coefficients differ. In the absence of right-handed couplings, parity violation is maximal. 

To describe $\beta$ decay of the nucleon we define the hadronic matrix elements \cite{Her01}
\begin{subequations}
\begin{eqnarray}\label{eq:matrixel}
	\langle p | \bar{u} \gamma_\mu d | n \rangle & = & g_V ( q^2 ) \bar{p} \gamma_\mu n \ ,
	\\
	\langle p | \bar{u} \gamma_\mu \gamma_5 d | n \rangle & = & g_A ( q^2 ) \bar{p} \gamma_\mu \gamma_5 n  \ ,
	\\
	\langle p | \bar{u} d | n \rangle & = & g_S ( q^2 ) \bar{p} n  \ ,
	\\
		\langle p | \bar{u} \gamma_5 d | n \rangle & = & g_P ( q^2 ) \bar{p}\gamma_5 n  \ ,
		\\
	\langle p | \bar{u} \sigma_{\mu \nu} d | n \rangle & = & g_T ( q^2 ) \bar{p} \sigma_{\mu \nu} n  \ ,
\end{eqnarray}
\end{subequations}
modifying the effective Lagrangian in Eq.~\eqref{eq:lageff} accordingly. As before, the vector charge is $g_V\equiv g_V(0)=1$. The other couplings, $g_A, g_S, g_P,$ and $g_T$ can be calculated theoretically by using lattice QCD. Estimates for $g_A$ on the lattice are currently not competitive with the experimental value $|g_A(0)|=1.2723(23)$ determined from neutron $\beta$ decay \cite{pdg2014}. The scalar, pseudoscalar, and tensor constants, $g_{S}, g_{P}$, and $g_T$, are determined theoretically. They are further discussed in Sec.~\ref{sec:constraints}. 

Searches for exotic coupling also include searches for right-handed $V+A$ currents. Such currents are predicted for instance by left-right (LR) models, which add an $SU(2)_R$ gauge symmetry to the SM. This extends the SM with an additional gauge boson $W_R$, which mixes with the original SM $W$ boson $W_L$. The weak eigenstates can be expressed in the mass eigenstates $W_1$ and $W_2$ as
\begin{subequations}\label{eq:wr}
\begin{align}
W_L {}&= W_1 \cos\xi + W_2 \sin\xi \ , \\
W_R {}& = e^{i\omega} (-W_1 \sin\xi + W_2 \cos\xi) \ ,
\end{align}
\end{subequations}
where $\xi$ is the mixing angle and $\omega$ a CP-violating phase. The coupling of $W_R$ to quarks and leptons introduces the right-handed coupling $g_R$ and the right-handed CKM element $V_{ud}^R$, the equivalents of the SM parameters. The expressions for $a_{LR}, a_{RL},$ and $a_{RR}$ in terms of these parameters are given in \textcite{Her01}. A specific class of LR models are the symmetric LR models, in which P or C symmetry of the Lagrangian is imposed, which implies $g_L=g_R$. We focus on bounds for such models in Sec.~\ref{sec:LHC}.

\subsection{Lorentz violation}\label{sec:introLIV}
The study of Lorentz violation is motivated by the possibility of spontaneous breaking of Lorentz invariance predicted by theories of quantum gravity \cite{Kos89, Lib09, Lib13}. The natural energy scale for these theories of quantum gravity is the Planck scale, which lies 17 orders of magnitude higher than the electroweak scale. This precludes the direct detection of Planck-scale physics, but the effects of Lorentz violation at the Planck scale can become manifest at much lower energies, providing a ``window on quantum gravity.''   
At low energy, Lorentz violation can be systematically described by the Standard Model Extension (SME) \cite{kossme}, by using an EFT approach. The SME contains all possible Lorentz-violating terms that obey the SM gauge symmetries, which include CPT-violating terms, since Lorentz violation allows for the breaking of CPT invariance. In fact, CPT violation can only occur if Lorentz symmetry is also broken \cite{Gre02}. 

Spontaneous Lorentz violation arises as Lorentz-tensor fields acquire a vacuum-expectation value (VEV), resulting in Lorentz-violating tensor coefficients in the SME Lagrangian. These coefficients can be understood as constant background tensor fields. Due to these tensor fields, the Lagrangian is no longer invariant under particle or active Lorentz transformations, i.e. boosts or rotations of the particles, because the background fields do not transform under the Lorentz group \cite{kossme}. However, the low-energy theory remains invariant under observer Lorentz transformations,  i.e. boosts or rotations of the observer's inertial frame. Because Lorentz symmetry is spontaneously broken, the underlying fundamental theory at the Planck scale remains Lorentz invariant, implying that important features such as energy-momentum conservation and microcausality are still valid. 
A possible experimental signature of Lorentz violation is a sidereal variation of observables, which arise as the laboratory moves through the Lorentz-violating background field when Earth rotates (other examples are given in {\it e.g.} \textcite{Mat05}). 


Schematically, terms in $\mathcal{L}_{\textrm{BSM}}$ in Eq.~\eqref{eq:Lagrangian} can be written as \cite{Col96}
\begin{equation}\label{eq:LNPLIV}
\mathcal{L}_{\textrm{NP}} = \lambda^{(3)} \left\langle T \right\rangle\cdot \bar{\psi}\Gamma \psi +  \frac{\lambda^{(4)}}{\Lambda}\left\langle T \right\rangle\cdot \bar{\psi}\Gamma (i\partial) \psi + \frac{\lambda^{(4+k)}}{\Lambda^k} \left\langle T \right\rangle\cdot\mathcal{O}^{(4+k)}  \ , 
\end{equation}
where we summed over repeated indices and where $\lambda^{(i)}$ are dimensionless constants, $ \left\langle T \right\rangle$ is the expectation value of tensor $T$, $\Gamma = 1, \gamma_5, \gamma_{\mu}, \gamma_{\mu}\gamma_5, \sigma_{\mu\nu}$ represents the gamma-matrix structure, and $\mathcal{O}^{(4+k)}$ are higher-dimensional operators. Furthermore, $\Lambda$ represent the scale of the fundamental theory, which would naturally be the Planck scale. The higher-dimensional operators are suppressed by powers of this high scale. The first two terms in Eq.~\eqref{eq:LNPLIV} have mass dimension 3 and 4, respectively. These terms are described in the original SME papers by \textcite{kossme} and are now referred to as the minimal Standard-Model Extension (mSME). For our present discussion we limit ourselves to the mSME, although higher-dimensional coefficients have also been described (\textcite{kos13, neutrino2012, kos09, Bol07}).

From an EFT point of view, the introduced Lorentz-violating dimension-3 and dimension-4 operators are unnatural. Naively, one would expect the dimension-3 operators to scale linearly with the large scale $\Lambda$, while the coefficients of the dimension-4 operators should be of order unity. The experimental bounds on these dimension-3 and dimension-4 operators are much smaller, of course. This problem does not occur for higher-dimensional operators, which are naturally suppressed by the scale $\Lambda$. 
To evade these naturalness problems, the current limits on dimension-3 and -4 coefficients require either large fine-tuning, or a symmetry that forbids these coefficients. However, even if dimension-3 and -4 operators are forbidden at tree level, they will be induced by quantum corrections generated by higher-dimensional non-renormalizable operators. These corrections scale quadratically with the cutoff scale, which might be as large as $\Lambda$. This can be circumvented by introducing new physics between the weak scale and the Planck scale. In that case, radiative corrections scale with a significantly lower cutoff scale (see $\textit{e.g}$ \textcite{Mat08}). Such a scenario occurs in supersymmetry (SUSY) \cite{Bol05,Gro05}. SUSY restricts Lorentz-violating operators to dimension 5 and higher, and forbids those of dimension 3 and 4. 
Dimension-3 and dimension-4 operators are generated by loop corrections if SUSY is broken. This would naturally lead to a suppression of $m^2/\Lambda$ and $m/\Lambda$ for dimension-3 and dimension-4 operators, respectively, where $m$ is the SUSY-breaking scale \cite{Bol05,Gro05}. In the mSME, it is assumed that dimension-3 and dimension-4 operators are suppressed by some unspecified higher-scale mechanism, and the experimental constraints are studied without any assumptions on the nature of this suppression mechanism \cite{kossme, kospara}. 

The SME contains a large number of coefficients that parametrize possible Lorentz violation. We list the relevant coefficients for $\beta$ decay, which are the lepton, Higgs, and gauge terms. The Lorentz-violating terms for leptons are \cite{kossme}
\begin{eqnarray}\label{eq:leptonsme}
\mathcal{L}_{\textrm{lepton}}  & = & \bar{L}_A\left[i(c^{\textrm{LV}}_L)_{\mu\nu AB}\gamma^{\mu}D^\nu-(a^{\textrm{LV}}_L)_{\mu AB}\gamma^{\mu}\right]L_B \nonumber  \\
&&+ \bar{R}_A\left[i(c^{\textrm{LV}}_R)_{\mu\nu AB} \gamma^{\mu} D^{\nu} - (a^{\textrm{LV}}_R)_{\mu AB}  \gamma^{\mu} \right]R_B  \ , 
\end{eqnarray}
where $L$ denotes the $SU(2)_L$ doublet and $R$ denotes the singlet, defined in Eq.~\eqref{eq:LandR}. The subscripts $A,B$ are flavor indices, and $D_{\mu}$ is the covariant derivative. This introduces the Lorentz-violating coefficients $a^{\textrm{LV}}_{L,R}$ and $c^{\textrm{LV}}_{L,R}$, which are CPT-odd and CPT-even, respectively. We have introduced the superscript LV for these coefficients, in order not to confuse them with the coefficients in Eq.~\eqref{eq:lageff}. 

Before electroweak symmetry breaking, the Higgs and gauge sector are described by \cite{kossme}
\begin{eqnarray}\label{eq:smegauge}
\mathcal{L}_{\textrm{Higgs+gauge}} & = &\left[\frac{1}{2}k_{\phi\phi}^{\mu\nu}(D_{\mu}\phi)^{\dagger} D_{\nu}\phi + \textrm{H.c.}\right] + \left[i\; k_{\phi}^{\mu} \phi^{\dagger} D_{\mu} \phi + \textrm{H.c.} \right] \nonumber \\
&&  -\frac{1}{2} k_{\phi B}^{\mu\nu}\phi^{\dagger}\phi B_{\mu\nu}-\frac{1}{2} k_{\phi W }^{\mu\nu} \phi^{\dagger}W_{\mu\nu} \phi   -\frac{1}{2} (k_G)_{\kappa\lambda\mu\nu} \textrm{Tr}(G^{\kappa\lambda}G^{\mu\nu})\nonumber \\
&&-\frac{1}{2} (k_W)_{\kappa\lambda\mu\nu} \textrm{Tr}(W^{\kappa\lambda}W^{\mu\nu}) -\frac{1}{4} (k_B)_{\kappa\lambda\mu\nu} B^{\kappa\lambda}B^{\mu\nu} \ ,
\end{eqnarray}
where $G^{\mu\nu}, W^{\mu\nu}$, and $B^{\mu\nu}$ are the $SU(3)_c, SU(2)_L,$ and $U(1)_Y$ field-strength tensors, respectively, and $\phi$ is the Higgs doublet. The coefficient $k_{\phi}$ is CPT-odd, and the only coefficient with dimension of mass. The other coefficients are CPT-even and dimensionless. The coefficient $k_{\phi\phi}$ has symmetric real and antisymmetric imaginary components. The $k_{\phi W}$ and $k_{\phi B}$ coefficients are real and antisymmetric. The gauge couplings $k_G, k_W,$ and $k_B$ are real and have the symmetry properties of the Riemann tensor \cite{kossme}. 

The SME parameters have been studied in a wide range of experiments \cite{kospara}. The electromagnetic and gravity sector have been studied extensively, whereas the number of searches in the weak interaction is rather low. This changed recently \cite{Noo13a, Mul13, Noo13b}, and the search for Lorentz violation has been extended to weak decays, in particular $\beta$ decay. $\beta$ decay places strong constraints on Lorentz-violating coefficients in the Higgs and gauge sector. In addition, $\beta$ decay has a unique sensitivity to some coefficients in the neutrino sector \cite{Dia13}. We discuss these constraints in Sec.~\ref{sec:LIV}. 

\section{Observables in $\beta$ decay}\label{sec:obs}
%
\subsection{Correlation coefficients in $\beta$ decay}
In $\beta$ decay, the correlations between different observables, such as the $\beta$ momentum and the nuclear spin, can be measured. The amount of correlation is expressed in terms of correlation coefficients.  These correlation coefficients depend on SM couplings and possible new $V$, $A$, $S$, $P$, and $T$ interactions. Using the general effective Lagrangian in Eq.~\eqref{eq:lageff}, we can write the decay-rate distribution for polarized nuclei as \cite{Jac57}
\begin{align}\label{eq:decayrate}
    \omega(\langle {}& {\vec{J}} \rangle|
    E_e, \Omega_e,\Omega_\nu)dE_e d\Omega_e d\Omega_\nu  \nonumber \\
   = {}&
   \frac{F(\pm Z,E_e)}{(2\pi)^5}p_e E_e(E_0 - E_e)^2 dE_e d\Omega_e d\Omega_\nu\ \nonumber
    \\  &\times     \bar{\xi} \left\{ 1 + a \frac{\vec{p}_e\cdot \vec{p}_{\nu}}{E_e E_\nu} + b \frac{m_e}{E_e} + c\left [\frac{1}{3}\frac{\vec{p}_e\cdot\vec{p}_{\nu}}{E_eE_\nu} -
   \frac{(\vec{p}_e\cdot\vec{ j})(\vec{p}_{\nu}\cdot\vec{j})}{E_eE_\nu}\right ]
   \left[\frac{J(J+1)-3\langle(\vec{ J}\cdot\vec{ j})^2\rangle}{J(2J-1)}\right] \right. \nonumber \\
   &\left. + \frac{\langle {\vec{J}}\rangle}{J}\cdot \left[ A\,\frac{\vec{p}_e}{E_e} + B\,\frac{\vec{p}_{\nu}}{E_\nu} +
   D\,\frac{\vec{p}_e\times\vec{p}_{\nu}}{E_eE_\nu}\right]
   \right\} \ ,
\end{align}
where $E_{e(\nu)}$, $\Omega_{e(\nu)},$  and $p_{e(\nu)}$ denote the total $\beta(\nu)$ energy, direction, and momentum, respectively, $E_0$ is the energy available to the electron and the neutrino, $\langle {\vec{J}}\rangle$ is the expectation value of the spin of the initial nuclear state, and $\vec{j}$ is the unit vector in this direction; $F(\pm Z,E_e)$ is the Fermi function which modifies the phase space of the electron due to the Coulomb field of the nucleus. Also affecting the phase space is the Fierz interference term, factorized with the coefficient $b$. This term is zero in the SM. We defined $\bar{\xi}\equiv G_F^2 V_{ud}^2/2 \xi$, where $\xi$ gives the strength of the interaction.  The remaining terms describe the $\beta$-correlation coefficients: the $\beta$-neutrino asymmetry $a$,  the P-odd ``Wu-parameter,'' the $\beta$-asymmetry $A$, the neutrino asymmetry $B$, and the triple-correlation coefficient $D$. The $c$ coefficient vanishes for non-oriented nuclei and for nuclei with  $J=1/2$, such as the neutron. The $c$ coefficient has not been taken into account in any experiment to date. However, in future experiments, which use laser beams to trap and cool samples, the expectation value $\langle(\vec{ J}\cdot\vec{ j})^2\rangle$ may be affected, such that the $c$ coefficient can play a role. 

\begin{table}
\begin{center}
    \begin{tabular}{ | l | l | l | l |} 
    \hline\hline
    Coefficient & Correlation & P & T \\ \hline
     $a$ ($\beta\nu$ angular correlation) & $\vec{p}_e\cdot\vec{p}_{\nu}/E_eE_{\nu}$ & Even & Even \\ \hline
       $b$ (Fierz interference term) & $m_e/E_e$ & Even & Even \\ \hline
  $A$ ($\beta$ asymmetry) & $\vec{J}\cdot\vec{p}_e/E_e$ & Odd & Even\\ \hline
     $B$ ($\nu$ asymmetry) & $\vec{J}\cdot\vec{p}_{\nu}/E_{\nu}$ & Odd & Even\\ \hline
		  $G$ (Longitudinal polarization) &$\vec{\sigma}_e \cdot\vec{p}_e/E_{e}$ & Odd & Even\\ \hline
		  $N$  & $\vec{J}\cdot \vec{\sigma}_e $ & Even & Even\\ \hline
			$Q$  & $\vec{\sigma}_e\cdot \vec{p}_e \vec{J}\cdot\vec{p}_{e}/E_{e}$ & Even & Even\\ \hline
      $D$ (triple correlation) & $\vec{J}\cdot(\vec{p}_e\times\vec{p}_{\nu})/E_eE_{\nu}$ & Even & Odd\\ \hline
      $R$ (triple correlation) & $\vec{\sigma}_e\cdot(\vec{J}\times\vec{p}_e)/E_e$ & Odd & Odd\\ \hline
    \hline
    \end{tabular}
    \caption{Overview of symmetry properties under parity (P) transformations and time reversal (T) of the most relevant correlation coefficients in allowed $\beta$ decay. }
		\label{tab:sym}
\end{center}
\end{table}

The decay rate integrated over neutrino direction, but taking into account electron polarization, is \cite{Jac57} 
\begin{align}\label{eq:withr}
    \omega(\langle {\vec{J}} \rangle , \vec{\sigma}_e |
    E_e, \Omega_e)dE_e d\Omega_e  ={}&\frac{F(\pm Z,E_e)}{(2\pi)^4}p_e E_e(E_0 - E_e)^2 dE_e d\Omega_e \nonumber \\  
    {}& \times     \bar{\xi} \left\{ 1 + b\frac{m_e}{E_e} + \frac{\vec{p}_e}{E_e}\cdot\left(A\frac{\langle {\vec{J}}\rangle}{J}+G\vec{\sigma}_e\right) \right. \nonumber\\ 
 {}& \left.  +\vec{\sigma}_e\cdot\left[N\frac{\langle {\vec{J}} \rangle}{J}+Q\frac{\vec{p}_e}{E_e+m}\left(\frac{\langle {\vec{J}} \rangle}{J}\cdot\frac{\vec{p}_e}{E_e}\right)+R\frac{\langle {\vec{J}}\rangle}{J}\times \frac{\vec{p}_e}{E_e}\right]\right\} \ ,    
\end{align}
where $\vec\sigma_e$ is the spin vector of the $\beta$ particle. This introduces the longitudinal $\beta$ polarization $G$, the spin-correlation coefficients $N$ and $Q$, and the triple-correlation coefficient $R$. The symmetry properties of the correlation coefficients are listed in Table~\ref{tab:sym}. The $A$, $B$, and $G$ coefficients are associated with parity violation. Depending on the type of transition they can have SM values close to $\pm 1$, which is characteristic for maximal parity violation. The triple-correlation coefficients $D$ and $R$ are T-odd and unmeasurably small in the SM \cite{PhysRevD.56.80}. 

Integrating the decay rate over all kinematical variables gives the inverse lifetime,
\begin{equation}\label{eq:Hardybasic}
\frac{1}{\tau}=\frac{m_e^5}{2\pi^3}\mathit{f}\bar{\xi}\left( 1+b\left\langle \frac{m_e}{E_e}\right\rangle\right) \ ,
\end{equation}
where $\mathit{f}$ contains the integration over the modified phase space and $\langle m_e/E_e\rangle$ is the average inverse energy in units of the electron mass.

In Appendix \ref{sec:app} we list the relevant correlation coefficients in terms of the couplings defined in Eq.~\eqref{eq:lageff} and the Fermi/Gamow-Teller matrix elements. 
The different correlation coefficients contain combinations of the complex $V$, $A$, $S$, $P$, and $T$ couplings. Given the current experimental precision, we have neglected Coulomb corrections. These corrections mainly introduce additional imaginary couplings (except for the $D$ and $R$ coefficients) \cite{Jac57}. 

We proceed by discussing how $\beta$-decay correlation experiments, combined with lifetime measurements, are used to obtain precise values for the SM $V$ and $A$ coupling strengths. In Sec.~\ref{sec:constraints} we discuss constraints on exotic couplings.
\subsection{Standard Model parameters in $\beta$ decay}\label{sec:smpara}
The correlation coefficients in Appendix \ref{sec:app} reduce to the SM expressions when putting the scalar and tensor couplings to zero, $A_{LL, LR, RR, RL}=0$ and $\alpha_{L(R)}=0$, and by using only $V-A$ couplings, $a_{LR, RR, RL}=0$. The Fierz-interference coefficient $b$ is zero in the SM. 
The lifetime in Eq.~\eqref{eq:Hardybasic} can be derived from the $\mathit{f}t$ value, using the measured half-life $t$ instead of $\tau$. In the SM, 

\begin{equation}\label{eq:Hardyfull}
\frac{1}{ft}=\frac{m_e^5}{2\pi^3 \ln(2)} G_F^2\; V_{ud}^2\; g_V^2|M_F|^2(1+|\rho|^2) \ .
\end{equation}
The SM value for $G_F$ is obtained from muon decay \cite{Web10}. It is important to note that if one considers non-SM contributions these may influence muon decay as well.  
In principle, $g_A$ is calculable using lattice QCD, but as mentioned before, current lattice calculations are not as accurate as values derived from experiments and henceforth $\lambda=|g_A|/g_V$ is considered a free parameter. In general, $M_F$ and $M_{GT}$ need to be derived from nuclear model calculations. For superallowed Fermi transitions $\rho=0$ and $M_{F}=\sqrt{2}$, in the isospin limit. \textcite{hardytowner2009} analyzed all available superallowed Fermi transitions, and derived a value for the $ud$ CKM matrix element. Since the $ft$ values of superallowed transitions should be equal, a large number of measurements could be combined, leading to the most precise value of $V_{ud} = 0.97425(22)$ \cite{hardytowner2009}. In the analysis, details of the isotope-dependent nuclear-structure corrections on the matrix element $M_F$ (e.g. isospin breaking) and the phase-space modifications are also considered. The superallowed transitions also give the best bound on the Fierz coefficient $b$ in Eq.~\eqref{eq:Hardybasic} by considering the energy dependence of the lifetime (Sec.~\ref{sec:scalar}). 

The parameters $\lambda$ and $V_{ud}$ can also be determined from $\beta$-decay correlations in neutron decay and from the neutron lifetime \cite{Abe08,Nic09,Dub11,Wie11}. The best current values are $\lambda=1.2723(23)$ \cite{pdg2014} and $V_{ud}=0.9742(12)$ \cite{Dub11}. The latter is more than five times less precise, see also Fig. 22 in \textcite{Dub11}. The strong Gamow-Teller dependence of neutron decay and the precision of the neutron-decay parameters is such that neutron decay also plays an important role in searches for tensor currents, as we will discuss in Sec.~\ref{sec:neutron}.

Another class of nuclei for which the nuclear structure is relatively well known are the mirror nuclei \cite{Severijns:2008ep}. Like neutron decay, mirror decays are mixed Fermi-Gamow-Teller transitions. Extraction of $V_{ud}$ from lifetime measurements requires knowledge of the mixing parameter $\rho$, such that an additional measurement of at least one of the correlation coefficients is necessary. \textcite{NaviliatCuncic:2008xt} find $V_{ud} =0.9719(17)$, using 5 available transitions. The important structure corrections to Eq.~\eqref{eq:Hardyfull} for mirror nuclei have been evaluated \cite{Severijns:2008ep}, in analogy to the work of \textcite{hardytowner2009} for superallowed Fermi decays. 
This new class of nuclei will broaden the spectrum of data and remove any possible bias in selecting only superallowed Fermi transitions in the determination of $V_{ud}$. Measurements with this motivation were undertaken. For example, \textcite{Shi14}, \textcite{Bro13}, and \textcite{Tri12} have measured the lifetime of two relevant mirror nuclei, $^{19}$Ne and $^{37}$K. We will not review the status of this field here, but comment on their relevance in limiting left-handed tensor couplings via the Fierz-interference term in the next section. It demonstrates that the contribution of nuclear physics to high-precision SM data goes hand in hand with the searches for new physics in $\beta$ decay.

\section{Constraints on exotic couplings}\label{sec:constraints}
$\beta$ decay played an important role in establishing the $V-A$ structure of the SM, initially eliminating to a large extent the possible contributions of scalar and tensor interactions. Modern searches in nuclear $\beta$ decay consider again scalar and tensor currents as possible very small deviations from the SM due to new physics (see e.g. \textcite{Severijnstest, symmetrytest}). 

The searches in $\beta$ decay are part of a much wider search in subatomic physics for new physics. Comparison between different searches has become possible in an EFT framework by using the effective Lagrangian in Eq.~\eqref{eq:lageff}. At the quark level the relations between different observables are clean, but at the nucleon level they involve the nuclear form factors $g_A, g_S, g_P,$ and $g_T$. Accurate values for these parameters are necessary in order to compare different limits. Recently, significant progress on the accuracy of both $g_S$ and $g_T$ has been reported. First results for $g_P$ are also available. The most precise value for $g_T$ is calculated with lattice QCD. Two recent results are from \textcite{Gre12}, $g_T=1.038(16)$, and \textcite{Bha14}, $g_T=1.047(61)$.

The calculation method used in these works gives a much larger uncertainty for $g_S$. Estimates range from $g_S= 0.72(32)$ \cite{Bha14} to $g_S=1.08(32)$ \cite{Gre12}. A value for $g_S$ can also be derived using the CVC relation and lattice calculations \cite{Gon13}, 
\begin{equation}\label{eq:gs}
g_S(0) = \frac{\delta M^{\textrm{QCD}}}{\delta m_q} = 1.02(11) \ ,
\end{equation}
where both $\delta M^{\textrm{QCD}}=(M_n-M_p)^{\textrm{QCD}}$ \cite{Gon13} and $\delta m_q = m_d-m_u$ \cite{FLAG} are obtained separately from lattice calculations. However, the determination of $g_S$ with Eq.~\eqref{eq:gs} might underestimate the error, because correlations between the numerator and denominator are neglected. Such errors could be avoided by calculating the ratio in Eq.~\eqref{eq:gs} directly on the lattice. Further efforts to reduce the error for $g_S$ directly on the lattice are being pursued \cite{Bha14,Bha12}. 

The pseudoscalar constant $g_P$ can be calculated by using the PCAC relation. Combined with lattice QCD results \cite{Gon13} one finds
\begin{equation}\label{eq:gp}
g_P(0) = \frac{\bar{M}_N}{\bar{m}_q}g_A =349(9)\ ,
\end{equation}
where $\bar{M}=(M_p+M_n)/2$ is the average nucleon mass and $\bar{m}_q=(m_u+m_d)/2=3.42(9)$ MeV is the average light-quark mass determined on the lattice \cite{FLAG}. According to the PDG, $\bar{m}_q=3.5^{+0.7}_{-0.2}$ MeV \cite{pdg2012}, which gives a much larger error, $g_P=340^{+68}_{-19}$. 
Nevertheless, this shows that the pseudoscalar form factor is of order $\mathcal{O}(10^{2})$. In $\beta$ decay, pseudoscalar terms are generally neglected, because they only occur as higher-order recoil corrections. This surpresses pseudoscalar interactions compared to scalar and tensor interactions. The large value of $g_P$ cancels this suppression to a large extent, and $\beta$-decay experiments may be sensitive to pseudoscalar couplings after all. There are, however, already strong constraints on pseudoscalar couplings from pion decay, as we discuss in Sec.~\ref{sec:pseudo}.

In the remainder of this Section we comment on searches for exotic couplings in $\beta$ decay (Sec.~\ref{sec:betabounds}), but considering only real couplings. We compare these results with constraints from the Large Hadron Collider (LHC) experiments (Sec.~\ref{sec:LHC}) and due to the nonzero mass of the neutrino (Sec.~\ref{sec:neutrino}). 
Bounds on 
imaginary couplings are discussed separately in Sec.~\ref{sec:time}.

\subsection{Constraints from $\beta$ decay}\label{sec:betabounds}
In nuclear $\beta$ decays, exotic couplings are mainly searched for in either pure Fermi or pure Gamow-Teller decays. Pure Fermi transitions depend on vector and possibly scalar couplings, while pure Gamow-Teller transitions depend on  axial-vector  and possibly tensor couplings. The use of mixed transitions is necessary when searching for interference terms. Preferred are isotopes with a relatively simple nuclear structure,  e.g. mirror nuclei, or the neutron. We discuss the constraints from Fermi, Gamow-Teller, and mixed decays separately, focusing on the best current experimental data. We discuss the constraints on scalar and tensor couplings, while assuming no additional vector or axial-vector interactions. For a fit of the data including these interactions we refer to \textcite{Severijnstest}, where also a review of the experimental techniques is given. We discuss $V+A$ couplings in Sec.~\ref{sec:LHC}. 

Most $\beta$-correlation coefficients are measured by constructing asymmetry ratios. For example, the $\beta$ asymmetry is measured from the quantity
\begin{equation}\label{eq:asymmetry}
A_{\textrm{measured}} =\frac{N(\uparrow)-N(\downarrow)}{N(\uparrow)+N(\downarrow)} \ ,
\end{equation}
where $N(\uparrow)$ and $N(\downarrow)$ are the decay rates derived from measuring $\beta$ particles in a particular detector while the polarization, $P$, of the nucleus changes sign. The arrows indicate the direction of polarization. The rates $N(\uparrow), N(\downarrow)$ correspond to the integration of Eq.~\eqref{eq:decayrate} over all unobserved degrees of freedom, which removes the dependence on the neutrino direction. In the numerator only the P-odd term remains, while in the denominator the odd term drops out. However, the Fierz interference term remains in the sum ${N(\uparrow)+N(\downarrow)}$, so that 
\begin{eqnarray}\label{eq:Ameasured}
A_{\textrm{measured}}&=&\frac{\int_{\Delta\Omega}\int_{E_\textrm{min}}^{E_0}F(\pm Z, E_e)p_e (E_0 - E_e)^2 A|P|(p_e/E_e)\cos\theta_e dE_e d\Omega_e}{\int_{\Delta\Omega}\int_{E_\textrm{min}}^{E_0}F(\pm Z, E_e)p_e (E_0 - E_e)^2 (1+b/E_e) dE_e d\Omega_e}\nonumber\\
&\ &\nonumber\\ 
&=&\frac{ A|P|\langle\beta_e \cos\theta_e \rangle}{1+b\left\langle  \frac{m_e}{E_e}\right\rangle}  \ .
\end{eqnarray}
This implies that actually not the coefficient $A$ is measured, but
\begin{equation}\label{eq:tilde}
\tilde{A} = \frac{A}{1+b\left\langle \frac{m_e}{E_e}\right\rangle} \ .
\end{equation}
 The inverse average energy is approximated by 
\begin{equation}\label{eq:energydep}
\left\langle \frac{m_e}{E_e}\right\rangle = \frac{\int^{E_0}_{E_{\textrm{min}}}  F(\pm Z, E_e)p_e (E_0 - E_e)^2 dE_e}{\int^{E_0}_{E_{\textrm{min}}}  F(\pm Z, E_e)p_e (E_0 - E_e)^2 E_e dE_e} \ ,
\end{equation}
which depends on the specific isotope and the experimental setup. In principle, the average energy could also depend on the angular distribution $(\theta_e)$. This makes it preferable that the analysis of $\left\langle m_e/E_e\right\rangle$ is done and published together with the observed correlation coefficients. At present, many of the values for $\left\langle m_e/E_e\right\rangle$ are derived by using the $\beta$-energy threshold $E_{min}$ \cite{Severijnstest, Pat13, Wau13}. 

For the measured quantity $\tilde{X}$, $X= a, A, B, G,$ etc., Eq.~\eqref{eq:tilde} applies. Except for $B$ and $N$, the numerator of Eq.~\eqref{eq:tilde} depends only on the square of the coupling constants, while $b$ has a linear dependence on left-handed couplings. In such cases one is most sensitive to $b$, and the measurement of $\tilde{X}$ provides in the first place a measurement of the Fierz coefficient $b$. Therefore, the exact value of the $\left\langle m_e/E_e\right\rangle$ will become increasingly important with increasing experimental precision.




\subsubsection{Nuclear scalar searches}\label{sec:scalar}
Throughout the discussion of limits on scalar and tensor couplings, we will assume conventional left-handed vector couplings for the $V$-$A$ part, such that $a_{LL}=1$, and $a_{LR, RL, RR}=0$.  These and the other couplings are defined in Eq.~\eqref{eq:lageff}. The notation is chosen such that the difference between the left-handed and right-handed coupling of the neutrino is emphasized, i.e. for the scalar couplings $A_L=A_{LL} + A_{LR}$ (left-handed neutrino coupling) and  $A_R=A_{RR} + A_{RL}$ (right-handed neutrino coupling). Further details on the notation and some relevant expressions can be found in Appendix \ref{sec:app}. 

For pure Fermi transitions
\begin{eqnarray}
\xi&=&2|M_F|^2g_V^2\left\{1+\left(\frac{g_S}{g_V}\right)^2\left[A_L^2+A_R^2\right]\right\} \ ,\\
\xi b_F &=& \pm 4\gamma |M_F|^2g_V g_S A_L \ ,
\end{eqnarray}
from Eq.~\eqref{eq:xibar} and Eq.~\eqref{eq:bxi}, where $b_F$ is the Fermi part of the Fierz coefficient $b$, the upper (lower) sign is for $\beta^- (\beta^+)$ decays and $\gamma=\sqrt{1-Z^2\alpha^2}$, with $Z$ the atomic number of the daughter nucleus and $\alpha$ the fine-structure constant. For the positron-emitting superallowed $0^+\rightarrow 0^+$ Fermi decays
\begin{equation}\label{eq:ftwithscalar}
\frac{1}{ft_F}=\frac{m_e^5}{2\pi^3\ln(2)}G_F^2\;V_{ud}^2\;g_V^2 \;|M_F|^2\left\{1+\left(\frac{g_S}{g_V}\right)^2\left[A_L^2+A_R^2\right]-2\gamma\left\langle \frac{m_e}{E_e}\right\rangle\frac{g_S}{g_V}A_L\right\} \ .
\end{equation}
\textcite{hardytowner2009} obtained an average of all $ft$ values, $\overline{\mathcal{F}t}$, after the appropriate corrections for radiative and nuclear-structure effects. The current best value of $V_{ud}$ is derived from $\overline{\mathcal{F}t}$, assuming no exotic couplings. 
Allowing for scalar terms one can exploit  \cite{hardytowner2005C} the different values of $\langle m_e/E_e \rangle$ to put a stringent limit on
 the Fermi Fierz-interference coefficient \cite{hardytowner2009}, 
\begin{equation}\label{eq:bf}
b_F = -0.0022(26) =-2\frac{ \frac{g_s}{g_V} A_L } {1+\frac{g_S^2}{g_V^2}(A_L^2 + A_R^2)}\simeq -2\frac{g_S}{g_V}A_L \ .
\end{equation}
Although $b_F$ is not sensitive to right-handed scalar currents, the value of $\overline{\mathcal{F}t}$ is sensitive to these. In fact, the bound on right-handed couplings is more than an order of magnitude larger than that of left-handed couplings, such that both contributions to the $\overline{\mathcal{F}t}$ values are of the same order, as can be seen in Eq.~\eqref{eq:ftwithscalar}. Therefore, in searches for BSM physics one may not assume $V_{ud}$ as given by the PDG when such a search concerns also right-handed scalar terms. In the correlation coefficients, the value of $V_{ud}$ mostly drops out, but in limits derived from measured lifetimes the actual value of $V_{ud}$ is required.  

Constraints on right-handed scalar couplings can be extracted from the $\beta$-$\nu$-correlation coefficient $a$ defined in Eq.~\eqref{eq:xia}. We define $\delta_-=|a_{SM}-a_{exp}^{-}|$ as the lower bound and $\delta_+=|a_{exp}^{+}-a_{SM}|$ as the upper bound, where the experimental value, at 90\% confidence level (C.L.), lies between $a_{exp}^{-}$ and $a_{exp}^+$. Limits from $a$ then give 
\begin{equation}
2\left(\frac{g_S}{g_V }\right)^2 [A_L^2+A_R^2]<\delta_- \ ,
\end{equation} which gives a circular bound in the $A_L,A_R$ plane. Thus, the 
bound on $A_L$ and $A_R$ would be the same, 
\begin{equation}\label{eq:bound_a_wrong}
\left|\frac{g_S }{g_V}A_{L(R)}\right|<\sqrt{\frac{\delta_-}{2}} \ .
\end{equation} 
In practice experiments normalize the correlation to the total number of counts, and the absolute normalization is not measured. This means that in fact $\tilde{a}$ is measured, as discussed below Eq.~\eqref{eq:asymmetry}. In this way the Fierz-interference term $b$ enters. The bounds remain circular, but the bound on $A_L$ changes to
\begin{equation}\label{eq:bounds_a}
 \frac{-\delta_-}{2\gamma\langle m_e/E_e\rangle}<\frac{g_S}{g_V}A_L<\frac{\delta_+}{2\gamma\langle m_e/E_e\rangle} 
\end{equation} 
for $\beta^+$ and with opposite signs for $\beta^-$. 

\begin{figure}
	\centering
		\includegraphics[width=0.50\textwidth]{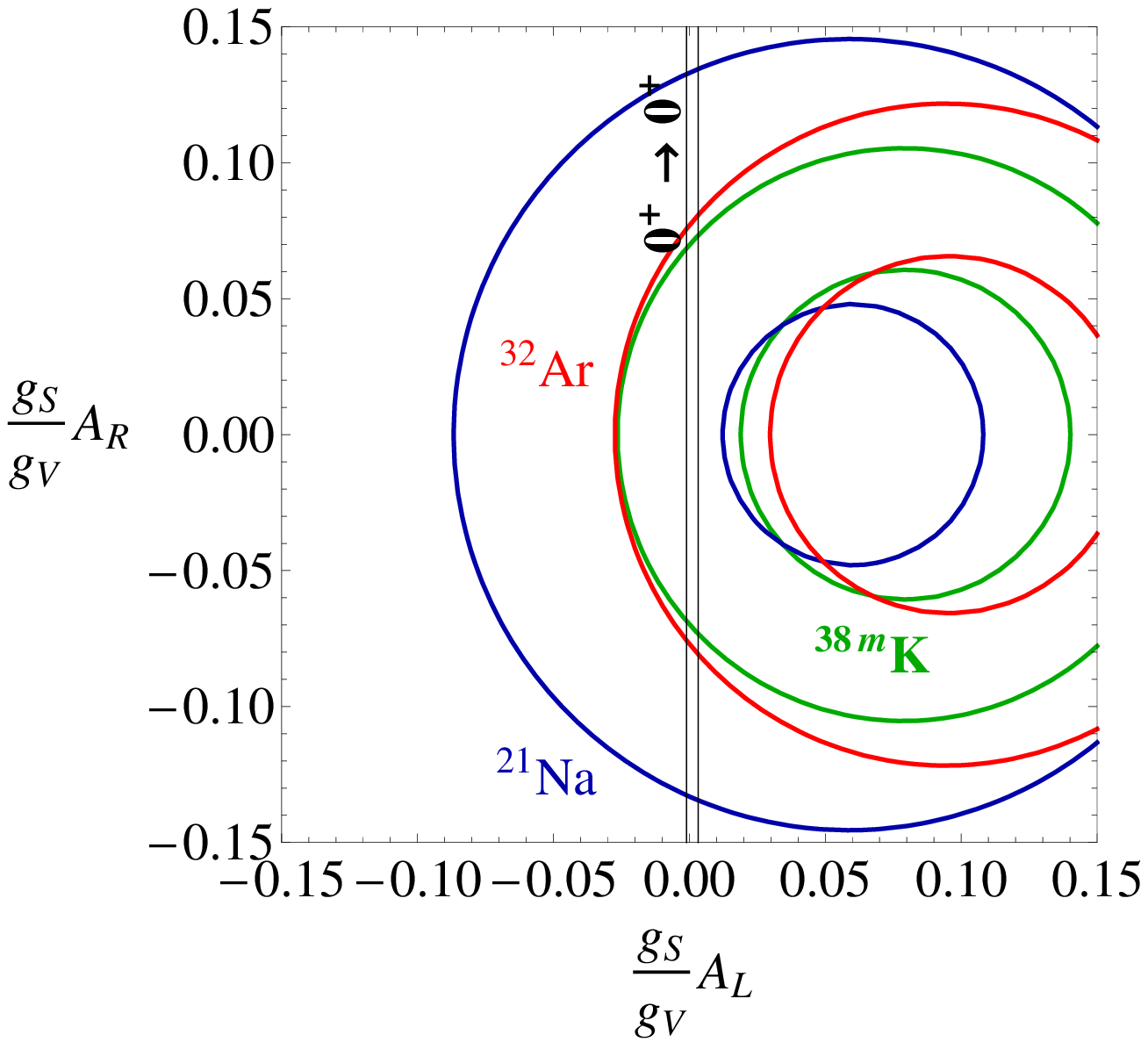}
	
		\caption{(Color online) Bounds on left- and right-handed scalar couplings (90\% C.L.). The narrow $0^+ \rightarrow 0^+$ band is from superallowed Fermi transitions Eq.~\eqref{eq:bf} 
		\cite{hardytowner2009}.	The ring-shaped boundaries are derived from $\beta$-$\nu$ correlation measurements in $^{38m}$K \cite{gorelovK} and $^{32}$Ar \cite{AdelbergerAr}, cf.\ Eq.~\eqref{eq:bounds_a}. Also the bound from the mirror nucleus $^{21}$Na \cite{vetterna21} is given, neglecting tensor contributions.}  
			\label{fig:scalarart}
			\end{figure}

Figure \ref{fig:scalarart} shows the bounds from the best current experiments. The superallowed Fermi decays only constrain left-handed couplings and give a narrow vertical band \cite{hardytowner2009}. The right-handed coupling $A_R$ is constrained only by the $\beta$-$\nu$ correlations, and depends on the square root of the experimental error $\delta_-$. The most sensitive $\beta$-$\nu$ correlation measurements are from $^{38m}$K \cite{gorelovK} and $^{32}$Ar \cite{AdelbergerAr}. We also include the recent measurement of the mirror nucleus $^{21}$Na \cite{vetterna21}, a mixed transition, where we have put tensor contributions to zero. In an earlier review this was erroneously shown with a bound as in Eq.~\eqref{eq:bound_a_wrong} \cite{symmetrytest}. We show it because it is the first mixed transition available with such competitive precision. 
The best current bounds on real scalar couplings from pure Fermi decays are found by minimalizing the $\chi^2$-distribution of the $b_F$ from Eq.~\eqref{eq:bf} and the measurements of the $\beta$-$\nu$ correlation in $^{38m}$K \cite{gorelovK} and $^{32}$Ar \cite{AdelbergerAr} (Table~\ref{tab:betadata}). At 90\% C.L.,
\begin{subequations}\label{eq:scalarchi}
\begin{align} 
-0.1\times 10^ {-2}< \frac{g_S}{g_V} A_L {}& < 0.3\times10^{-2} \ ,   \\
-6 \times 10^{-2} < \frac{g_S}{g_V} A_R {}& < 6 \times 10^{-2} \ .
\end{align}
\end{subequations}
For $A_L$ the bound comes from the strong limit on the Fierz-interference term. The limit on $A_R$ is less strong. Improving the bound on right-handed scalar couplings substantially is a daunting task: exploiting the forward-backward symmetry in the $\beta$-$\nu$ correlation would require collecting $10^{14}$ events to reach a bound $<10^{-3}$ on $g_S A_R$.

\subsubsection{Nuclear tensor searches}\label{sec:nucten}
The nuclear Gamow-Teller matrix element $M_{GT}$ can only be evaluated in the context of a nuclear model, because the spin of a nucleus is an observable, but the orbital angular momentum of a valence nucleon is not. For this reason $M_{GT}$ cannot be evaluated sufficiently robustly to put a bound on the left-handed tensor couplings from  $ft$ values, as was done for the scalar coupling by using the superallowed Fermi decays. However, the Fierz-interference term will enter most observables via the normalization requirement discussed previously, cf. Eq.~\eqref{eq:tilde}.
The $\beta$-asymmetry coefficient $A$ in Gamow-Teller decays is a good example of this, where
\begin{align}\label{eq:Anoright}
\tilde{A} {}&= \frac{A_{GT}}{1+b_{GT} \left\langle \dfrac{m_e}{E_e}\right\rangle}\nonumber \\
{}& \simeq \pm\lambda_{J'J}\left[-1+8\dfrac{g_T^2}{g_A^2}\alpha_L^2- 4 \dfrac{g_T}{|g_A|} \alpha_L \gamma \left\langle \dfrac{m_e}{E_e}\right\rangle \right] \ ,
\end{align}
from Eq.~\eqref{eq:Axi}. Thus in the absence of Coulomb corrections one finds that $\tilde{A}$ becomes independent of $\alpha_R$ and therefore only limits on $\alpha_L$ can be obtained from $\tilde{A}$. Defining the experimental bounds of $\tilde{A}- A_{\textrm{SM}}$ as before gives
\begin{equation}\label{eq:bounds_A}
 \frac{-\delta_-}{4\gamma\langle m_e/E_e\rangle}<\frac{g_T }{|g_A|}\alpha_L <\frac{\delta_+}{4\gamma\langle m_e/E_e\rangle } \ . 
\end{equation}  
To obtain a bound on $\alpha_R$ one can exploit the $\beta$-$\nu$ correlation $a$. The result is similar to the result for $a$ in Fermi decay. For $\beta^-$ Gamow-Teller decay $a_{\textrm{SM}}=-1/3$ and the bounds are
\begin{eqnarray}\label{eq:bounds_aGT}
|\frac{g_T }{g_A}\alpha_R|&<&\sqrt {\frac{3\delta_-}{8}} \ , \nonumber\\   
  -\frac{3\delta_-}{4\gamma\langle m_e/E_e\rangle}& <& \frac{g_T}{|g_A| } \alpha_L < \frac{3\delta_+}{4\gamma\langle m_e/E_e\rangle} \ .
\end{eqnarray} 

The limits on tensor interactions can be improved by combining scalar and tensor searches. In particular, the left-handed tensor couplings can be further constrained by using the measurements of the Fermi and Gamow-Teller-transition ratio of the longitudinal $\beta$ polarization. These measurements were performed in the first place to study the manifest left-right symmetric model \cite{Wichers,Carnoy}, see also Sec.~\ref{sec:LHC}. The ratio of longitudinal polarizations ($P$, see Appendix~\ref{sec:app}) of the emitted positrons was measured in the systems $^{26}\textrm{Al}^m/^{30}\textrm{P}$ \cite{Wichers} and $^{14}\textrm{O}/^{10}\textrm{C}$ \cite{Carnoy}, where the first nucleus decays via a Fermi and the second a Gamow-Teller transition. The two transitions have nearly identical endpoint energies, which eliminates systematic errors. The measured ratio is 
\begin{equation}\label{eq:pfgt}
\frac{P_F}{P_{GT}}\simeq \frac{\tilde{G}_F}{\tilde{G}_{GT}}\simeq 1-2 \left\langle \frac{m_e}{E_e}\right\rangle \left(\frac{g_S}{g_V}A_L +2\frac{g_T}{|g_A|}\alpha_L\right) \ .
\end{equation}
Combining these measurements with the bounds on $b_F$ in Eq.~\eqref{eq:bf} gives a more precise left-handed tensor bound, but it does not constrain right-handed couplings.


\begin{figure}[t]
	\centering
		\includegraphics[width=0.50\textwidth]{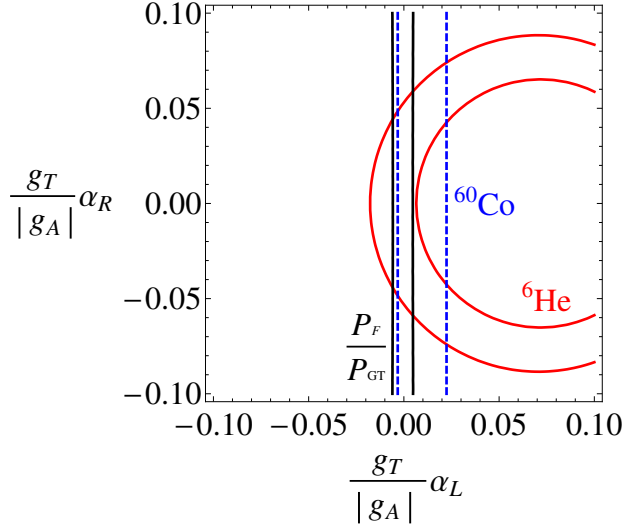}
		\caption{(Color online) Bounds on left- and right-handed tensor couplings (90\% C.L.). The measurement of the $\beta$-$\nu$ correlation in $^6$He \cite{gluckHe,Heliummeas} gives a ring-shaped boundary. The boundary of measurements of the $\beta$-asymmetry in the pure Gamow-Teller-decay of $^{60}$Co \cite{WautersCo} is given by dashed lines, the measurement only constrains left-handed couplings (Eq.~\eqref{eq:Anoright}). The strongest bounds on left-handed couplings are from measurements of the $\beta$-longitudinal polarization $P_F/P_{GT}$ in Eq.~\eqref{eq:pfgt} \cite{Wichers,Carnoy}, combined with the constraint on $b_F$.}
				\label{fig:tensorart}
\end{figure}
Figure \ref{fig:tensorart} shows the best constraints on tensor couplings. We use the $P_F/P_{GT}$ values \cite{Wichers,Carnoy}, the $\beta$-$\nu$ correlation in $^6$He \cite{gluckHe,Heliummeas}, and the $\beta$ asymmetry in $^{60}$Co \cite{WautersCo} (see Tab.~\ref{tab:betadata}) to find the best bounds for nuclear searches, using $\chi^2$ minimalization. For the $P_F/P_{GT}$ values we have included the limits on scalar couplings in Eq.~\eqref{eq:scalarchi}. The combined fit for real tensor couplings gives, at 90\% C.L.,   
\begin{subequations}\label{eq:tensorchi}
\begin{align}
-0.3\times 10^{-2} < \frac{g_T}{|g_A|} \alpha_{L}{}& < 0.6 \times 10^{-2}\ ,\\ 
-6 \times 10^ {-2} <\frac{g_T}{|g_A|} \alpha_{R} {}& < 6 \times 10^{-2} \ .
\end{align} 
\end{subequations} 
Reducing the limits will require increased statistics and experimental improvements (Sec.~\ref{sec:efforts}). Further constraints from $\beta$ decay come from mixed decays which we discuss next.

\subsubsection{Tensor constraints from neutron and mirror nuclei}\label{sec:neutron}
Mirror transitions are mixed transitions and therefore sensitive to both scalar and tensor interactions. Mirror decays might be used to improve the bounds of pure Fermi and Gamow-Teller transitions discussed above. At this point only the neutron can be considered. The prospects of using mirror nuclei are discussed at the end of this subsection. 
 The neutron can serve as a laboratory for studying a range of fundamental interactions \cite{Abe08, Dub11, Nic09}). In neutron $\beta$ decay, the main focus lies on determining the SM parameters $V_{ud} $ and $\lambda=g_A/g_V$.  Non-SM values are included by allowing $\lambda$ to be complex and/or by allowing for scalar ($A_L,A_R$) and/or tensor ($\alpha_L,\alpha_R$) interactions. We still consider only real couplings, and defer to Sec.~\ref{sec:D} and Sec.~\ref{sec:R} for complex $\lambda$ and scalar and tensor couplings, respectively.  
To clarify the role of possible left- and right-handed scalar and tensor contributions, we keep the simplifying assumptions that the $V$ and $A$ couplings are those of the SM. For neutron decay, with $M_{GT}=\sqrt{3}$ and $M_F=1$, the $ft$ value is given by 
 \begin{align}\label{eq:neutronlifetime}
 1/ft_n= {}& \frac{m_e^5}{2\pi^3 \ln(2)} G_F^2 V_{ud}^2g_V^2 \nonumber \\ 
{}& \left\{1+\left[\frac{g_S}{g_V}\right]^2\left[A_L^2+A_R^2\right]+2\gamma\left\langle \frac{m_e}{E_e}\right\rangle\frac{g_S}{g_V}A_L \right.\nonumber \\ 
{} &\left. +3\lambda^2\left(1+\left[\frac{g_T}{g_A}\right]^2\left[\alpha_L^2 +\alpha_R^2\right]-  4\gamma\left\langle \frac{m_e}{E_e}\right\rangle \frac{g_T}{|g_A|}\alpha_L\right)  \right\} \ .
\end{align}
The current value recommended for the lifetime is $\tau_n = 880.3(1.1)$ s \cite{pdg2014}, which is nearly 6 seconds lower, but with the same error, as the recommended value of 2008. Of course, this affects the SM values for $V_{ud}$ and $\lambda$, but cross-checks with other correlation coefficients are possible, allowing for consistency of the SM parameters \cite{Wie11}. Including scalar and tensor contributions increases the number of degrees of freedom and such cross-checks are no longer possible. 
The observable $ft_n$ is most sensitive to $\alpha_L$, because of the partial Gamow-Teller nature of neutron decay. 
One can combine various correlation coefficients from neutron decay to extract $\lambda$, while allowing for non-SM contributions. In combination with the experimental results from the superallowed Fermi transitions ($b_F$ and $\overline{\mathcal{F}t}$), improved bounds on tensor contributions can be obtained. 
For example, with the  recent limits on $A$ from UCNA and PERKEOII \cite{Mun12, Men13} and neglecting right-handed neutrinos ($A_R=0,\ \alpha_R=0$), it is possible to obtain an analytical bound on $\alpha_L$ \cite{Pat13}. Allowing for right-handed neutrinos requires a fitting procedure. 

A complete set of neutron correlation data has been compiled by \textcite{Dub11}. More recent results are obtained with the PERKEOII setup \cite{Mun12} and from the UCNA collaboration \cite{Men13}.
Combined with the bounds from pure Fermi and Gamow-Teller transitions a fit can be made to obtain all relevant parameters ($\lambda, A_L, A_R, \alpha_L, \textrm{and} \alpha_R$) in a consistent way. This was recently done by \textcite{Wau13}, to extract both left-handed and right-handed tensor-coupling limits. Their fitting method entails a grid search. For all $\alpha_L$ and $\alpha_R$ values,  a value of $\chi^2$ was obtained by minimizing $\chi^2$ for the other 3 parameters. With this 2D $\chi^2$ surface a contour plot can be made, by plotting the equal $\Delta \chi^2 \equiv \chi^2-\chi_0^2$ lines, where $\chi_0^2$ is the minimal $\chi^2$.

 \begin{figure}[t]
 	\centering
 	\includegraphics[width=0.7\textwidth]{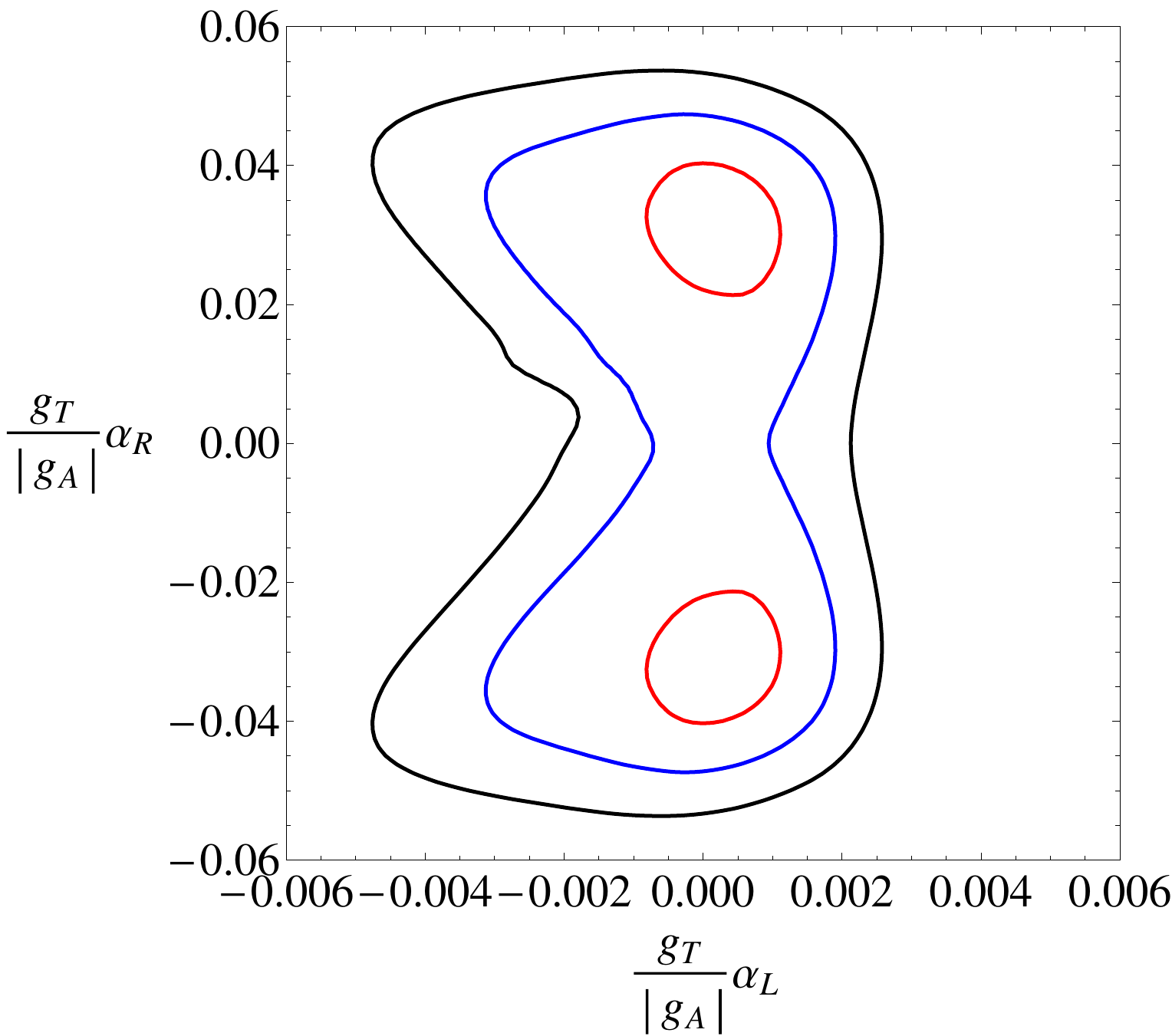}
 	\caption{(Color online) Contour plot of the 1, 2, and 3 $\sigma$ contours, derived from the selection of available data listed in Table~\ref{tab:betadata}. In the fitting procedure we minimized $A_L, A_R,$ and $\lambda$. Notice the scale difference of the two axis.}
 	\label{fig:totalplot}
 \end{figure}

Figure~\ref{fig:totalplot} shows the contour plot for the 1, 2, and 3 $\sigma$ ($\Delta \chi^2 = 1, 4,$ and $9$) bounds obtained with this method and by using the most relevant experiments listed in Table~\ref{tab:betadata}. It is important to note that the neutron lifetime requires the value of $V_{ud}$. The most precise value for $V_{ud}$ is obtained from the $\mathcal{F}t$ of superallowed decays \cite{Har13}, under the assumption of no scalar interactions. We have corrected for this by using Eq.~\eqref{eq:taucomplete} for the neutron lifetime. For the neutron lifetime we use the average value of the PDG \cite{pdg2012}. For the correlation coefficients the averages of the PDG cannot be used, because these are obtained by assuming only SM interaction. The possible different dependence on the Fierz-interference term is therefore not included. We consider the different values of $A$ separately, for which we have calculated the energy dependence with Eq.~\eqref{eq:energydep}. We have included the measurement of $B$, although for neutron decay this coefficient actually has a reduced sensitivity to the Fierz term $b$ and to $\lambda$, see Eq.~\eqref{eq:B}.



 
We find at $90\% \;\textrm{C.L}$\footnote{Bounds are extracted by scanning the 2D $\chi^2 +1.64^2$ surface for scalar $(A_{L,R})$ and tensor $(\alpha_{L,R})$, while for $\lambda$ we used the 1D probability density.} 
 \begin{subequations}\label{eq:cons}
  \begin{align}
-0.3 \times 10^{-2}< {}& \frac{g_T}{|g_A|}\alpha_L < 0.06 \times 10^{-2} \;  \ , \\ 
  -4.6\times 10^{-2} < {}&\frac{g_T}{|g_A|}\alpha_R < 4.6 \times 10^{-2}\;  \ , \\ 
   -0.1 \times 10^{-2}< {}& \frac{g_S}{g_V}A_L < 0.3 \times 10^{-2} \;  \ , \\
   -5\times 10^{-2} < {}&\frac{g_S}{g_V}A_R < 6\times 10^{-2} \;  \ , \\
   1.2659 <{}& \lambda < 1.2746 \; \ .  
   \end{align}\end{subequations}
The extracted value of $\lambda$ has a much larger error compared to $\lambda =1.2723(23)$ from PDG. 
The scalar bounds are the same as the bounds in Eq.~\eqref{eq:scalarchi}, but the tensor bounds are improved because of the inclusion of the neutron data. Especially the positive bound for $\alpha_R$ is reduced as compared to Eq.~\eqref{eq:tensorchi}. This is caused by the large spread in experimental values for $A$. Using only the two most recent values of the PERKEOII setup \cite{Mun12} and from the UCNA collaboration \cite{Men13} gives $-0.3 \times 10^{-2}<  g_T\alpha_L/|g_A| < 0.2 \times 10^{-2}$. For the tensor bounds, the neutron lifetime has a large influence \cite{Wau13}. We therefore anticipate that the error in the neutron lifetime and the spread in $A$ will soon give the dominant error on the limit on tensor couplings. 
 
 {\renewcommand{\arraystretch}{1.5}%
 \begin{table}[t]
 	\begin{center}
 		\begin{tabular}{l l l l l l l}
 			\hline\hline
 			 Isotope & Parameter & Decay & $\left\langle m_e/E_e \right\rangle$ & Value & Error & Reference \\ \hline
 			$^6$He  & $\tilde{a}_{GT}$ & $\beta^-$, GT & 0.286  & -0.3308 & 0.003 & \textcite{He93}\\ 
 			& & & & & & \textcite{gluckHe}\\ 
 			$^{14}$O/$^{10}$C & $P_F/P_{GT}$ (Eq.~\eqref{eq:pfgt}) & $\beta^+$ & 0.292 & 0.9996 & 0.0037 & \textcite{Carnoy} \\
 			$^{26m}$Al/$^ {30}$P & $P_F/P_{GT}$ (Eq.~\eqref{eq:pfgt}) & $\beta^+$ & 0.216 & 1.003 & 0.004 & \textcite{Wichers}\\
 			$^ {32}$Ar & $\tilde{a}_F$ & $\beta^+$, F & 0.191 & 0.9989 & 0.0065 & \textcite{AdelbergerAr} \\
 			$^ {38m}$K & $\tilde{a}_F$ & $\beta^+$, F & 0.133 & 0.9981 & 0.0045 & \textcite{gorelovK} \\
 			$^{60}$Co & $\tilde{A}_{GT}$ & $\beta^-$, GT & 0.704 & -1.027 & 0.022 & \textcite{WautersCo} \\
 			$0^+\rightarrow0^+$ & $b_F$ & $\beta^+$, F & 0.2560 & -0.0022 & 0.0026 & \textcite{hardytowner2009} \\
 			$n$ & $\tau$ (Eq.~\eqref{eq:taucomplete}) & $\beta^-$, F/GT & 0.655 & 880 s. & 0.9 s. & \textcite{pdg2012} \\
 			$n$  & $\tilde{A}_n$ & $\beta^-$, F/GT & 0.56 & -0.11952 & 0.00110 & \textcite{Men13} \\
 			 			$n$  & $\tilde{A}_n$ & $\beta^-$, F/GT & 0.534 & -0.11926 & 0.00050 & \textcite{Mun12} \\
						$n$  & $\tilde{A}_n$ & $\beta^-$, F/GT & 0.582 & -0.1160 & 0.0015 & \textcite{Lia97} \\
			 			$n$  & $\tilde{A}_n$ & $\beta^-$, F/GT & 0.558 & -0.1135 & 0.0014 & \textcite{Yer97}\\
						& & & & & & \textcite{Ero91} \\
						 			$n$  & $\tilde{A}_n$ & $\beta^-$, F/GT & 0.551 & -0.1146 & 0.0019 & \textcite{Bop86} \\
 			$n$  & $\tilde{B}_n$ & $\beta^-$, F/GT & 0.594 & 0.9801 & 0.0046 & \textcite{Serebrov:1998aj} \\
				$n$  & $\tilde{B}_n$ & $\beta^-$, F/GT & 0.63 & 0.9802 & 0.0050 & \textcite{Bschu} \\
 			$n$  & $\tilde{a}_n$ & $\beta^-$, F/GT & 0.655 & -0.1054 & 0.0055 & \textcite{Byr02} \\
 		
 			\hline \hline
 		\end{tabular}
 		\caption{Experimental values used to construct Fig.~\ref{fig:totalplot}. The values for $\left\langle m_e/E_e \right\rangle$ are mostly not calculated by the experimental groups and are derived with Eq.~\eqref{eq:energydep}, except for the $0^+\rightarrow0^+$ decays, for which we use the value derived in \textcite{Pat13}. Averages from the PDG are only used for the $\tau$ \cite{pdg2012}, since different measurements of $\tilde{A}$ and $\tilde{B}$ might also have a different energy dependence, which is not taken into account in the PDG averages. We have taken all experimental values for $\tilde{A}$ used by the PDG. }
 		\label{tab:betadata}
 	\end{center}
 \end{table}

Recently, also mirror decays have been used to constrain tensor couplings. The strong constraint on $b_F$ from superallowed Fermi decays, can be combined with measurements on mirror nuclei, to derive a value for $b_{GT}$. In \textcite{Severijns:2008ep} a complete survey of $\mathcal{F}t$ values of the available mirror transitions is given. For $T=1/2$ transitions the relation between the $\mathcal{F}t$ values of the mirror and superallowed $0^+ \rightarrow 0^+$ is given by \cite{Severijns:2008ep}
\begin{equation}\label{eq:ftmirror}
\mathcal{F}t^{\textrm{mirror}} \equiv \frac{2 \mathcal{F}t^{0^+ \rightarrow 0^+}\left\langle 1+ \tfrac{g_S^2}{g_V^2}\left[A_L^2+A_R^2\right]-2\gamma\left\langle \frac{m_e}{E_e}\right\rangle^{0^+\rightarrow 0^ +} \tfrac{g_S}{g_V} A_L\right\rangle }{1+\tfrac{g_S^2}{g_V^2}\left[A_L^2+A_R^2\right]+\frac{f_A}{f_V}\rho^2 \left[1+4\alpha_L^2+4\alpha_R^2\right] \pm 2\gamma\left\langle \frac{m_e}{E_e}\right\rangle \left(\tfrac{g_S}{g_V} A_L-2\tfrac{g_T}{|g_A|} \alpha_L \rho^2\right)} \;,
\end{equation}
where $f_A/f_V=1.0143(29)$ is the ratio of the axial-vector and vector statistical rate functions \cite{Severijns:2008ep}. The inverse energy dependence of the superallowed Fermi decays is denoted by $\left\langle m_e/E_e\right\rangle^{0^+\rightarrow 0^ +}$ and calculated in \textcite{Pat13}. If $\rho$ is known, a value for $\alpha_{L}$ can be extracted from $\mathcal{F}t^{\textrm{mirror}}$. 

The mirror $\beta^+$ decay of $^{19}$Ne to $^{19}$F was recently studied to determine the lifetime of $^{19}$Ne \cite{Bro13}. In this work, the effectiveness of the method described above is shown. 
For mixed decays an independent measurement of $\rho$ is necessary. For $^{19}$Ne, $\rho=1.5995(45)$ \cite{Cal75}, was derived from the measurement of the $\beta$ asymmetry $A$.  
Neglecting quadratic couplings in Eq.~\eqref{eq:ftmirror}  
and using the extracted value $\mathcal{F}t=1719.8(13)$ s with $\left\langle m_e/E_e\right\rangle=0.387022(18)$ from \textcite{Bro13} a limit on $b_{GT}$ is derived.
For left-handed tensor couplings this gives at $90\% \; \textrm{C.L.}$ \cite{Bro13} 
\begin{equation}
-1.5 \times 10^{-2} < \frac{g_T}{|g_A|} \alpha_{L} < 0.12 \times 10^{-2}\; . 
\end{equation}
The bounds are only an order of magnitude less precise than the combined limits in Eq.~\eqref{eq:cons}, and show the potential for this kind of measurements for improving the existing bounds. 

\subsubsection{Tensor constraints from radiative pion $\beta$ decay}
In \textcite{Byc09} limits on tensor couplings are derived from radiative pion decay, $\pi^+ \rightarrow e^+ +\nu_e +\gamma$. These bounds can be translated into bounds on $\alpha_L$ \cite{Bha12} by using estimates for the pion form factor \cite{Mat07}. 
Assuming no right-handed couplings and using $g_T=1.047(61)$, a limit at 90\% C.L. is found, 
\begin{equation} 
-1.9 \times 10^{-3} < \frac{g_T}{|g_A|} \alpha_{L} < 2.3 \times 10^{-3} \ . 
\end{equation}  
These bounds are the strongest bounds on tensor couplings from a single decay experiment and show that future $\beta$-decay experiments should probe $\alpha_L<10^{-3}$ and beyond, in order to improve these existing limits.  
\subsubsection{Pseudoscalar constraints}\label{sec:pseudo}
Pseudoscalar interactions have so far been neglected in $\beta$-decay searches, since they are strongly suppressed because the nuclei are nonrelativistic. The suppression of these terms is $\mathcal{O}(1/M)$, where $M$ is the nucleon mass. However, in $\beta$ decay, the pseudoscalar interactions are always multiplied by $g_P$, the pseudoscalar form factor discussed in Eq.~\eqref{eq:gp}. The large value $g_P=349(9)$ \cite{Gon13} largely cancels this suppression, and $\beta$-decay experiments might be used to probe these interactions. There are, however, already strong constraints on pseudoscalar couplings from pion decay \cite{Her01,Her94,Bha12}. 

The ratio $R_\pi=\Gamma(\pi\rightarrow e\nu)/\Gamma(\pi\rightarrow \mu\nu)$ is sensitive to pseudoscalar couplings defined by
\begin{equation}
\mathcal{L} = \frac{G_F V_{ud}}{\sqrt{2}}\; \left[A_L^P \;\bar{e}(1-\gamma_5) \nu_e+A_R^P \;\bar{e}(1+\gamma_5) \nu_e \right] \; \bar{u}\gamma_5 d \ ,
\end{equation}
where we have neglected flavor-changing couplings, which can be found in \textcite{Bha12}. The ratio $R_\pi/R_\pi^{SM}$, where $R_\pi$ is the measured value, is sensitive to electron and muon pseudoscalar couplings, $A^{P(e)}$ and $A^{P(\mu)},$ respectively. If these couplings are such that $A^{P(e)}/m_e = A^{P(\mu)}/m_\mu$, their contributions to the ratio cancel and no bounds on pseudoscalar interactions can be obtained. Since there is no reason to assume such a cancellation, we can place bounds on pseudoscalar interactions, because these would show up as $R_\pi/R_\pi^{\textrm{SM}} \neq 1$. 
The current best value for this ratio is $R_\pi/R_\pi^{\textrm{SM}} = 0.996(3)$ \cite{Cir07,pdg2012}, which leads to (90\% C.L.) \cite{Bha12, Cir13}
\begin{subequations}\label{eq:pseudo}
\begin{align}
-1.4 \times 10^{-7} < A^P_L {}& < 5.5 \times 10^{-4} \label{eq:pseudol}\ , \\
-2.8 \times 10^{-4} < A^P_R {}& < 2.8 \times 10^{-4} \ . 
\end{align}
\end{subequations}

In $\beta$ decay the pseudoscalar term shows up in Gamow-Teller and mixed decays. The most relevant to experiments are its contributions to the Fierz interference term,
\begin{equation}
b_{GT} = \pm 4\frac{g_T}{|g_A|} \alpha_L \pm 2\frac{g_P}{|g_A|} A^P_L \frac{E_0-E_e}{M} \ ,
\end{equation} 
which enters with the usual $\left\langle m_e/E_e\right\rangle$ suppression. The $(E_0-E_e)/M$ term is responsible for the suppression of pseudoscalar contributions, however, because $g_P(E_0-E_e)/M \simeq 0.4$ pseudoscalar interactions are still suppressed compared to tensor interactions. Given the current limit on $\alpha_L$, improving the bounds in Eq.~\eqref{eq:pseudol} seems unlikely in the near future.

The pseudoscalar couplings in Eq.~\eqref{eq:pseudo} can also be translated into bounds on scalar and tensor couplings. If scalar and tensor interactions are present at the new physics scale $\Lambda$, they will mix via radiative loop corrections, and pseudoscalar couplings will radiatively be generated \cite{Cam05,Her94}. 
Current limits are at the level of \cite{Bha12,Cir13,Cir13b}
\begin{subequations}
\begin{align}
|A_L|\lesssim 8 \times 10^{-2} &{}\;\;\;\textrm{and} \;\;\; |A_R|\lesssim 5 \times 10^{-2} \ , \\
|\alpha_L|\lesssim 2\times 10^{-3} &{}\;\;\; \textrm{and} \;\;\; |\alpha_R|\lesssim 1.2 \times 10^{-3} \ ,
\end{align}
\end{subequations}
and depend logarithmically on the scale of new physics $\Lambda$, for which $\Lambda= 10$ TeV is used. These bounds are of the same order of magnitude as global-fit limits from $\beta$ decay in Eq.~\eqref{eq:cons}, except for the bound on $\alpha_R$, which is an order of magnitude better. However, because the constraints for right-handed currents rely on the flavor structure of new physics \cite{Cir13}, we do not further consider these bounds.   

\subsubsection{Left-handed scalar versus tensor}
\begin{figure}[h]
	\centering
			\includegraphics[width=0.50\textwidth]{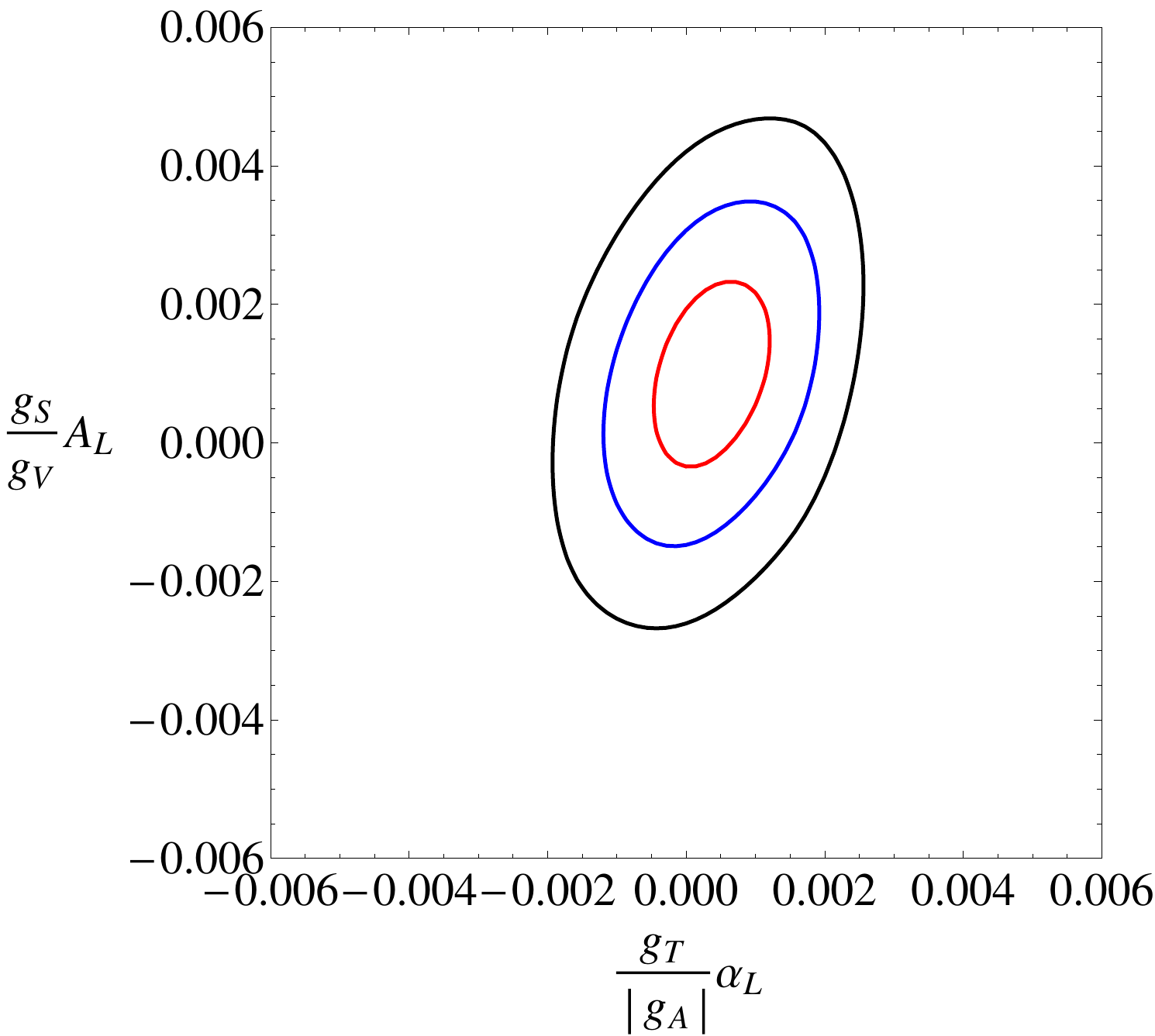}
		\caption{(Color online) Contour plot of the 1, 2,  and 3 $\sigma$ contours, derived from the selection of available data listed in Table~\ref{tab:betadata} without right-handed couplings,  i.e. $A_R=\alpha_R=0$.}
\label{fig:scalarvstensor1}
\end{figure}
In Sec.~\ref{sec:neutrino} we discuss exotic couplings involving right-handed neutrinos. If right-handed neutrinos are absent, or too heavy to be energetically allowed in $\beta$ decay, right-handed neutrino couplings,  i.e. $A_R$ and $\alpha_R$, can be neglected. The resulting reduction of parameter space allows us to use mixed decays to fit the correlations between left-handed tensor and scalar couplings. Figure~\ref{fig:scalarvstensor1} shows these correlations. For the complete set of data listed in Table~\ref{tab:betadata} we find at 90\% C.L. 
\begin{subequations}
\begin{align}
-0.1 \times 10^{-2} < \frac{g_S}{g_V} A_L {}& < 0.3 \times 10^{-2} \ , \\
-0.2 \times 10^{-2}< \frac{g_T}{|g_A|} \alpha_{L} {}& < 0.06 \times 10^{-2}\ ,\\ 
1.2715 < |\lambda|  {}& <  1.2744 \ .  
\end{align}
\end{subequations}
These bounds are not significantly different from the bounds from the complete fit in Eq.~\eqref{eq:cons}. For comparison: limits on right-handed couplings from neutron decay alone are found in \textcite{Kon10} and \textcite{Dub11}.

\subsection{Constraints from LHC experiments}\label{sec:LHC} 
Low-energy experiments are mostly viewed as complementary to high-energy collider searches for BSM physics. Experiments at LHC can place bounds on new physics by looking for the on-shell production of new particles, as done in searches for a $W_R$ boson (Eq.~\eqref{eq:wr}) or supersymmetric particles. 
We focus here on the effect of a $W_R$ boson, because this has been studied complementary by precision decay experiments and by LHC, e.g. \textcite{Dek14b}. At the LHC, $W_R$ is searched for by considering its possible decay channels. In the $W_R \rightarrow t\bar{b}$ channel, such direct searches at the CMS experiment constrain $M_R>2$ TeV \cite{Cha14}. Constraints from the $W_R \rightarrow e \nu$ channel are similar, but depend on assumptions for the right-handed neutrino. Constraints from neutral-kaon mixing give $M_R>3$ TeV \cite{Ber14}. 

In $\beta$ decay, strong limits come from CKM unitarity tests, for which the best bound is \cite{Har13}
\begin{equation}
|V_{ud}|^2+|V_{us}|^2+|V_{ub}|^2 = 1.00008 (56) \ ,
\end{equation} 
which uses the value of $V_{us}$ from \textcite{Vus}. The error has equal contributions from $V_{ud}$ and $V_{us}$. Following \textcite{hardytowner2009}, this leads to a constraint on $a_{LR}$, i.e. left-handed lepton couplings and right-handed quark couplings, of 
\begin{equation}\label{eq:alr}
-4 \times 10^{-3} < a_{LR} < 5 \times 10^{-3} \ ,
\end{equation} 
at 90\% C.L. The precision of both $V_{ud}$ and $V_{us}$ should improve simultaneously for such a test to remain significant. 


In $\beta$ decay, some correlation coefficients are sensitive to $a_{LR}, a_{RL},$ and $a_{RR}$, where the latter two are only present if light right-handed neutrinos are assumed. For example, the measurements of $P_F/P_{GT}$ \cite{Wichers, Carnoy} and $A_{GT}$ in $^{60}$Co \cite{WautersCo}, are used to constrain parameters of manifest LR-symmetric models. Such models have a P symmetry, such that for the CKM matrices $V_{ud}^L=\pm V_{ud}^R$. There is no additional spontaneous CP violation, so $\omega=0$. In this simplified model, $a_{RL}= \pm a_{LR}\sim-\xi$ and $a_{RR}=\delta=(M_1/M_2)^2$. Measurements of $P_F/P_{GT}$ limit the combination $\delta\cdot \xi$ and do not give additional bounds, because of the strong bound on $\xi$ from unitarity tests given in Eq.~\eqref{eq:alr}. Because $\xi$ is strongly constrained, $\beta$-decay experiments can only constrain $a_{RR}$ and thus the mass of the $W_R$. Derived limits are of the order of 200 GeV \cite{WautersCo,gorelovK}, an order of magnitude below the bound from the LHC experiments presented above. In fact, when assuming manifest LR symmetry, the strongest bound on $W_R$ comes from the $K_L$-$K_S$ mass difference, from which $W_R>20$ TeV was derived \cite{Mai14}. 

Besides constraining new physics by searching for direct on-shell production, it is also possible for colliders to constrain exotic couplings. When the mass of the non-SM particle exceeds the energy accessible at LHC, the new particles cannot be produced on-shell, but their effects can still be found in deviations from the SM predictions. In that way, the exotic interactions in Eq.~\eqref{eq:lageff} will also manifest themselves in proton-proton collisions. This makes it possible for LHC data to constrain the same tensor and scalar couplings relevant in $\beta$ decay \cite{Bha12, Cir13}. 

In particular, the $pp\rightarrow e + \textrm{MET}+ X$ channel is considered, where MET signifies Missing Transverse Energy. This channel is closely related to $\beta$ decay, since it involves the $\bar{u}d\rightarrow e\bar{\nu}$ process at quark level.
At the LHC, both the ATLAS and CMS detectors are used to search for new physics in this channel \cite{Atlas12,CMS12}, by searching for an excess of events predicted at a large lepton transverse mass cut $\bar{m}_T$. At large $\bar{m}_T$, the SM cross section approaches zero more rapidly than the cross sections for new physics, making the sensitivity to non-SM physics larger at high momenta. The total cross section is 
\begin{eqnarray}\label{eq:sigma}
\sigma(m_T>\bar{m}_T) &=& \sigma_{\textrm{SM}}(1+|a_{LR}|^2+|a_{RL}|^2) + \sigma_{R}|a_{RR}|^2 \notag \\
&& + \sigma_S (|A_{L}|^2 + |A_R|^2) + \tfrac{1}{4}\sigma_T (|\alpha_{L}|^2 + |\alpha_R|^2) \ ,
\end{eqnarray}
where $\sigma_{\textrm{SM}}$ is the SM cross section and $\sigma_{R, S,T}$ are the cross sections for new physics. The explicit form of $\sigma_{\textrm{SM}}$ and $\sigma_{R,S,T}$ is given, to lowest order in QCD corrections, in \textcite{Cir13}. The coefficients $a_{LR}$ and $a_{RL}$ cannot be constrained, because their contribution is proportional to $\sigma_{\textrm{SM}}$, and therefore small at large $\bar{m}_T$. 

With the expected number of background events and the number of actual observed events, one can place an upper limit on the number of new physics events, $n_s^{up}$ \cite{Bha12}. This translates into an upper limit for $\sigma$, and finally into bounds on exotic couplings. 
First bounds were derived by \textcite{Bha12}, updated bounds are given in \textcite{Nav13}.   

The bounds are derived by using the experimental data of \textcite{CMS14} at an integrated luminosity of $20\; \textrm{fb}^{-1}$ and at a center-of-mass energy of $\sqrt{s}=8$ TeV. \textcite{Nav13} also gives the combined limits for scalar and tensor couplings, assuming only left-handed couplings. In Table~\ref{tab:LHCbeta} we give the $90\%$ C.L. bounds, obtained by allowing one exotic interaction and putting all other couplings to zero. 
To compare these results with $\beta$-decay constraints, we use the values from the global fit in Eq.~\eqref{eq:cons} and the form factors $g_S=1.02(11)$ \cite{Gon13} and $g_T=1.047(61)$ \cite{Bha14}. Because the errors on the form factors are not Gaussian, we use the R-fit method described in \textcite{Bha12}, which treats all the values in a 1 $\sigma$ interval with equal probability. Therefore, only the lower bounds are important. We stress again that the reduction of the error in $g_S$ and $g_T$ is important to make meaningful comparisons between the different experiments. 
 \begin{table}[t]
 	\begin{center}
 		\begin{tabular}{c|c|c|c|c}
 			\hline\hline
 			&  $|A_L|$ & $|A_R|$ & $|\alpha_{L}|$ & $|\alpha_{R}|$  \\ \hline
 			$\beta$ decay  & $2.5\times 10^{-3}$ & $6\times 10^{-2}$ &  $3\times 10^{-3}$  & $4.6\times 10^{-2}$ \\  
 			LHC & $6\times 10^{-3}$ & $6\times 10^{-3}$ & $2\times 10^{-3}$ & $2\times 10^{-3}$  \\
 			Neutrino & - & $1\times 10^{-3}$ & - & $1\times 10^{-3}$ \\
 			\hline \hline
 		\end{tabular}
 		\caption{Comparison between $\beta$-decay limits on left- and right-handed scalar $A_L$ and $A_R$ and tensor couplings $\alpha_L$ and $\alpha_R$, constraints from LHC data \cite{Nav13}, and from the neutrino mass \cite{Ito2005}. 
 			 Constraints are at 90\% C.L., and all couplings are assumed to be real.}
 		\label{tab:LHCbeta}
 	\end{center}
 \end{table}

Table~\ref{tab:LHCbeta} shows that the LHC constraints on left-handed couplings are comparable to $\beta$-decay constraints, while for right-handed couplings the LHC constraints are an order of magnitude better than the $\beta$-decay limits. The current status is illustrated in Fig.~\ref{fig:scalarlhcneutrino} and Fig.~\ref{fig:tensorlhcneutrino}. \textcite{Nav13} also make a projection for the 14 TeV run at $50 \;\textrm{fb}^{-1}$, and find that the expected bounds are a factor 3 better.

\subsection{Neutrino-mass implications}\label{sec:neutrino}
Besides strong bounds from LHC experiments on right-handed interactions, there are also bounds from the neutrino mass. 
In the SM, neutrinos are assumed to be massless, but neutrino oscillations indicate the existence of at least two massive neutrinos. A direct upper limit on the neutrino mass comes from the shift of the end-point of the $\beta$ spectrum. Recent measurements of the $\beta$ spectrum of $^3$H give $m_\nu<2 $ eV (95\% C.L.) \cite{Ase11, Kra04}. The KATRIN experiment aims to improve these limits by an order of magnitude \cite{katrin}. Other bounds on the neutrino mass are derived from cosmological observations; WMAP \cite{Hin12} limits $\sum m_\nu < 0.44 $ eV and a recent study of Planck \cite{Planck}, in which Planck data is combined with neutrino oscillation data, gives a similar limit $m_\nu < 0.15$ eV, for three degenerate neutrinos. 

In Eq.~\eqref{eq:lageff}, the couplings $a_{RR}, a_{RL}, A_R,$ and $\alpha_R$ involve right-handed neutrinos. These couplings can only be generated if the decay to right-handed neutrinos is kinematically allowed, i.e. if right-handed neutrinos are light enough to be created in the decay. The possibility of these light right-handed neutrinos has been considered in various new physics scenarios as a possible dark-matter candidate. If right-handed neutrinos are very heavy, as is suggested in many see-saw mechanisms, we can omit all exotic couplings with first index $R$. 

\textcite{Prezeau:2004md} showed that the small neutrino mass also limits the presence of exotic couplings in low-energy experiments that involve a (light) right-handed neutrino \cite{Prezeau:2004md}. For $\beta$ decay this strongly constrains the couplings $A_R, \alpha_R,$ and $a_{RL}$ \cite{Ito2005}. Neutrino masses can be either Dirac ($\bar{\nu}_L m_D \nu_R$) or Majorana ($\frac{1}{2} \bar{\nu}_L m_\nu \nu_L^c$), where $\nu_L^c= i \gamma_2\gamma_0 \bar{\nu}_L^T$, or a combination of the two. However, the following results are general and apply to both types. 
\begin{figure}[t]
	\centering
\subfloat[Constrains $A_{RR}, A_{RL},$ and $\alpha_{R}$]{\label{fig:1}\includegraphics[width=0.40\textwidth]{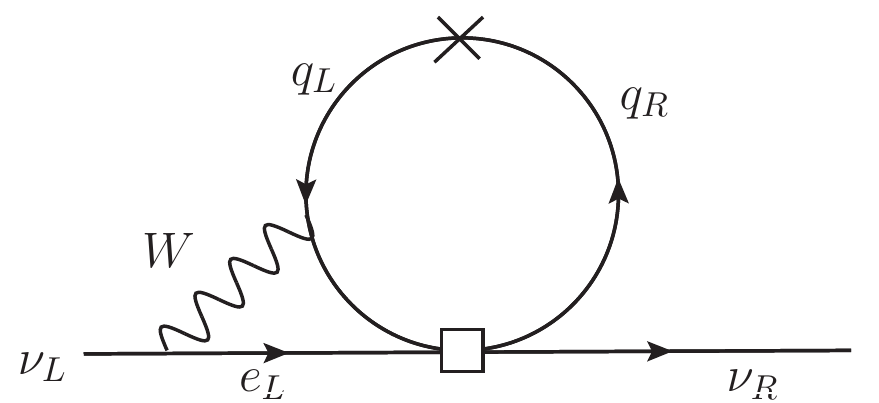}}
	%
\subfloat[Constrains $a_{RL}$]{\label{fig:2} \includegraphics[width=0.48\textwidth]{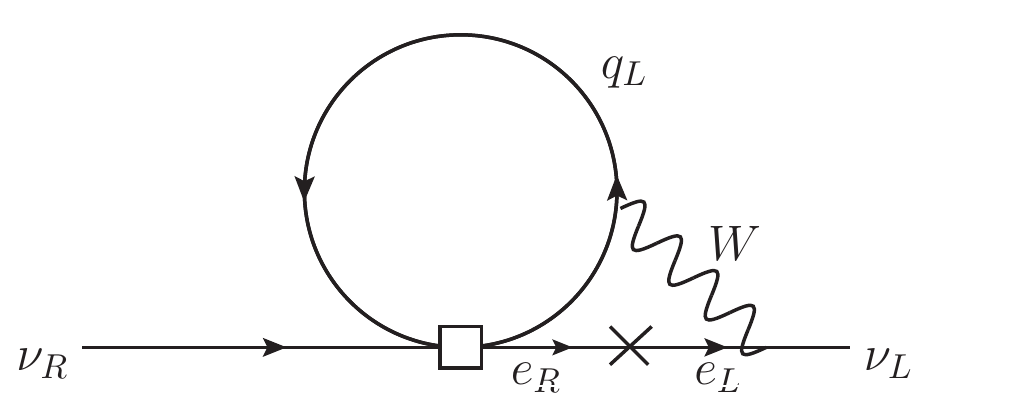}}
\caption{The two-loop contribution to the neutrino mass, where the box indicates the exotic couplings. The cross indicates a mass insertion, with (a) $m_q=4$ MeV (b) $m_e=0.511$ MeV \cite{Ito2005}. For Majorana neutrinos one can substitute $\nu_R \rightarrow \nu_L^c$. }
	\label{fig:neutrinomassgen}
\end{figure} 
Couplings to right-handed neutrinos contribute to the neutrino mass via loop interactions. Figure~\ref{fig:neutrinomassgen} shows the leading two-loop contribution to the neutrino mass, where the box indicates the non-SM coupling to right-handed particles. The cross indicates the mass insertion needed to couple two fermions with different chiralities. Here, the chirality-changing interactions are either proportional to (a) the quark or (b) the electron mass. In a power-counting scheme, one-loop contributions are in general less suppressed than two-loop contributions. However, the two-loop diagrams in Fig.~\ref{fig:neutrinomassgen} are enhanced by the $W$-boson mass, while the one-loop diagrams are only suppressed by the light-fermion mass. This makes the two-loop contribution dominant, as the additional loop suppression of $1/(4\pi)^2$ is diminished by the heavy $W$-boson mass. 

One can estimate the two-loop contribution to the neutrino mass by considering only the logarithmic part of Fig.~\ref{fig:neutrinomassgen}. The analytic parts are renormalization-scheme dependent and are therefore neglected \cite{Prezeau:2004md}. By using dimensional regularization the contribution to $\delta m_{\nu}$ is estimated as \cite{Ito2005} 
\begin{equation}\label{eq:mnuest}
\delta m_{\nu} \simeq  3g^2 G_F \bar{a} \frac{m_f M_W^2}{(4\pi)^4} \left(\ln{\frac{\mu^2}{M_W^2}}\right)^2 \ , 
\end{equation}
where $\bar{a} = \left\{A_{RL}, A_{RR}, \alpha_{R}, a_{RL}\right\}$ are the exotic couplings from Eq.~\eqref{eq:lageff}, $g=0.64$ is the gauge coupling, $m_f$ is the inserted fermion mass, and $\mu$ is the renormalization scale, which should exceed the heaviest mass in the interaction, $\mu>m_t$, where $m_t$ is the top-quark mass. Assuming that the loop corrections do not exceed the mass of the neutrino\footnote{There might be scenarios in which this is not obeyed, but these scenarios would have to be fine-tuned.}, i.e. $\delta m_{\nu} < m_{\nu}$, setting $m_q=4$ MeV, $\mu=1$ TeV, and $m_\nu < 0.15$ eV in Eq.~\eqref{eq:mnuest} gives
\begin{subequations}\label{eq:neutr}
\begin{align}
\left|a_{RL}\right| {}&\lesssim  10^{-2} \label{eq:neutrinocon1} \ , \\
\left|A_{RR}\right|,  \left|A_{RL}\right|, \left|\alpha_{R}\right| {}& \lesssim  10^{-3} \label{eq:neutrinocon}\ .
\end{align}
\end{subequations}
In Table \ref{tab:LHCbeta} we compare these limits with current right-handed $\beta$-decay bounds and bounds from LHC. The estimates from the neutrino mass are currently the strongest bounds on right-handed currents. They are more than an order of magnitude stronger than the $\beta$-decay bounds, and comparable to the LHC bounds. 
For the bounds in Eq.~\eqref{eq:neutr} we have used the updated neutrino mass from the Planck space observatory, which might further improve in the future. The given bounds are conservative estimates, but nevertheless they show the large impact of the neutrino mass on $\beta$-decay measurements. 
Even stronger constraints of $\mathcal{O}(10^{-5})$ from the neutrino mass have been derived in the unpublished thesis of \textcite{Wang:2007ay}.

\subsection{Conclusions and outlook}\label{sec:efforts}
We summarized the current status of the bounds on real right-handed vector, scalar, pseudoscalar, and tensor interactions in $\beta$ decay. We compared these bounds with those obtained from proton-proton collisions at the LHC experiments and the upper limit on the neutrino mass, mainly focusing on scalar and tensor interaction. The best current bounds are given in Table~\ref{tab:LHCbeta}. 
\begin{figure}[t]
	\centering
		\includegraphics[width=0.50\textwidth]{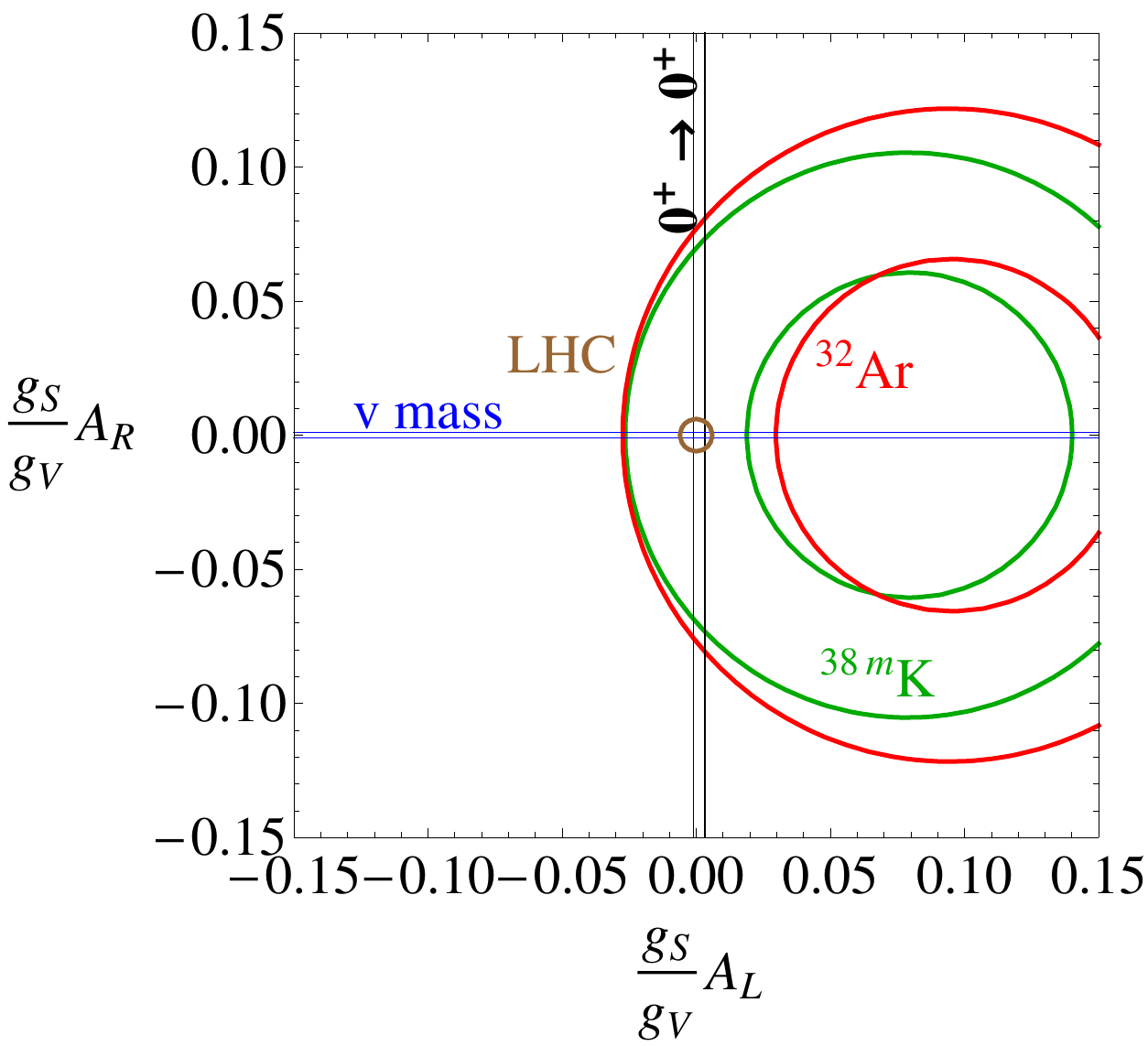}
		\caption{(Color online) Scalar bounds from nuclear $\beta$ decay as in Fig.~\ref{fig:scalarart} combined with limits derived from the neutrino mass (horizontal line) and constraints from the LHC experiments (circular bound).}
\label{fig:scalarlhcneutrino}
\end{figure}\begin{figure}[htpb]
	\centering
		\includegraphics[width=0.50\textwidth]{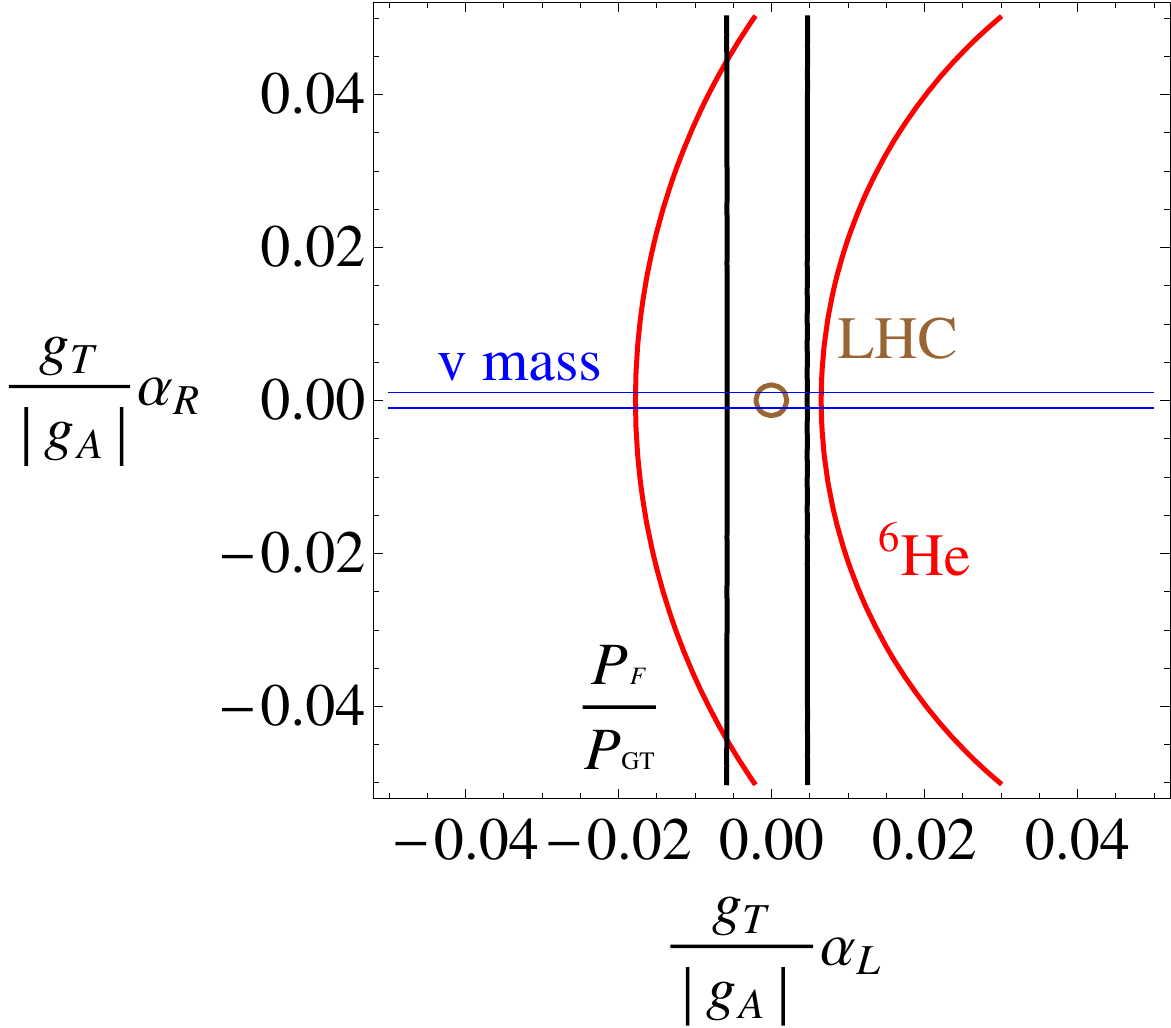}
		\caption{(Color online) Tensor bounds from nuclear $\beta$ decay as in Fig.~\ref{fig:tensorart} combined with limits derived from the neutrino mass (horizontal line) and constraints from the LHC experiments (circular bound). }
\label{fig:tensorlhcneutrino}
\end{figure}
We distinguished between bounds on left- and right-handed scalar and tensor interactions, where left or right denotes the chirality of the neutrino. The constraints on left-handed interactions are equally constrained by the LHC and $\beta$-decay experiments. On the other hand, $\beta$-decay experiments measuring right-handed interactions would have to improve orders of magnitude to compete with the bounds from the LHC experiments and the neutrino mass. This is illustrated in Fig.~\ref{fig:scalarlhcneutrino} for scalar interactions and in Fig.~\ref{fig:tensorlhcneutrino} for tensor interactions. Table~\ref{tab:pro} projects the competitive accuracies required for different $\beta$-decay parameters. For left-handed currents we give the necessary precision to compete with projected future LHC bounds \cite{Nav13}. For right-handed bounds, we give two accuracies. The first corresponds to the required sensitivity to compete with current LHC bounds, the number in brackets corresponds to the required precision to compete with the bounds from the neutrino mass (see Table~\ref{tab:LHCbeta}).  

The bounds on left-handed couplings are best pursued via measurements of the Fierz interference coefficient $b$. For left-handed scalar couplings $A_L$ the bound is most stringent because of the vast effort in the study of super-allowed Fermi transitions. These studies also provide the best current value for $V_{ud}$. The left-handed tensor coupling $\alpha_L$ requires a larger effort, for which several measurements need to be combined. The best current bounds are from the global fit in which neutron and nuclear data are combined. In this fit, especially the uncertainties in the neutron lifetime and the $A$ coefficient of the neutron have a significant impact. We pointed out that the large spread in the available $A$ measurements influences the obtained bound significantly. The Gamow-Teller part $b_{GT}$ of the Fierz interference term and $V_{ud}$ can also be constrained in mirror nuclei, in analogy to the superallowed Fermi transitions. However, this also requires the measurement of at least one correlation coefficient. Measurements with this aim are undertaken \cite{Ban13}. 

In Gamow-Teller transitions, measurements of the Fierz interference term $b_{GT}$ allow for bounds on the left-handed tensor terms. In Seattle, a $^6$He factory has been set up to study this term. The lifetime of $^6$He was already measured with high precision \cite{Kne11}, but the shell-model calculations are not sufficiently accurate as yet to search for tensor interactions. One straightforward, but not so simple, approach is to measure the decay spectrum precisely. This would give access to $b_{GT}$. These measurements would also have to consider contributions from the SM weak-magnetism (cf. Eq.~\eqref{eq:compva}). Measurements of $b_{GT}$ from electron-antineutrino correlation $\tilde{a}_{\beta\nu}$ and the spectrum are both ongoing and being set up \cite{Fle08,Fle11,Kne11a,Navpri,Sev14,Avi12}. If these measurements reach $b<10^{-3}$, they would allow for a strong limit on $\alpha_L$. Such a precision is necessary to compete with the projected bounds from the 14 TeV run of the LHC.
In neutron decay, many efforts are undertaken to improve the measurements of $a_{\beta\nu}$ and $A$ \cite{Mar09,Kon12, Bae08, Wie09, Poc08, Bae14}. For comparison, limits on the Fierz terms from neutron decay alone are found in \textcite{Kon10} and \textcite{Dub11}, including limits derived from the electron energy dependence of the $\beta$-asymmetry $A_{\textrm{exp}}(E)$ alone. 

Right-handed interactions, which imply the existence of a light right-handed neutrino, do not interfere with the SM interactions and can therefore only be measured directly,  i.e. via quadratic terms. This makes it difficult to reach the sensitivity obtained for left-handed couplings. In $\beta$ decay, the right-handed tensor coupling $\alpha_R$ can be constrained by measuring the $\beta$-$\nu$ correlation, $\tilde{a}_{\beta\nu}$. The best measurement in pure Gamow-Teller decays of $a_{\beta\nu}$ stems from the measurement in $^6$He \cite{Heliummeas}. Many efforts are undertaken to improve this limit in $^6$He \cite{Avi12, Kne11a, Cou12}.  
A dedicated effort to limit right-handed tensor couplings is ongoing in $^8$Li, for which the daughter nucleus $^8$Be breaks up into two $\alpha$ particles, $^8 \textrm{Li} \rightarrow e^-+\bar{\nu}+2\alpha$. The $a_{GT}$ coefficient can be measured by measuring the $\beta$-$\alpha$ correlation, and by taking advantage of the increased sensitivity due to the population of a $2^+$ state in $^8$Be. After putting the Fierz term $b=0$, such that only right-handed interactions are constrained \cite{Li13}, one finds
    \begin{equation}
 \frac{g_T}{|g_A|} |\alpha_R|  < 8 \times 10^{-2} \ .
\end{equation}
The bound reaches the precision of the combined fits, but when considering the LHC or neutrino bounds the experiment would have to improve by more than three orders of magnitude to compete (see Table~\ref{tab:pro}). 

When comparing tensor and scalar bounds from different fields, the form factors $g_S$ and $g_T$ are important. Lattice QCD calculations have made enormous progress, and will continue to do so in the next period. The lattice prediction of $g_A$ will hopefully reach the experimental precision soon, which would allow for a cross-check between the experimental value and the theoretical lattice value. 

Besides scalar and tensor searches, we also discussed searches for $V+A$ and pseudoscalar interactions. Pseudoscalar interactions are less suppressed than previously thought, due to the large value of $g_P$. However, strong bounds exists from radiative pion decay, and pseudoscalar interactions can still be neglected in the upcoming $\beta$-decay experiments. Strong constraints on $V+A$ currents are extracted from CKM unitarity tests, to which $\beta$-decay experiments contribute by providing the most accurate value of $V_{ud}$. Besides this, measurements of correlation coefficients can be used to constrain parameters of (manifest) left-right symmetric models. For these specific models, strong limits from the LHC experiments and the neutral-kaon mass difference exist. Therefore, the significance of $\beta$ experiments in these experiments is limited to specific models.

 \begin{table}[t]
 	\begin{center}
 		\begin{tabular}{l|l|l}
 			\hline\hline
 			 Parameter & Bound & Constraint at $ 90\% \;\textrm{C.L.}$ \\ \hline
 			     $b_{GT}$ & $10^{-3}$ & $ \alpha_L < 3 \times 10^{-4} $   \\
					$b_F$ & $10^{-3}$ & $ A_L < 5 \times 10^{-4} $   \\
		$a_{GT}$ & $ 10^{-4} $ ($5\times 10^{-6}$)   & $\alpha_R < 6\times 10^{-3}$ ($\alpha_R <   10^{-3}$)  \\
	 $a_{F}$ & $ 8\times 10^{-6}$ ($2\times10^{-6} $) & $A_R < 2\times 10^{-3}$ ($A_R<  10^{-3}$)  \\
 			\hline \hline
 		\end{tabular}
 		\caption{Required experimental precision  on $\beta$-decay parameters to remain competitive with LHC bounds, cf.~\textcite{Nav13}. Only the Fermi (F) and Gamow-Teller (GT) parts of the Fierz-interference term $b$ and the $\beta$-$\nu$-correlation $a$ are listed. The third column gives the corresponding limit on scalar couplings $A_L$ and $A_R$ and tensor couplings $\alpha_L$ and $\alpha_R$. The Fierz term is the leading term in most $\beta$-correlation experiments (Sec.~\ref{sec:betabounds}).  The indicated bounds for $b$  assumes  that future LHC data lead to bounds indicated in the last column. The $a$ parameter is the most direct way to obtain a bound on right-handed couplings, which should be the motivation to measure $a$. Here the current bounds of thr LHC are assumed, while the value in parentheses is the required accuracy when the bound derived from the limit of the neutrino mass is considered (Table~\ref{tab:LHCbeta}).}
 		
 		\label{tab:pro}
 	\end{center}
 \end{table}

\section{Limits on time-reversal violation}\label{sec:time}
So far we have only considered the real parts of the exotic couplings. In this Section we focus on their imaginary parts. A nonzero measurement of an imaginary coupling would imply that time-reversal (T) symmetry and, by the CPT-theorem, CP symmetry is violated\footnote{In any Lorentz-symmetric local field theory, CP violation is equivalent to T violation, according to the CPT theorem. For CPT-violation, see Sec.~\ref{sec:LIV}.}. Becaues of the matter-antimatter asymmetry of the universe, new sources of CP violation are expected \cite{Sakh}. Many models of BSM physics 
predict such additional sources of CP violation, see {e.g.} \textcite{Ibr08,Bra12, Dek14}. This makes T or CP violation one of the main portals to search for new physics. These searches range from experiments at the LHC to atomic-physics experiments. As such the observables can be quite diverse. With advances in theory, in particular via EFT methods, relations between the different observables have become more clear (cf. Sec.~\ref{sec:LHC} and Sec.~\ref{sec:neutrino}). 

In this section we focus on the connection between T-violating observables in $\beta$ decay and the bounds on electric dipole moments (EDMs).  
The P- and T-odd EDM measurements are a powerful probe of CP violation beyond the SM \cite{Pos05}. High-precision EDM searches have been made for the neutron, paramagnetic and diamagnetic atoms, and molecules. 
The EDM is a static observable, and, therefore, allows for very precise atomic-physics experiments. It is also a background-free observable, because the electroweak SM contributions to the EDM are strongly suppressed. Therefore, EDM experiments give strong limits on new T-violating physics.
BSM physics contributions to the EDM can be parametrized by dimension-6 operators \cite{Vri11,  Vri11a, Vri11b,Vri13}. At low energy this leads to a relation between the T-violating correlations in $\beta$ decay and EDMs. 

Many correlation coefficients in $\beta$ decay 
depend on the square of the underlying coupling constants. As such they depend only on the imaginary couplings squared, which are therefore difficult to access. A more direct way to probe imaginary couplings is to consider the T-odd triple correlations $\vec{J}\cdot(\vec{p}_e\times\vec{p}_\nu)$ and $\vec{\sigma}_e\cdot(\vec{J}\times\vec{p}_e)$ multiplied by the $D$ (Eq.~\eqref{eq:decayrate}) and $R$ (Eq.~\eqref{eq:withr}) coefficients, respectively. The first is P-even and T-odd, while the latter is P- and T-odd. They probe left-handed imaginary couplings, which are absent in the SM. 

Since the interactions contributing to $D$, $R$, and EDMs are generated by the same operators, a limit on the EDM also limits the $D$ and $R$ coefficients. We consider these relations and discuss the relative precision of the two types of experiments.

\subsection{Limits on triple-correlation coefficients in $\beta$ decay}
A finite $D$ coefficient arises from the interference between the imaginary parts of the left-handed vector couplings and is proportional to $\textrm{Im}\; a_{LR}$. The $R$ coefficient arises from the interference between the imaginary parts of scalar or tensor couplings and SM couplings, making this coefficient sensitive to both $\textrm{Im}\; A_L$ and $\textrm{Im}\; \alpha_{LL}$. 

The SM contributes to both the $R$ and $D$ coefficients through electromagnetic final-state-interactions (FSI) and through SM CP violation. The FSI are only motion-reversal odd, i.e. the initial and final state are no longer interchangeable, due to radiative corrections. In this way, FSI mimic time-reversal violation, but in fact are T-even. We will denote their contributions by $R_f$ and $D_f$, and write $D=D_t+D_f$ and $R=R_t+R_f$ \cite{Her05}, where $D_t$ and $R_t$ are the true T-violating contributions. The contributions from FSI are comparable to the current experimental precision and depend on the momentum of the $\beta$ particle. We will discuss their values for specific isotopes later. True T violation in the SM arises from the CP-violating phase of the CKM matrix and the QCD $\theta$-term. These sources only contribute at the level of $\mathcal{O}(10^{-12})$ \cite{PhysRevD.56.80}, much below the current experimental precision.

\subsubsection{$D$ coefficient}\label{sec:D}
To first order in exotic couplings, the $D_t$ coefficient can be expressed as \cite{Jac57a}
\begin{equation}
D_t = a_D \textrm{Im}\; a_{LR}\ ,
\end{equation}
from Eq.~\eqref{eq:Dap}, with
\begin{equation}\label{eq:ad}
a_D = \frac{4\delta_{J'J}\sqrt{\frac{J}{J+1}} \rho}{1+\rho^2}\ .
\end{equation}
The $D$ coefficient can only be accessed in mixed transitions, and has been measured in both neutron and $^{19}$Ne decay, which have $a_D=0.87$ and $a_D = -1.03$, respectively. 
For $^{19}$Ne the best measurement is $D= 1 (6)\times 10^{-4}$ \cite{ned}, and from neutron decay $D= -0.94 (2.10)\times 10^{-4}$ \cite{Chu12,mummdneutron}. 

The value of the FSI depends on the kinematics of the experiment. For $^{19}$Ne the FSI have been derived by \textcite{Cal67} as $D_f=2.6\times 10^{-4} p_e/p_e^{\textrm{max}}$, which is of the same order as the experimental precision. For neutron decay the FSI were also calculated in chiral perturbation theory by \textcite{Ando:2009jk}. Their derivation reproduces the original result of \textcite{Cal67}. However, \textcite{Ando:2009jk} include higher-order corrections, which are of order $\mathcal{O}(10^{-7})$, allowing for an accurate expression for the FSI,
\begin{equation}\label{eq:Dfsi}
D_f= (0.228 \frac{p_e^{\textrm{max}}}{p_e} + 1.083 \frac{p_e}{p_e^{\textrm{max}}})\times 10^{-5} - 5.88\frac{p_e^{\textrm{max}}}{p_e} \times 10^{-8} \ ,
\end{equation}
where the first two terms are the \textcite{Cal67} terms, and the last term represents the higher-order corrections. Equation~\eqref{eq:Dfsi} is accurate to better than 1\%. For the current best neutron experiment the FSI are estimated at $D_f\simeq 1.2 \times 10^{-5}$ \cite{Chu12}. The uncertainty in $D_f$ stems from the uncertainty of the $\beta$ momentum in the experiment. The T-violating part of the neutron $D$ measurement gives at 90\% C.L.
\begin{equation}
|D_t|< 4 \times 10^{-4} \; \ , 
\end{equation}
and with $a_D=0.87$,
\begin{equation}
|\textrm{Im}\;a_{LR}|< 4 \times 10^{-4} \; \ . 
\end{equation}

Given the current experimental precision, it is clear that the FSI become increasingly more important. In this respect, neutron experiments are favored over nuclei, because the FSI can be calculated with a higher precision. Eventually the accuracy to which the FSI are known will limit measurements of true T violation. 

\subsubsection{$R$ coefficient}\label{sec:R}
Neglecting quadratic non-SM couplings, the $R_t$ coefficient is given by \cite{Jac57a} 
\begin{equation} 
R_t = \frac{(a_D\mp b_D)}{|g_A|} g_T \textrm{Im}\;\alpha_{L} - \frac{a_D}{2g_V} g_S \textrm{Im}\;A_{L} \ ,
\end{equation}
from Eq.~\eqref{eq:Rap}, where the upper (lower) sign is for $\beta^- (\beta^+)$ decay, $a_D$ is given in Eq.~\eqref{eq:ad}, and 
\begin{equation}
b_D = \frac{4\lambda_{J'J} \rho^2}{1+\rho^2 } \ ,
\end{equation}
with $\lambda_{J'J}$ as given in Appendix \ref{sec:app}. The $R$ coefficient can be measured in both mixed or pure Gamow-Teller transitions, where the latter limits Im $\alpha_L$. The leading contributions to the FSI are given by the Coulomb corrections calculated by \textcite{Jac57},
\begin{equation} \label{eq:Rfsi} 
R_f = \frac{Z\alpha  m_e}{2p_e} (\mp a_D+ b_D) \ .
\end{equation}

The $R$ coefficient has been measured in the pure Gamow-Teller decay of $^8$Li, where $a_D=0$ and $b_D=4/3$. The FSI give $R_f\simeq 7 \times 10^{-4}$, leading to $R_t=(0.9\pm2.2) \times 10^{-3}$ \cite{PhysRevLett.90.202301}. This constrains at 90\% \textrm{C.L.}
\begin{equation}\label{eq:alphall}
g_T|\textrm{Im}\; \alpha_{L}| < 3 \times 10^{-3}\ .
\end{equation}
The best measurement of $R$ in a mixed decay has been obtained for neutron decay, for which $a_D=0.87$ and $b_D=2.2$. \textcite{kozelaneutron2012} find $R=(4\pm12\pm5)\times 10^{-3} $. The FSI are calculated with Eq.~\eqref{eq:Rfsi}. By using the energy distribution seen by the experimental setup one obtains $R_f\simeq 6\times 10^{-4}$ \cite{kozelaneutron2012}. The error in $R_f$ is less than 10\%. $R_f$ can be neglected given the current experimental precision. At 90\% C.L. 
\begin{equation}
-1.1 g_T\textrm{Im}\;\alpha_{L} - 0.44  g_S \textrm{Im}\;A_{L} < 2.4\times 10^{-2}\ .
\end{equation}
With the constraint given in Eq.~\eqref{eq:alphall} one finds at 90\% C.L.
\begin{equation}
 g_S |\textrm{Im}\; A_{L}| < 6\times 10^{-2}\;\ . 
\end{equation}

\subsubsection{Alternative correlations}
The measurement of the $D$ coefficient requires the detection of the recoiling nucleus instead of detecting the neutrino. This imposes strong experimental constraints on any measurement scheme. Current schemes consider atomic trapping in a magneto-optical trap, which has led to the best value for the $\beta$-$\nu$ correlation $a$. Measuring $D$ requires a modification of this trap technique, to allow for a polarized sample. It will be extremely challenging to achieve high statistical precision and systematical accuracy with this technique. An alternative lies in the $\beta$-$\gamma$ correlations of polarized nuclei \cite{Mor57,Cur57}, where the photon with momentum $\vec{k}$ is emitted from the state populated by the $\beta$ decay. In this way one measures the correlation proportional to
\begin{equation}\label{eq:Ecoef}
E {\vec{J}}\cdot({\vec{p}_e}\times\vec{ k})(\vec{J}\cdot \vec{ k}) \ ,
\end{equation}
when the emission is due to an $E1$ transition. The correlation coefficient $E\propto \textrm{Im}\; a_{LR}$ is nonzero only for mixed decays. \textcite{You95} have identified $^{36}$K as a promising candidate for such a measurement, since this isotope allows for the comparison between a mixed and a Gamow-Teller transition. The latter is insensitive to T violation and can be used to test the experimental setup and reduce systematic errors. Secondary beams of high intensity can be produced, stopped, and polarized in a buffer gas allowing to measure $\beta$-$\gamma$ correlations \cite{Mul13} with high precision. 
Correlations alternative to measuring $R$ are also possible (the $L$ and $M$ coefficients \cite{Jac57a, Ebe57}) but, similar to $R$, will always require to measure the polarization of the $\beta$ particle, which is an inefficient process. 

In radiative $\beta$ decay, it is possible to have triple-correlation coefficients without nuclear or electron spin \cite{Bra01,Gar12,Gar13}, such as
\begin{equation}
K \vec{k} \cdot (\vec{p}_\nu \times \vec{p}_e) \ .
\end{equation}
This coefficient has not been measured, but \textcite{Dek15} showed that EDMs provide extremely strong constraints on the coefficient $K$.
\subsection{EDM limits}
Limits exist for the neutron EDM, the electron EDM, and several atomic EDMs. The best current bounds are listed in Table~\ref{tab:EDM}, where the limits from molecular YbF and ThO are expressed as a limit on the electron EDM $d_e$. The last column of Table~\ref{tab:EDM} indicates if a connection to the triple-correlation coefficients $D$ and $R$ exists \cite{Ng:2011ui, Khr91}. 


\begin{table}[t]
\begin{center}
\begin{tabular}{l|l|l|l}
\hline\hline
  EDM & e cm (90\% C.L.)  &Reference & Connection to $\beta$ decay  \\ \hline
$n$  & $2.9 \times 10^{-26} $ & \textcite{Bakernedm} & $D$\\  
$^{199}$Hg & $2.6\times 10^{-29}$ & \textcite{Gri09} & $D$, $R$\\
$^{205}$Tl &  $ 0.9\times 10^{-24}$ & \textcite{Reg02} &$R$\\
YbF & $|d_e| < 10.5 \times 10^{-28}$ & \textcite{Hud11} &$R$ \\
ThO & $|d_e| < 8.7 \times 10^{-29}$ & \textcite{Bar13} &$R$ \\
\hline \hline
	\end{tabular}
	\caption{The current best EDM limits of the neutron, diamagnetic Hg, paramagnetic Tl, and molecular YbF and ThO. The neutron EDM and Hg can be connected to the $D$ coefficient (and $E$ coefficient). Other EDM measurements, except the neutron, can be connected to the $R$ coefficient. The limit from molecular YbF and ThO are expressed as a constraint on the electron EDM $d_e$.}
	\label{tab:EDM}
\end{center}
\end{table}

\subsubsection{Limits on $D$ from EDM limits}
Any new vector interaction that contributes to Im $a_{LR}$ (and thus to $D_t$) also contributes to nuclear EDMs \cite{Ng:2011ui,Her05}. This makes it possible to translate bounds on the EDMs of the neutron and diamagnetic atoms into bounds on Im $a_{LR}$. The $D$ coefficient is P-even and T-odd, while the EDM is both P- and T-odd. Nevertheless, loop corrections, containing the $W$ boson, allow for a relation between these observables. 


The relevant CP-odd dimension-six operator is \cite{Ng:2011ui}
\begin{equation}\label{eq:operator} 
\mathcal{L}^{(\textrm{eff})} = \frac{c }{\Lambda^2} \bar{u}_R\gamma^{\mu}d_R\; \tilde{\varphi}^\dagger iD_{\mu}\varphi+\textrm{h.c.}
\end{equation}
where $c$ is a complex coefficient, $\Lambda$ is the scale of new physics, $D_\mu$ is the covariant derivative, and $\varphi$ is the Higgs doublet with $\tilde{\varphi}^I=\epsilon^{IJ}\varphi^{J*}$, where $\epsilon^{IJ}$ is the antisymmetric tensor. Fig.~\ref{fig:Dcoef} shows the energy evolution of this operator. First, electroweak symmetry breaking generates the coupling of the $W$ boson to right-handed quarks,
\begin{equation}
\mathcal{L}^{(\textrm{eff})}  =  \frac{g v^2 }{2\sqrt{2}\Lambda^2} (c\;\bar{u}_R \gamma^{\mu} d_R W^{+}_{\mu} + c^*\bar{d}_R \gamma^{\mu} u_R W^{-}_{\mu}) \ ,
\end{equation}
where $\varphi$ acquired its vacuum expectation value $v/\sqrt{2}$ and $g$ is the $SU(2)_L$ coupling constant.
The $W$ boson can couple to a lepton current or a quark current. At lower energy, the $W$ boson is integrated out. This generates a P- and T-odd four-quark coupling and the lepton-quark coupling $a_{LR}$ in $\beta$ decay. The effective Lagrangian is
\begin{equation}\label{eq:lagc}
\mathcal{L}^{(\textrm{eff})} =  -\frac{c}{\Lambda^2}\left(\bar{u}_R\gamma^{\mu}d_R\bar{e}_L\gamma_{\mu}\nu_{eL}+ V_{ud} \bar{u}_R\gamma^{\mu}d_R\bar{d}_L\gamma_{\mu}u_L\right) + \textrm{h.c.}\ ,
\end{equation}
which shows that the two couplings $c$ and $a_{LR}$ have a common origin. 
They are related by 
\begin{equation}
\textrm{Im}\;a_{LR}=\frac{\textrm{Im}\;c}{2\sqrt{2}G_F\Lambda^2}\ .
\end{equation}
When evolving to the QCD scale, the second term in Eq.~\eqref{eq:lagc} is affected by QCD renormalization. However, this only has a small numerical effect \cite{Dek14}, which can be neglected given the uncertainties coming from the calculation of the neutron EDM. 

\begin{figure}
	\centering
		\includegraphics[width=0.50\textwidth]{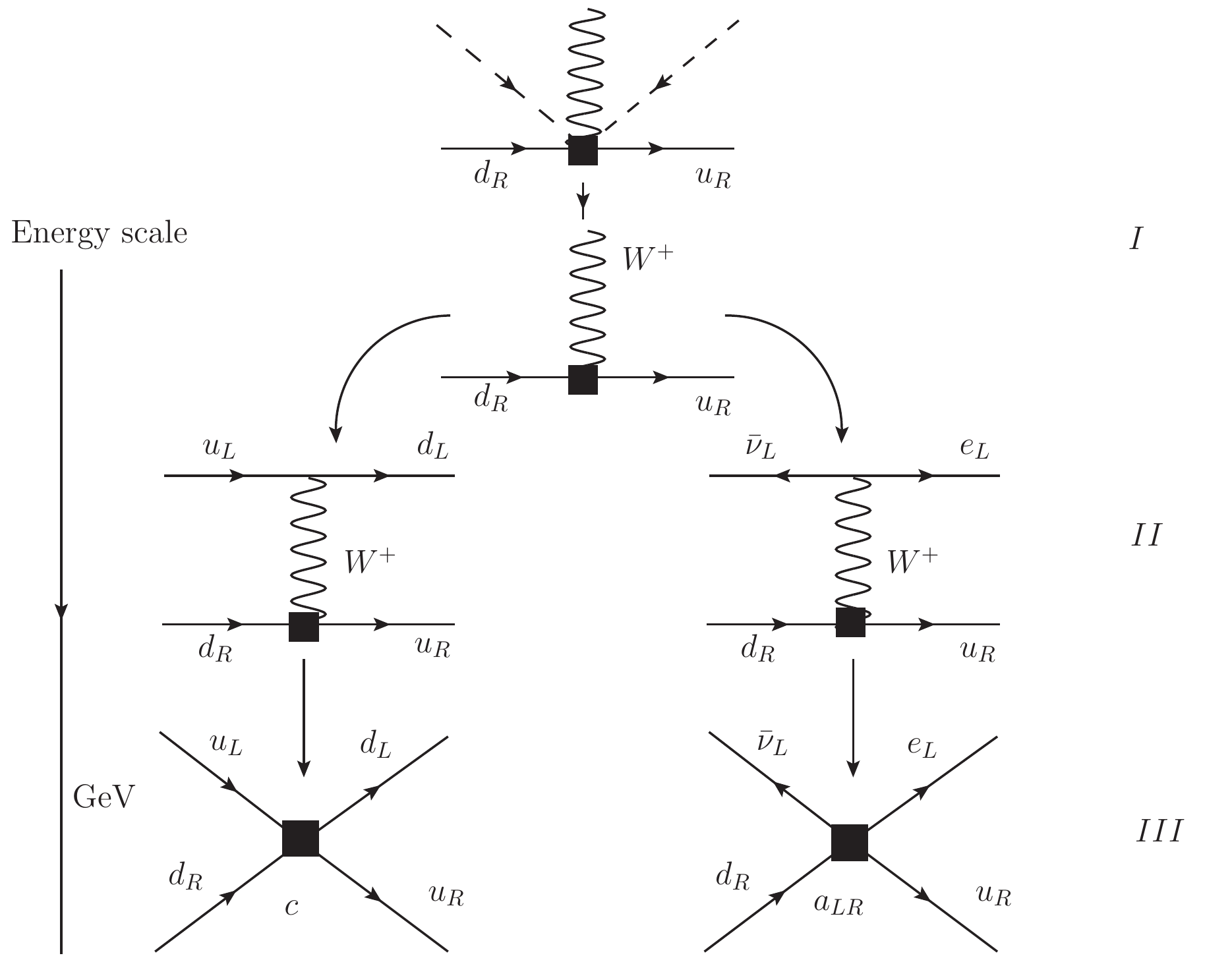}
	\caption{Generation of the four-fermion operators that contribute to the EDM (left) and $\beta$-decay (right). The boxes denote the four-fermion couplings, $c$ and $a_{LR}$ respectively. The coupling of the $W$ boson to the right-handed quarks is generated by the dimension-6 operator in Eq.~\eqref{eq:operator}.}
	\label{fig:Dcoef}
\end{figure}

Bounds on $\textrm{Im}\;c$ thus lead to an upper limit on $\textrm{Im}\;a_{LR}$.
The dependence of the EDM on $\textrm{Im}\;c$ involves theoretical calculations at different energy scales. Especially for diamagnetic atoms such as $^{199}$Hg, differences in nuclear calculations lead to a large uncertainty in the interpretation of the bounds on atomic EDMs. Therefore, we do not consider bounds from $^{199}$Hg.
No such problem occurs for the neutron, and \textcite{Vri13,Sen14} estimated the link between the neutron EDM and Im $c$ as
\begin{equation}\label{eq:dnlink}
d_n=-1\times 10^{-20} \frac{\textrm{Im}\;c}{2\sqrt{2}G_F\Lambda^2} e \ \textrm{cm} \ .
\end{equation}
This result differs by an order of magnitude from the result used in \textcite{Ng:2011ui}, which was obtained from \textcite{An, He93}. 
In \textcite{Vri13,Sen14} it was pointed out that, due to use of a relativistic meson-nucleon field theory, \textcite{An, He93} overestimate the neutron EDM by an order of magnitude. 

The current bound on the neutron EDM $|d_n|<2.9\times 10^{-26} \; e$ cm \cite{Bakernedm} and Eq.~\eqref{eq:dnlink} gives at 90\% C.L. 
\begin{equation}\label{eq:edmDresult}
|\textrm{Im}\;a_{LR}| < 3\times10^{-6} \ .
\end{equation}
This bound is at least two orders of magnitude below the bound obtained from $\beta$ decay. Improving this bound in $\beta$ decay requires a measurement of $D_t<10^{-6}$, which is an order of magnitude below the contribution of the FSI. 
 
\begin{figure}[t]
	\centering
\subfloat[Tree-level contribution to $\beta$ decay]{\label{fig:lepto}\includegraphics[width=0.42\textwidth]{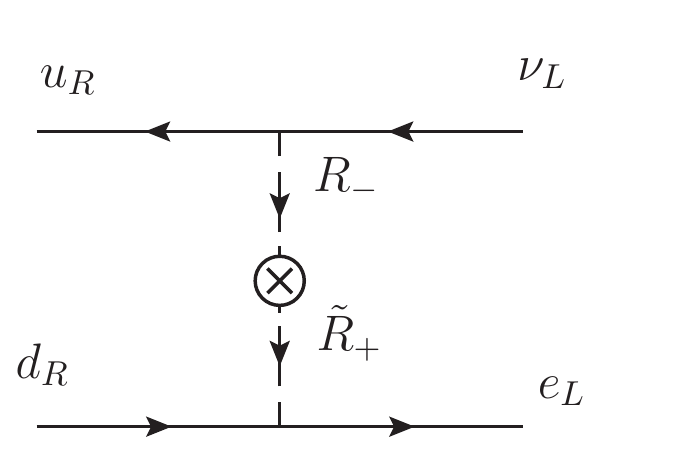}}
	%
\subfloat[Loop contribution to the neutron EDM]{\label{fig:leptoloop} \includegraphics[width=0.4\textwidth]{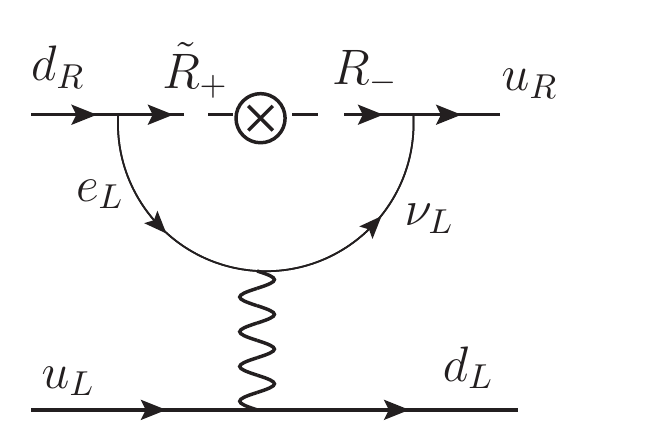}}
\caption{Example of scalar LQ exchange that contributes to (a) $\beta$ decay at loop level, and to (b) the neutron EDM via an electroweak loop. The scalar LQ are denoted by $R_-$ and $\tilde{R}_+$, where $\pm$ refers to the weak isospin component \cite{Ng:2011ui}.}
	\label{fig:leptoquarks}
\end{figure} 
The result above is obtained in a model-independent EFT approach, by introducing dimension-6 operators. The constraints apply to left-right symmetric models, exotic fermion models, and the R-parity violating MSSM \cite{Ng:2011ui}. Evasion of the bounds in Eq.~\eqref{eq:edmDresult} is only possible in either a strongly fine-tuned model or in a model in which the dimension-6 operators do not exist or do not contribute to either EDMs or $\beta$ decay. An example of the latter are leptoquarks (LQs). LQs are particles with both baryon and lepton number, which can be either vector or scalar particles depending on their spin. These were previously considered ``EDM-safe,'' but in fact they are not \cite{Ng:2011ui}. LQs can contribute to $\beta$ decay at tree level, for example via the exchange of scalar LQs as depicted in Fig.~\ref{fig:lepto}. Leptoquarks also contribute to EDMs, but only the $W$ exchange (Fig.~\ref{fig:leptoloop}). \textcite{Ng:2011ui} show that these loop contributions are not suppressed by the light quark masses $m_{u,d}^2$, as was previously argued \cite{Her01}. Therefore, the constraints from EDMs in the LQ scenario are much more stringent than previously thought. 

Estimates of the limit on $D_t$ in this scenario depend on the LQ mass and on whether light right-handed neutrinos exist. Assuming the existence of light right-handed neutrinos, \textcite{Ng:2011ui} found
\begin{equation}
\textrm{Im}\;a_{LR} = D_t/a_D < 3 \times 10^{-4} \left(\frac{300\; \textrm{GeV}}{m_{LQ}}\right)^2 \ ,
\end{equation}
while without them
\begin{equation}
\textrm{Im}\;a_{LR}= D_t/a_D < 7 \times 10^{-5} \left(\frac{300 \;\textrm{GeV}}{m_{LQ}}\right)^2 \ . 
\end{equation}
\textcite{Ng:2011ui} conservatively take $m_{LQ} = 300$ GeV, which would give, assuming the existence of light right-handed neutrinos, $D_t<3\times 10^{-4}$, a limit of the same order as the current $\beta$-decay bounds. Nevertheless, improving the current $\beta$ decay limit seems a difficult task, since there are many experiments ongoing or planned that aim to improve the bounds on the neutron EDM \cite{Ito07, Grin09, Alt09, Alt12, Bak11, Ser09}. In addition, strong bounds on the scalar LQ mass exist from the ATLAS and CMS experiments at the LHC. The bounds on their masses range from 607 GeV to 830 GeV, depending on the assumed LQ branching ratio \cite{pdg2014}, which suggests much stronger bounds on $D_t$.



\subsubsection{Limits on $R$ from EDM limits}
The $R$ coefficient and the EDM are both P- and T-odd. EDM measurements in atoms and molecules limit both the electron EDM and BSM scalar and tensor electron-nucleon interactions. \textcite{Khr91} showed the relation between these electron-nucleon interactions and the electron-quark interaction of $\beta$ decay. The scalar and tensor electron-nucleon interactions are defined by 
\begin{equation}\label{eq:barnn}
\mathcal{L} = \sum_N \frac{G_f}{\sqrt{2}}\left[C_S \bar{N}N \bar{e}i\gamma_5e + C_T \bar{N}\sigma_{\mu\nu}N\bar{e}i\gamma_5\sigma^{\mu\nu}e\right] \ ,
\end{equation}
where $C_S \;(C_T)$ is the scalar (tensor) coupling and we have neglected pseudoscalar couplings. In \textcite{Khr91,Khr97} it was shown that the limits on $C_S$ and $C_T$ can be related to both $\textrm{Im}\; A_{L}$ and $\textrm{Im}\; \alpha_{LL}$, the couplings contributing to the $R$ coefficient.  

The best current limit on nucleon scalar couplings is due to the EDM limit on molecular ThO, $|C_S|<5.9\times 10^{-9} \;(90\% \; \textrm{C.L.})$ \cite{Bar13}. The best bound on the nucleon tensor coupling, $|C_T|<1.3\times 10^{-9}\; (90\% \; \textrm{C.L.})$, 
is derived from the EDM limit on atomic Hg \cite{Gri09,Gin03}. These couplings must be translated to quark couplings in order to compare them to the $\beta$-decay couplings in Eq.~\eqref{eq:lageff}. At the quark level, scalar and tensor couplings in the electron-quark ($e$-$q$) interaction are described by \cite{Her03}
\begin{align}\label{eq:hameq}
\mathcal{L} = \sum_{q=u,d} \frac{G_F}{\sqrt{2}} \left[k_{Sq} (\bar{e}i\gamma_5e \bar{q}q) + k_{Tq} (\bar{e}i\gamma_5\sigma_{\mu\nu} e \bar{q}\sigma^{\mu\nu}q) \right]\ , 
\end{align}
where $k_{Sq} (k_{Tq})$ is the scalar (tensor) coupling in the $e$-$q$ interaction. 
The nucleon couplings can be translated into quark couplings by using the calculations in \textcite{Her03, Her05}, which show that nucleon and quark couplings are of the same order of magnitude. Conservatively, we find that the $k_{Sq}$ and $k_{Tq}$ couplings are $<10^{-8}\;(90\%\; \textrm{C.L.})$ 

Figure~\ref{fig:Rcoef} shows that the electroweak corrections to the exotic $\beta$-decay couplings contribute to the EDM $e$-$u$ couplings, $k_{Su}$ and $k_{Sd}$. The effective P- and T-odd $e$-$u$ interaction in Fig.~\ref{fig:Rcoef} is estimated as \cite{Khr91}
\begin{equation}\label{eq:effedm}
\frac{-G_F}{\sqrt{2}} \frac{\alpha}{4\pi} \textrm{ln}\left( \frac{\mu^2}{M_W^2}\right) V_{ud}\; \textrm{Im}\left( 2 A_L + 24 \alpha_L \right)\left[\bar{e}i\gamma_5e \bar{u}u +\tfrac{1}{2} \bar{e}i\gamma_5 \sigma_{\mu\nu} e \bar{u}\sigma^{\mu\nu}u \right]   \ ,
\end{equation}
where $\mu$ is the renormalization scale. Limits on the scalar electron-nucleon interaction $C_S$ thus limit both $A_L$ and $\alpha_L$. The effective $e$-$d$ interaction only contains $A_L$, and gives similar constraints.  
\begin{figure}
	\centering
	\includegraphics[width=0.70\textwidth]{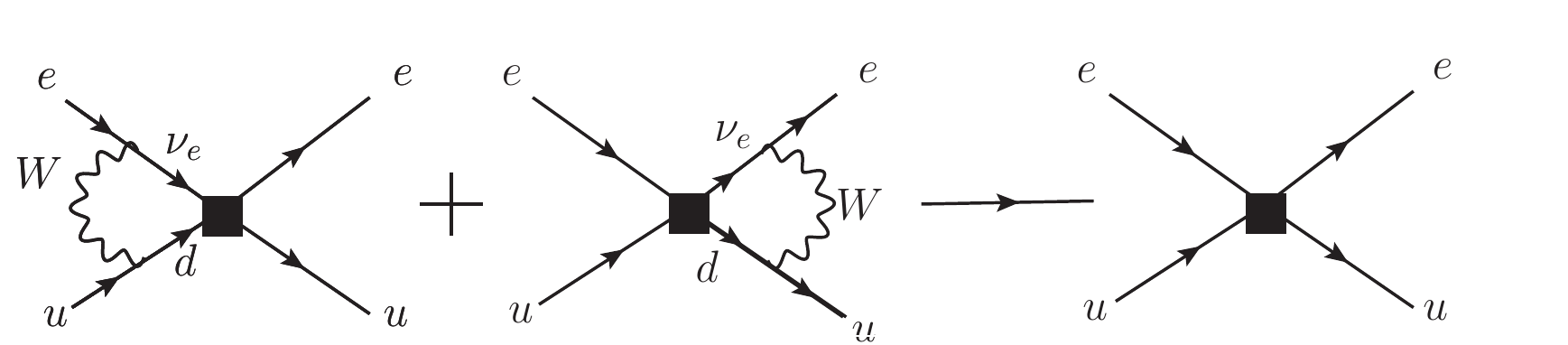}
	\caption{Contribution of $\beta$-decay coupling to the effective P- and T-odd electron-quark coupling through the exchange of the $W$ boson.}
	\label{fig:Rcoef}
\end{figure}

Comparing Eq.~\eqref{eq:hameq} and Eq.~\eqref{eq:effedm} we arrive at an expression for $k_{Su}$ and $k_{Tu}$. By using $k_{Su}<10^{-8}\;(90\% \;\textrm{C.L.})$ and the conservative 
assumption that $\ln(\mu^2/m_W^2) = 1$ as in \textcite{Khr91}, we estimate that at 90 \% C.L.
\begin{subequations}
\begin{align}
 |\textrm{Im}\;A_{L}| &   < 10^{-5} \ , \\ 
 |\textrm{Im}\; \alpha_{L}| &  <  10^{-6} \ . 
\end{align}
\end{subequations}
Both bounds are at least two orders of magnitude better than those obtained from the $R$ coefficient in $\beta$ decay. 


\subsection{Conclusion}
Table \ref{tab:EDMbeta} summarizes the limits on imaginary couplings. Bounds obtained from EDMs are several orders of magnitude better than current bounds from T-violating $\beta$-decay coefficients. 
The many ongoing efforts in the EDM field will strengthen the EDM bounds even further. 
\begin{table}[t]
\begin{center}
\begin{tabular}{l|l|l|l}
\hline\hline
&  $\textrm{Im}\; a_{LR}$ & $\textrm{Im}\; A_L$ & $\textrm{Im}\; \alpha_{L}$  \\ \hline
$\beta$ decay  &  $4 \times 10^{-4}$  & $6\times 10^{-2}$ & $3 \times 10^{-3}$  \\  
EDM & $3\times 10^{-6}$ & $10^{-5}$ & $10^{-6}$  \\
 & $(3 \times 10^{-4})$ & & \\ 
\hline \hline
	\end{tabular}
	\caption{Comparison between $\beta$-decay limits on imaginary couplings and constraints from EDMs. The bound in parentheses is derived in a model with leptoquarks and right-handed neutrinos. For the $\beta$-decay coefficients we use $g_S=1.02(11)$ \cite{Gon13} and $g_T=1.047(61)$ \cite{Bha14} and the R-fit mentioned described in Sec.~\ref{sec:LHC}. Constraints are at 90\% C.L.}
	\label{tab:EDMbeta}
\end{center}
\end{table}

The $D$ coefficient should be measured with a precision of $10^{-6}$ to improve the current EDM limits. Such a measurement is below the FSI interactions, and would require precise knowledge of the FSI for the used isotope. Measurements of the $D$ coefficient are considered as part of a larger effort to measure 11 coefficients ( $R$) in neutron decay \cite{Bod11}. Measurements of $D$ are also considered in nuclear decays \cite{Beh14, Lie14}. 
The $E$ coefficient in Eq.~\eqref{eq:Ecoef} depends on the same BSM coupling as the $D$ coefficient and is thus subject to the same EDM constraints.

It might be possible that the connection between EDMs and $\beta$ decay is diminished in a specific new-physics model, when such a model is strongly fine-tuned. For the $D$ coefficient examples are leptoquark models. Conservatively, this model relaxes the EDM constraint by maximally two orders of magnitude to $|D_t|<3\times 10^{-4}$ \cite{Ng:2011ui}. This is of the same order as current $\beta$ decay limits. Direct bounds on leptoquarks from the LHC experiments already suggest a stronger bound. Besides that, new bounds on the neutron EDM are also expected before any new $D$ measurement could realistically be done. This would further improve the bounds in Table~\ref{tab:EDMbeta}.  

Improving the current bounds on Im $\alpha_L$ requires a measurement of $R_t<10^{-6}$, which is an improvement of the current result by more than 3 orders of magnitude. An $R$ measurement in $^8$Li is ongoing at the Mott polarimeter for T-violation (MTV) \cite{Tot14}. Specific models may again weaken the connection between $\beta$ decay and EDMs. Such models would have to be strongly fine-tuned. For example, \textcite{Her05} showed that in R-parity violating SUSY \cite{Her05} such a cancellation would have to occur over 3 orders of magnitude. Such a severe cancellation is highly unnatural. Besides EDM limits there are also strong limits from the ratio $R_\pi=\Gamma(\pi\rightarrow e\nu)/\Gamma(\pi\rightarrow \mu\nu)$, which give Im $A_L<4\times 10^{-4}$ \cite{Her95,Her05}. 

Our EFT approach only applies when new physics can be parametrized by the heavy scale of new physics. If new particles are very light, the EFT approach does not apply anymore. 
However, the absence of experimental evidence for such light degrees of freedom supports the validity of the EFT approach. We therefore conclude that new measurements of the $D$ and $R$ coefficients should take the EDM bounds into account, and stress that the bounds can only be evaded in specific and strongly fine-tuned models.

\section{Lorentz violation}\label{sec:LIV}
We will now review the new field of searches for the violation of Lorentz symmetry in the weak interaction. Recently, it was found that $\beta$ decay offers unique possibilities to test Lorentz and/or CPT-invariance in the weak interaction, in both the gauge and the neutrino sector. We discuss these two sectors separately. 

\subsection{Gauge sector}
In the gauge sector, Lorentz violation can be studied in a general theoretical framework, developed to study allowed and forbidden $\beta$ decay and orbital electron capture \cite{Noo13a, Noo13b, Vos15a}. This framework considers a broad class of Lorentz-violating effects on the $W$ boson, by adding a general tensor $\chi^{\mu\nu}$ to the Minkowski metric. At low energies, this modifies the $W$-boson propagator to
\begin{equation}\label{eq:wprop}
\langle W^{\mu+}W^{\nu -} \rangle = \frac{-i (g^{\mu\nu}+\chi^{\mu\nu})}{M_W^2} \ ,
\end{equation}
where $g^{\mu\nu}$ is the Minkowski metric and $M_W$ is the $W$-boson mass. 
Vertex corrections are described by  
\begin{equation}
-i \Gamma  = -i g (g^{\mu\nu}+\chi^{\mu\nu}) \ .
\end{equation}
However, such vertex modifications also requires the modification of the electron and neutrino spinors \cite{Noo13a}. We restrict ourselves to propagator corrections, for which hermiticity of the Lagrangian implies that $\chi^*_{\mu\nu}(p) = \chi_{\nu\mu}(-p)$. In terms of the SME discussed in Sec.~\ref{sec:introLIV}, one finds, at lowest order,
\begin{equation}
\chi^{\mu\nu} = -k_{\phi\phi}^{\mu\nu} - \tfrac{i}{2g}k_{\phi W}^{\mu\nu} + 2k_W^{\rho\mu\sigma\nu} \frac{q_\rho q_\sigma}{M_W^2} \ ,
\end{equation}
where $q$ is the momentum of the $W$ boson and $g$ is the $SU(2)$ coupling constant. 

Bounds on $\chi$ have been derived from allowed \cite{Wil13, Mul13, Bod14} and forbidden $\beta$ decay \cite{Noo13b}, pion decay \cite{Alt13, Noo14}, muon decay \cite{Noo14a}, and nonleptonic kaon decay \cite{Vos14}. Here we discuss allowed and forbidden $\beta$ decay.



%
%
\subsubsection{Allowed $\beta$ decay}

For allowed $\beta$ decay, \textcite{Noo13a} derived the Lorentz-violating differential decay rate using the modified $W$-boson propagator in Eq.~\eqref{eq:wprop}. The complete expression is given in Eq.~\eqref{eq:appliv}. Lorentz violation gives many additional correlations, since the observables (momentum and spin) can now also couple to the tensor $\chi$. In $\beta$ decay, a variety of correlations can be used to access different (combinations of) $\chi$ components. The necessary expressions can be derived by integrating over one or more kinematic variables. 
Momentum-dependent terms are always suppressed by some power of a heavy mass ($M_W$ in the least-suppressed case), and can therefore be neglected given the current experimental precision. Neglecting momentum-dependent contributions to the propagator, the relation $\chi^*_{\mu\nu}(p) = \chi_{\nu\mu}(-p)$ implies that $\chi$ can only be real and symmetric or imaginary and antisymmetric, {\it i.e.} $\chi_r^{0l} = \chi_r^{l0}$, $\chi_i^{0l}=-\chi_i^{l0}$, $\chi^{\mu\mu}_i=0$, $\chi^{lk}_r=\chi^{kl}_r$ and $\chi^{lk}_i=-\chi_i^{kl}$. The subscripts $r$ and $i$ denote the real and imaginary parts of $\chi$, respectively. This leaves 15 independent CPT-even components of $\chi^{\mu\nu}$ that need to be measured.

With this simplification and in the absence of tensor polarizations, the decay rate 
is \cite{Noo13a, Vos15b} 
\begin{eqnarray}
dW & = &  \frac{ F(\pm Z, E_e)}{(2\pi)^5} |\vec{p}_e|E_e(E_e-E_0)^2dE_e d\Omega_e  d\Omega_\nu \bar{\xi}\; \left\{1 + (2a-c')\chi_r^{00} +\left(-(2a-c')\chi_r^{0l} + 2\breve{g}\tilde{\chi}_i^{l}\right)\frac{p_e^l}{E_e}\right. \notag \\
&& +\left. \frac{p_\nu^jp_e^l}{E_eE_\nu} \left[(a+c'+2\breve{a}\chi_r^{00}) \delta_{jl} -4\breve{g}\chi_r^{jl}\right] - (2a-c') \chi_i^{0s} \frac{(\vec{p}_\nu\times\vec{p}_e)^s}{E_eE_\nu} \right. \notag \\
&& + \left. \frac{\langle {J^k}\rangle}{J} \left(-2\breve{L}\tilde{\chi}_i^k + \frac{p_e^l}{E_e} \left[(A+B\chi_r^{00})\delta_{kl}-B\chi_r^{kl}\right] \right)+ A\chi_i^{0s}\frac{(\vec{p}_e \times \langle {\vec{J}}\rangle )^s}{JE_e} \right. \notag \\
&& + \left. \frac{p_\nu^j}{E_\nu}\left((-2a+c')\chi_r^{0j} -2\breve{g} \tilde{\chi}_i^j \right) + \frac{\langle {J^k}\rangle p_\nu^j}{J E_\nu} \left[(B+A\chi_r^{00})\delta_{kj}-A\chi_r^{kj}\right] - B\chi_i^{0s}\frac{( \vec{p}_\nu\times \langle{\vec{J}}\rangle)^s}{JE_\nu} \right\} \ , \notag \\
\label{eq:decayrateliv}
\end{eqnarray}
%
where $\langle {\vec{J}}\rangle$ is the expectation value of the spin of the parent nucleus, $\tilde{\chi}^l = \epsilon^{lmk}\chi^{mk}$, and Latin indices run over spatial directions. The last line of Eq.~\eqref{eq:decayrateliv} contains only the neutrino momentum or the neutrino momentum and the nuclear polarization, and can therefore mostly be ignored. In fact, the neutrino correlations give access to a similar combination of $\chi$ components as the electron correlations. The latter are considerably easier to obtain, and we will further only consider the electron correlations\footnote{In electron capture, the neutrino correlations play an important role \cite{Vos15a}.}. 
The coefficients $\bar{\xi}, a, A,$ and $B$ are the standard $\beta$-decay coefficients listed in Appendix \ref{sec:app}, the coefficient $c'$ is a modified $c$ coefficient. The coefficients with a breve $(\;\breve{}\;)$ multiply Lorentz-violating coefficients. They are given by 
\begin{align}
c'& = \frac{\rho^2}{1+\rho^2} \bar{\Lambda}_{J'J} \notag \ , \\
\breve{g}  &=  \frac{\tfrac{1}{3}\rho^2}{1+\rho^2} +  \tfrac{1}{2}c' \notag \ , \\ 
\breve{L} & =    \pm \frac{\tfrac{1}{2}\lambda_{J'J} \rho^2}{1+\rho^2} \notag \ ,\\
\breve{a}& =   \frac{1+\tfrac{1}{3}\rho^2}{1+\rho^2}+ \tfrac{1}{2}c' \ , 
\end{align}
where the upper(lower) sign refers to $\beta^{\mp}$ decay, and $\lambda_{J'J}$ and $\bar{\Lambda}_{J'J}= \Lambda_{J'J}\frac{\left\langle (\vec{J}\cdot\vec{j})^2\right\rangle -\tfrac{1}{3} J(J+1)}{J(2J-1)}$ are the standard $\beta$-decay coefficients given in Eq.~\eqref{eq:lab} and Eq.~\eqref{eq:lab2}, respectively. The coefficient $c'$ vanishes for non-oriented nuclei and for nuclei with $J'= J = \tfrac{1}{2}$. 


The effect of Lorentz violation in $\beta$ decay can already be studied by measuring the dependence of the decay rate as a function of the direction of the emitted $\beta$ particles. The modified Fermi decay rate integrated over neutrino energy and direction and summed over electron spin is
\begin{equation}\label{eq:fermi}
dW_{F} = dW^0 \left(1+2\chi^{00}_r - 2 \chi_r^{0l} \frac{p_e^l}{E_e}\right) \ ,
\end{equation}
while for Gamow-Teller transitions of randomly-oriented nuclei
\begin{equation}\label{eq:gt}
dW_{GT} = dW^0\left(1-\frac{2}{3} \chi^{00}_r + \frac{2}{3}(\chi_r^{l0}
+\tilde{\chi}_i^l) \frac{p_e^l}{E_e}\right) \ ,
\end{equation}
where
\begin{equation}
dW^0= \frac{1}{8\pi^4} p_e E_e (E_0-E_e)^2 F(\pm Z, E_e) dE_e\,d\Omega_e \bar{\xi} \ .
\end{equation}
The component $\tilde{\chi}_i$ can also be accessed by measuring the Gamow-Teller decays of polarized nuclei as a function of the spin direction,
\begin{align}
dW_{GT} {}&= dW^0\left(1-\frac{2}{3} \chi^{00}_r  \mp \lambda_{J'J} \tilde{\chi}^l \frac{\langle {J^l}\rangle}{J} \right)\ .
\end{align}
%
As an example of a mixed decay, one has for the neutron $a=-0.11, A=-0.12, B=0.98$, and $\breve{g} =\breve{L}= \frac{\lambda^2}{1+3\lambda^2} = 0.27$. Integrated over the neutrino direction\footnote{This formula corrects Eq.~(38) in \textcite{Noo13a} (see also Appendix \ref{sec:appb}). }
 \begin{eqnarray}
dW & = &  dW^0 \left\{1 -0.21\chi_r^{00} + (0.21 \chi_r^{0l} + 0.55 \tilde{\chi}_i^{l})\frac{p_e^l}{E_e}\right. \notag \\
&& +\left.\frac{\langle {J^k}\rangle}{J}\left[-0.55\tilde{\chi}_i^k + (-0.12+0.98\chi_r^{00})\frac{p_e^k}{E_e}-0.98\chi_r^{lk}\frac{p_e^l}{E_e}\right]  -0.12\chi_i^{0s}\frac{( \vec{p_e}\times \langle \vec{J}\rangle)^s}{J E_e}\right\} \ .
\label{neutrondecayrate}
\end{eqnarray}

Equation~\eqref{eq:decayrateliv} depends on SM parameters, which are often not known better than at the 1\%-0.1\% level. This dependence on SM coefficients can be avoided by measuring asymmetries that do not depend on the accuracy of the SM coefficients. The Lorentz-violating part of Eq.~\eqref{eq:fermi} can, for example, be accessed by measuring the decay asymmetry of a Fermi transition with the $\beta$ particles measured in opposite directions, 
\begin{equation}\label{eq:asfermi}
A_{F} = \frac{W_F^+ - W_F^-}{W_F^+ + W_F^-} = -2 \beta \chi^{0l}_r\;\hat{p}^l_e \ ,
\end{equation}
where $\beta= |\vec{p}_e|/E_e$ and $W_F^{\pm}$ is the rate of $\beta$ particles measured in the $\pm$ $\hat{p}_e$-direction. Similarly, the decay asymmetry in Gamow-Teller decays is 
\begin{equation}\label{eq:asgt}
A_{GT} = \frac{W_{GT}^+ -   W_{GT}^-}{W_{GT}^+ +   W_{GT}^-} = \tfrac{2}{3}\beta (\chi_r^{0l} + \tilde{\chi}_i^l) \hat{p}^l_e \ .
\end{equation}
The coefficient $\tilde{\chi}$ can also be obtained by measuring the spin asymmetry in a pure Gamow-Teller transition
\begin{equation}\label{eq:asspin}
A_{J} = \frac{W_{GT}^\uparrow -   W_{GT}^\downarrow}{W_{GT}^\uparrow +   W_{GT}^\downarrow} = P A \tilde{\chi}_i^k\; j^k \ ,
\end{equation}
where $W_{GT}^{\uparrow (\downarrow)}$ are the integrated decay rates independent of $\beta$ direction, but in the inverted polarization direction $\vec{j}$, and $P$ is the degree of nuclear polarization. $A$ is the $\beta$ asymmetry coefficient (for Gamow-Teller decays $A=\mp\lambda_{J'J}$). 
The remaining components of $\chi$ require more complicated measurements that involve at least two observables. The decay asymmetry between the spin and the $\beta$ particles can, for example, be measured from
\begin{equation}\label{eq:aschi}
A_{J\beta} = \frac{W_{L}^\uparrow W_{R}^\downarrow -   W_{R}^\uparrow W_{L}^\downarrow}{W_{L}^\uparrow W_{R}^\downarrow +   W_{R}^\uparrow W_{L}^\downarrow} = - 2 P \beta (A \chi_i^{0s} \epsilon^{slk} + B\chi_r^{lk})  j^l \hat{p}_e^k  \ ,
\end{equation}
where $W_{L,R}$ is obtained by measuring the $\beta$ particles in the opposite $\hat{p}_e$ direction, while the nuclei are polarized in the $\uparrow (\downarrow)$ opposite $\vec{j}$-direction. Similarly, $\chi_i^{0s}$ can also be obtained by measuring the decay asymmetry between the neutrino and electron in perpendicular directions. 

The spatial directions of $\chi$ are defined in the laboratory frame and their absolute orientation will depend on the orientation of Earth. It is therefore necessary to choose a standard absolute reference frame, for which the Sun-centered inertial reference frame is commonly chosen \cite{kospara}. The movement of this reference frame can safely be ignored. The transformation of $\chi^{\mu\nu}$ in the laboratory frame to the Sun-centered frame, in which we denote $\chi^{\mu\nu}$ by $X^{\mu\nu}$, is \cite{Noo13a}
\begin{equation}
\chi^{\mu\nu} = R^{\mu}_{\;\;\rho}R^{\nu}_{\;\;\sigma}X^{\rho\sigma} \ .
\end{equation}
The transformation matrix is
\begin{equation}
R(\zeta,t) = \left(
\begin{array}{cccc}
1 & 0 & 0 & 0 \\
0 & \cos\zeta \cos\Omega t & \cos\zeta \sin\Omega t & - \sin\zeta \\
0 & -\sin\Omega t & \cos\Omega t & 0 \\
0 & \sin\zeta\cos\Omega t & \sin\zeta\sin\Omega t & \cos\zeta
\end{array}
\right) \ ,
\end{equation}
where $\zeta$ is the colatitude of the experiment and $\Omega$ is Earth's sidereal rotation frequency. In the laboratory frame, $\hat{x}$ points in the north to south direction, $\hat{y}$ points west to east, and $\hat{z}$ is perpendicular to Earth's surface. 
The coefficients $\chi^{0l}_r$ and $\tilde{\chi}_i^l$ can be transformed to $X^{0l}_r$ and $\tilde{X}_i^l$, respectively. This transformation shows that the asymmetries $A_F$, $A_{GT}$, and $A_J$ can oscillate with the rotational frequency of Earth. These sidereal variations of the signal are a unique signature of Lorentz violation, and can therefore be separated from other limits on BSM physics. 
A generic example of how sidereal oscillations can be observed is shown in Fig.~\ref{fig:suncentered}, for $X^{0l}_r=0.1$. 
\begin{figure}
	\centering
		\includegraphics[width=0.60\textwidth]{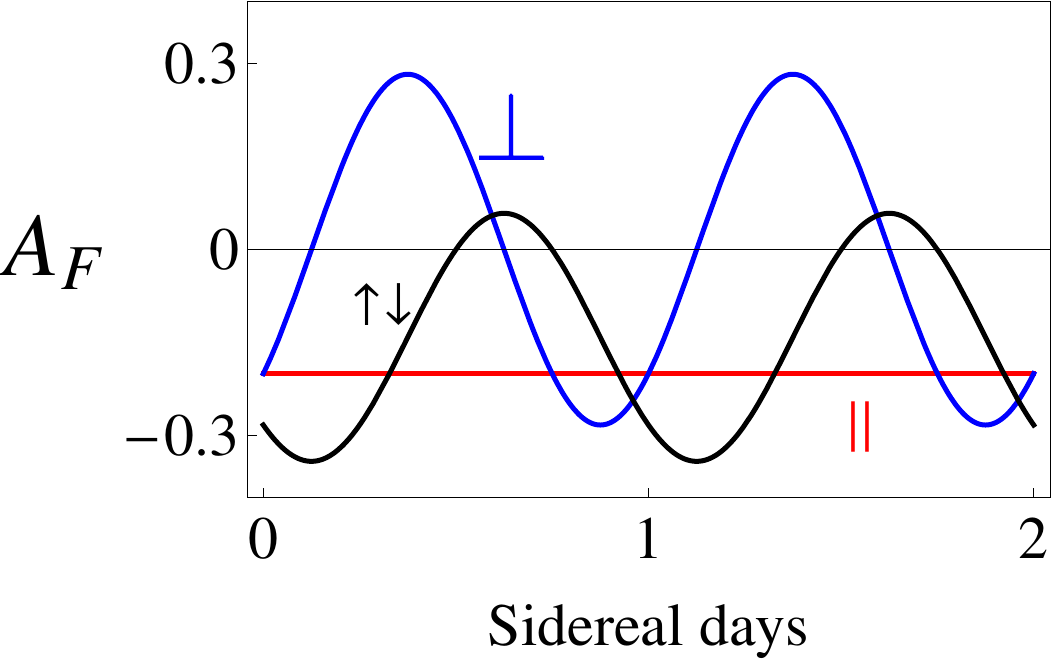}
	\caption{(Color online) Illustration of the oscillation of the asymmetry $A_F$ in Eq.~\eqref{eq:asfermi}, for $X^{0l}_r=0.1$ and $\zeta=45^\circ$. Three different detections directions of the $\beta$ particles are depicted. When $\beta$ particles are detected parallel $(\parallel)$ to Earth's rotation axis there is no sidereal variation (red line). The blue line shows the asymmetry when the $\beta$ particles are detected in the east-west direction $(\perp)$ direction and black line when they are detected perpendicular to the Earth's surface $(\uparrow\downarrow)$. Both show a sidereal variation, the latter with a constant offset. }
	\label{fig:suncentered}
\end{figure}
This example also shows that if the $\beta$ particles are detected parallel $(\parallel)$ to Earth's rotation axis, the asymmetry will have no sidereal dependence (red line). The blue line shows the case where the $\beta$ particles are detected in the east-west $(\perp)$direction. It has no offset because it is measured perpendicular to Earth's rotation axis. The black line gives the asymmetry for $\beta$ particles detected in the up-down $(\uparrow\downarrow)$ direction perpendicular to the Earth's surface ($\hat{z}$ direction in the labframe). It shows a sidereal oscillation on a constant offset. Detection of the $\beta$ particles perpendicular to the rotation axis is preferred, since an offset could be the result of systematic errors in the measurement. 

Tensor contributions involving $\chi^{jk}$ lead to terms that may oscillate with twice Earth's rotational frequency. Figure \ref{fig:asymmetries} illustrates three possible scenarios for an asymmetry that depends on $\chi^{lk}j^l \hat{p}_e^k$.  The red line shows the modulations when the polarization is in the up-down direction, while the $\beta$ particles are detected in the east-west direction. The blue line shows the modulations in the same polarization direction, but when the $\beta$ particles are detected parallel to Earth's rotation axis. It shows an oscillation with the period of the sidereal rotational frequency on top of a constant offset. The black line shows an oscillation with twice the period of the sidereal frequency. It arises when both the polarization and the $\beta$ particles are detected in the east-west direction.

\begin{figure}
	\centering
		\includegraphics[width=0.60\textwidth]{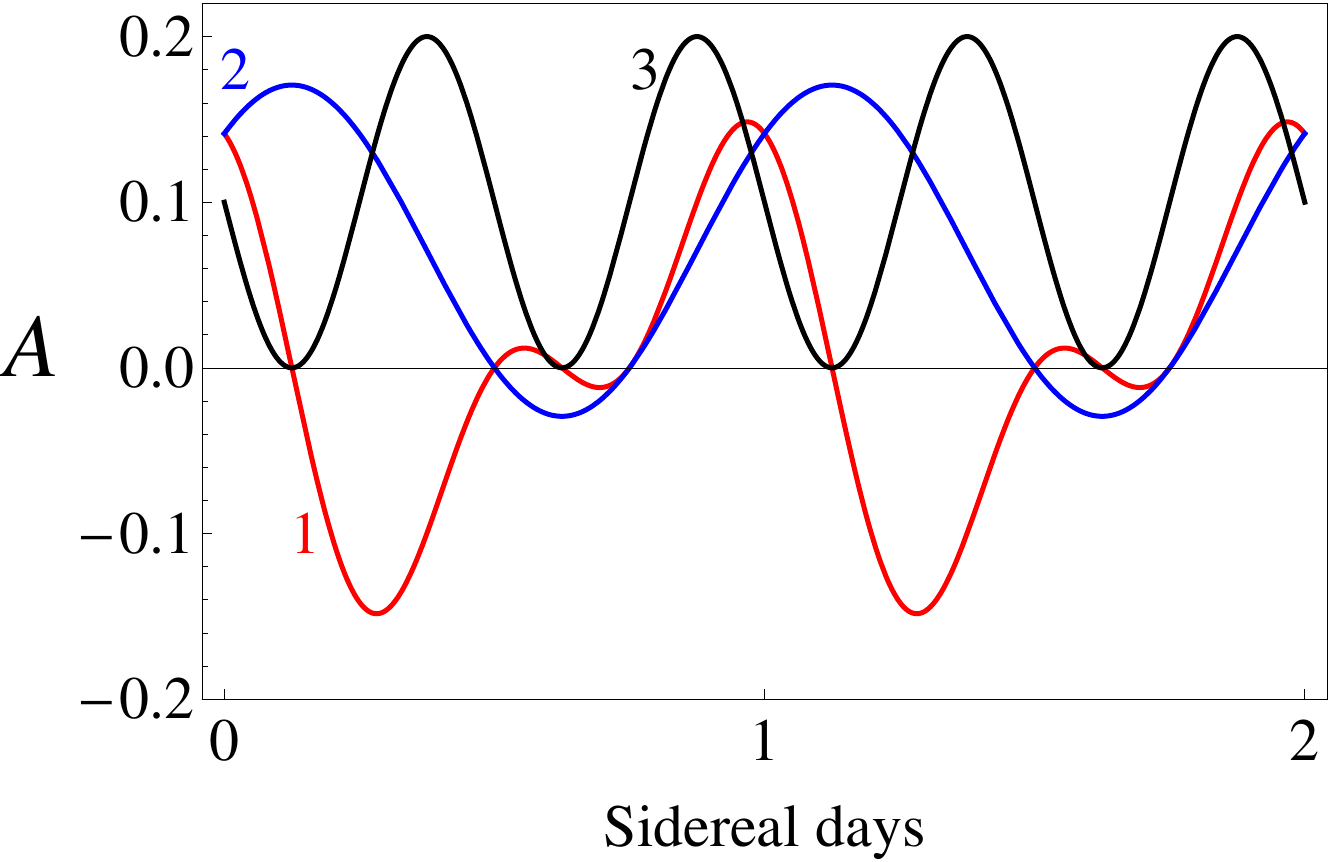}
	\caption{(Color online) Illustration of the possible sidereal variations of tensor Lorentz violation parametrized as $\chi^{lk}_r j^l \hat{p}_e^k$, with $X^{lk}_r = 0.1$. The red line (line 1) shows the modulations when $\vec{j}$ is in the $\hat{z}$ (up-down) direction and $\hat{p}_e$ in the $\hat{y}$ (east-west) direction. The blue line (2) is for $\vec{j}$ in the $\hat{z}$ direction and the $\beta$ particles detected parallel to Earth's rotation axis. The black line (3) shows the modulations when both $\vec{j}$ and $\hat{p}_e$ are in the east-west direction. }
	\label{fig:asymmetries}
\end{figure}

In allowed $\beta$ decay, Lorentz violation was for the first time tested in polarized $^{20}$Na \cite{Mul13}, by measuring the spin asymmetry $A_J$ (Eq.~\eqref{eq:asspin}). 
$^{20}$Na first decays with a $\beta^+$ $2^+\rightarrow 2^+$ Gamow-Teller transition, followed by a $\gamma$ decay of the daughter nucleus. The parity-odd $\beta$ decay was used to determine the polarization $P$ by measuring the $\beta$ asymmetry \cite{Mul13}. 
The parity-even $\gamma$ decay was used to measure the lifetime $\tau^{\uparrow(\downarrow)}$ and to determine the $\gamma$ asymmetry 
\begin{equation}
A_{\gamma} = \frac{\tau^\downarrow-\tau^\uparrow}{  \tau^\uparrow+\tau^\downarrow} = P A \vec{\tilde{\chi}}_i\cdot \vec{j} \ ,
\end{equation}
where the polarization direction is in the $\vec{j}$ direction. 
To reduce systematic errors, the polarization direction is preferably in the $\hat{y}$ (east-west) direction. The analysis of the setup in this direction places bounds of order $\mathcal{O}(10^{-3})$ \cite{Syt15}.  

Lorentz violation has also been searched for in polarized neutron decay \cite{Bod14}. Two different asymmetries, that depend on the nuclear polarization and the $\beta$ direction, were measured and are currently being analyzed. The asymmetries depend on combinations of $\vec{\tilde{\chi}}_i$ and $\vec{\chi}_r$ and preliminary bounds are $\mathcal{O}(10^{-2})$ \cite{Bod14}. This setup probably also allows for a measurement of $A_{J\beta}$ defined in Eq.~\eqref{eq:aschi}. Such a measurement would measure the so-far unconstrained coefficients $\chi_i^{0l}$. 
\subsubsection{Forbidden $\beta$ decay}\label{sec:forb}

\begin{figure}
	\centering
		\includegraphics[width=0.75\textwidth]{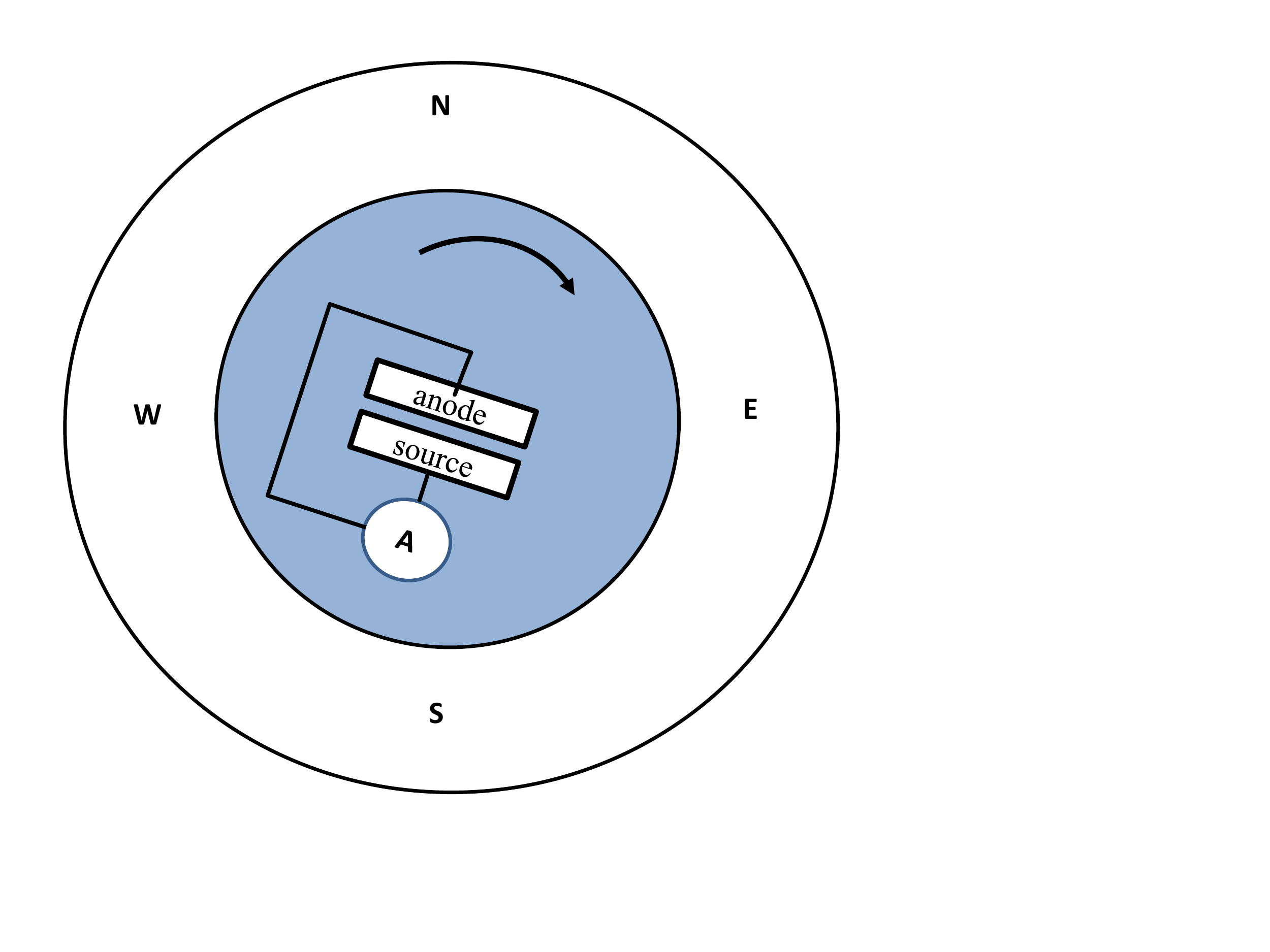}
	\caption{Schematic setup of the rotating $^{90}$Y experiment of \textcite{New76}.}
	\label{fig:setup}
\end{figure}
``Forbidden'' (slow) transitions are suppressed with respect to allowed transitions, because the lepton pair carries away angular momentum. Theoretically, the simplest of these transitions are the unique first-forbidden transitions ($\Delta J=2$), since they depend on only one nuclear matrix element. Because Lorentz violation includes rotational violation, it also implies the violation of angular-momentum conservation. Forbidden $\beta$ decays are then more sensitive to rotational invariance violation in the weak interaction. In the 1970s, two experiments were performed with this motivation.  \textcite{New76} searched for anisotropies in the angular distribution of $\beta$ particles in first-forbidden $^{90}$Y decay. \textcite{Ull78} searched for sidereal modulations of the count rates for first-forbidden $^{137}$Cs $\beta$ decay and second-forbidden $^{99}$Tc $\beta$ decay. The strongest bounds were found in the experiment by \textcite{New76}. In this experiment the $\beta$-decay distribution of $^{90}$Y from a high-intensity source was measured in a rotating setup. Schematically, the setup is depicted in Fig.~\ref{fig:setup}. The rotation of the setup allowed for the determination of three decay asymmetries 
\begin{equation}
\delta_{NS} = 2\frac{W_N - W_S}{W_N+W_S} \ , \; \; \; \; \delta_{EW} = 2\frac{W_E- W_W}{W_E+W_W} \ ,
\end{equation}
and 
\begin{equation}
\delta_{2\nu} = 2 \frac {W_N+W_S-W_E - W_W}{W_N+W_S+W_E+W_W} \ ,
\end{equation}
where $N,S,E,W$ are north, south, east, and west and $W$ is the decay rate measured by the $\beta$ particles in that direction. These asymmetries were fitted with 
\begin{equation}
\delta = a_0 + a_1 \sin(\Omega t + \phi_1) + a_2 \sin(2\Omega t + \phi_2) \ ,
\end{equation}
to search for a sidereal time dependence and to reduce systematic errors. The extracted bounds on $a_0, a_1,$ and $a_2$ are $\mathcal{O}(10^{-8})$ \cite{New76}. \textcite{Noo13b} reinterpreted the data from \textcite{New76, Ull78}, after extending the allowed $\beta$-decay framework to include higher-order terms in the multipole expansion,  i.e. all possible forbidden decays. 
The modified $W$-boson propagator gives an unconventional contraction of the nucleon and lepton currents, such that angular momentum is no longer conserved. In the Lorentz-symmetric case, rotational invariance implies that $\Delta J \leq J_{\textrm{lep}}$, where $\Delta J$ is the spin change of the nucleus and $J_{\textrm{lep}}$ is the total angular momentum of the leptons. In contrast, when contracting with $\chi^{0l}$, transitions with $\Delta J = J_{\textrm{lep}}+1$ are possible, and when contracting with $\chi^{lk}$ also $\Delta J = J_{\textrm{lep}}+2$ transitions are allowed. It is thus possible to have transitions in which the leptons carry away less angular momentum than in the Lorentz-symmetric case. Because the suppression of the forbidden decays is proportional to the angular momentum of the leptons, the Lorentz-violating terms are enhanced compared to the Lorentz-symmetric case. 

For unique first-forbidden transitions \cite{Noo13b}
\begin{equation}\label{eq:enfor}
\frac{dW}{d\Omega_e dE_e} \propto p_e^2 + p_\nu^2 + p_e^2 \frac{\alpha Z}{p_e R} \left[\tfrac{3}{10} \tfrac{p_e}{E_e}\left(\chi_r^{ij} \hat{p}^i \hat{p}^j - \tfrac{1}{3} \chi_r^{00}\right) - \tfrac{1}{2}\tilde{\chi}_i^l \hat{p}^l + \chi_r^{l0} \hat{p}^l\right] \ ,
\end{equation}
where $\alpha Z/p_eR \simeq \mathcal{O}(10^1)$. Equation~\eqref{eq:enfor} shows that the Lorentz-violating contributions are enhanced. Higher-order forbidden decays do not have additional enhancement compared to the simpler first-forbidden transitions. \textcite{Noo13b} translated the bounds from \textcite{New76} using Eq.~\eqref{eq:enfor}. This led to strong limits on several combinations of $\chi^{\mu\nu}$. Assuming no cancellations between coefficients, this results in the limits \cite{Noo13b} 
\begin{equation}
\begin{matrix}
\chi^{\mu\nu}_r = \begin{bmatrix} 10^{-6} & 10^{-7} & 10^{-7} & 10^{-8} \\  10^{-7} & 10^{-6} & 10^{-6} & 10^{-6} \\10^{-7} & 10^{-6} & 10^{-6} & 10^{-6} \\10^{-8} & 10^{-6} & 10^{-6} & 10^{-6} \\
\end{bmatrix}, & \textrm{and} & \chi^{\mu\nu}_i = \begin{bmatrix}  \times & - & - & - \\ - & \times & 10^{-8} & 10^{-7} \\ - & 10^{-8} & \times & 10^{-7} \\ - & 10^{-7} & 10^{-7} & \times \\
\end{bmatrix}  \ .\end{matrix}
\end{equation}
These are the strongest constraints on $\chi^{\mu\nu}$. The only coefficients not constrained by forbidden decays are $\chi_i^{0l}$. These coefficients can be studied in allowed $\beta$ decay by considering Eq.~\eqref{eq:aschi} or equivalent correlations. The bounds on $\chi$ were also translated into bounds on the SME parameters \cite{Noo13b}, providing strong direct bounds on the SME parameters $k_{\phi\phi}$ and $k_{\phi W}$ defined in Eq.~\eqref{eq:smegauge}. 

\subsubsection{Conclusion and outlook}
We have discussed the efforts to search for Lorentz violation in the weak interaction in forbidden and allowed $\beta$ decay. The bounds from forbidden $\beta$ decay are several orders of magnitude stronger than the current bounds in allowed $\beta$ decay, due to the intense sources that were used \cite{New76, Ull78}. In allowed $\beta$ decay, Lorentz-violating effects are not enhanced and matching the statistical precision of forbidden $\beta$-decay experiments would require long-running experiments with high-intensity sources. An interesting alternative lies in orbital electron capture, where it is possible to use such high-intensity sources \cite{Vos15a}. 

Allowed $\beta$ decay offers various correlations in which Lorentz violation could be probed. Observables can be chosen such that they give direct constraints on $\chi$ compared to the combination of coefficients constrained by forbidden decays. Two relatively simple experiments that probe the $\beta$ asymmetry in Fermi and Gamow-Teller decays, $A_F$ and $A_{GT}$ respectively, would give direct bounds on $\chi_r^{0l}$ and $\tilde{\chi}_i$. These asymmetries could be studied parallel to the efforts to measure the $\beta$-spectrum shape discussed in Sec.~\ref{sec:efforts} \cite{Vos15b}.  
Another interesting possibility is to exploit the $\gamma^2$ enhancement of decay asymmetries by considering fast-moving nuclei \cite{Vos15b, Vos14, Alt13}. The total decay rate in the rest frame of the nucleus is proportional to $\chi_r^{00}$ (see Eq.~\eqref{eq:fermi} and Eq.~\eqref{eq:gt}). For a fast-moving nucleus, the expression can be related to the Sun-centered frame with a boost. If the nucleus is moving ultra-relativistically in the $\hat{v}$ direction, 
\begin{equation}
\chi_r^{00} = \gamma_r^2 \left(X^{TT}_r + 2X^{TL}_r \hat{v}^L+ X^{LK}_r \hat{v}^L \hat{v}^K \right) \ ,
\end{equation}
where $\gamma_r$ is the Lorentz-boost factor and $T, L,$ and $K$ are coordinates in the Sun-centered reference frame. This relation was, for example, used to extract bounds of $\mathcal{O}(10^{-4})$ from pion decay \cite{Alt13}. For allowed $\beta$ decay, $\beta$-beam facilities, currently considered for producing neutrino beams \cite{Lin10}, could be exploited. 
  
So far the coefficients $\chi_{i}^{0l}$ remain unconstrained. In Fermi decays, this coefficient could be measured by considering the correlation $\chi_i^{0l}(\vec{p}_e\times \vec{p}_\nu)^l$. The coefficients can also be constrained by measuring the polarized $\beta$ asymmetry $A_{J\beta}$ in Eq.~\eqref{eq:aschi}. Such an asymmetry could probably be explored in the neutron-decay measurement pursued by \textcite{Bod14}.

\subsection{Neutrino sector}
A different possibility to study Lorentz violation in $\beta$ decay lies in the neutrino sector of the SME \cite{neutrino2012, Kos04}. Most interesting for $\beta$ decay are the modified versions of $a^{\textrm{LV}}$ and $c^{\textrm{LV}}$ defined in Eq.~\eqref{eq:leptonsme}. 

Unlike the gauge sector, the neutrino sector has been studied extensively in several experiments. Strong bounds exist from neutrino oscillations and time-of-flight measurements \cite{kospara}. However, there are four operators that do not show up in oscillations and have no effect on the neutrino group velocity. These operators are called ``countershaded'' \cite{Tas09}. Recently, \textcite{Dia13} 
showed that $\beta$ decay has a unique sensitivity to these operators. The four countershaded coefficients are denoted by $a_{\textrm{of}}^{(3)}$. The operators are dimension 3 and CPT-odd. These coefficients modify the neutrino dispersion relation and the available phase space of the neutrino, which affects $\beta$ decay in two ways, in the $\beta$ end-point and in the correlation coefficients.

\subsubsection{End-point in $\beta$ decay}
The $\beta$-spectrum end-point is very sensitive to the neutrino phase space and to the neutrino mass (see also Sec.~\ref{sec:neutrino}). Independent of the neutrino mass, the countershaded neutrino coefficients also shift the end-point, as can be seen from the modified decay rate \cite{Dia13, Dia14}
\begin{equation}
\frac{dW}{dT} \sim (\Delta T + \delta T_{\textrm{LV}})^2 - \tfrac{1}{2} m_\nu^2  \ ,
\end{equation}
where $\Delta T = T_0 -T_e$, $T_e=E_e-m_e$ is the electron kinetic energy, and $T_0$ is the end-point energy for $m_\nu = 0$. $\delta T_{\textrm{LV}}$ is the Lorentz-violating modification, which depends on sidereal time. Independent of the neutrino mass, a bound on the countershaded coefficients can be set by using the available data of the Troitsk \cite{Kra04} and Mainz \cite{Ase11} experiments, see \textcite{Dia13}. Since these experiments collected data over a long period of time, all the oscillations average out and only the time-averaged Lorentz-violating coefficients can be constrained. Therefore, only two of the four countershaded coefficients could be bounded. Conservatively, the analysis of \textcite{Dia13} gives bounds of order $\mathcal{O}(10^{-8})$ GeV. These limits improve and complement previous limits. A dedicated analysis of the data of the Troitsk, Mainz, or the expected KATRIN \cite{katrin} experiments could improve these results. If the data analysis also takes into account the sidereal time, bounds on all the countershaded coefficients could be set.

\subsubsection{Correlation coefficients}
The Lorentz-violating neutrino coefficients of Eq.~\eqref{eq:leptonsme} also modify the neutrino spinor solutions. Near the end point, this modification can be neglected because the phase space dominates. However, the derivation of the complete modified decay rate requires both the modified spinors and the phase-space modification. The modified neutrino phase space is $d^3\vec{p}_\nu \simeq (E_\nu^2 - 2 E_\nu a_{\textrm{of}}^{(3)}) dE_\nu d\Omega_\nu$. The modification of the spinors requires the replacement of $\vec{p}_\nu$ by $\tilde{\vec{p}}_\nu =  ( \vec{p}_\nu + \vec{a}_\textrm{of}^{(3)} - \dot{a}_\textrm{of}^{(3)}\hat{p}_\nu)$, where $\dot{a}_\textrm{of}^{(3)}$ is the isotropic component. The modified neutron decay rate is 
  \begin{eqnarray}\label{eq:moddecaya}
  \frac{dW} {d\Omega_e \;d\Omega_\nu\; dT} &\simeq &  F(Z,E) |\vec{p}_e| E_e (E_\nu^2 + 2 E_\nu \delta T_{\textrm{LV}}) \nonumber \\
&&	\times \left(1 + a \vec{\beta}\cdot \tilde{p}_\nu + A \frac{\langle \vec{J}\rangle}{J }\cdot\frac{\vec{p}_e}{E_e} + B \frac{\langle \vec{J}\rangle}{J} \cdot \frac{\tilde{p}_\nu}{E_\nu} )\right) \ .
  \end{eqnarray}
The neutrino coefficients modify the decay rate in a similar way as $\chi$ does, since there are now additional correlations between $\vec{J}$ and $\vec{p}_e$ and $a_{\textrm{of}}^{(3)}$. 

The countershaded coefficients could, for example, affect the $\beta$-$\nu$ correlation. 
The $\beta$-$\nu$ correlation can be measured as an asymmetry, defined by  
\begin{equation}\label{eq:tildea}
\tilde{a} = \frac{N_+-N_-}{N_++N_-} \ ,
\end{equation}
where $N_+ (N_-)$ is the number of decays in which the neutrino and electron are emitted (anti)parallel. The Lorentz-violating neutrino coefficients modify this correlation coefficient to \cite{Dia14}  
  \begin{equation}
 \tilde{a} = a |\vec{\beta}| + \sqrt{\frac{3}{\pi}}\frac{(a\beta^2+a|\vec{\beta}|)}{E_\nu} (a_{\textrm{of}}^{(3)})^{\textrm{lab}}_{10} \ ,
  \end{equation}
where the coefficients should be transformed to the Sun-centered frame and would depend on the sidereal frequency of Earth.

No experiment has searched for these variations, but \textcite{Dia13} estimate that a $0.1\%$ measurement of $a$ would limit the countershaded coefficients at the level of $10^{-8}$ GeV. Similar, for a $0.1\%$ measurement of the correlation coefficient $B$, the limits are estimated at $\mathcal{O}(10^{-6})$ GeV. A dedicated experiment measuring either $a$ or $B$ would thus provide interesting new bounds on Lorentz-violating parameters in the neutrino sector. Note that $\chi$ and $a_{\textrm{of}}^{(3)}$ have a similar influence on the decay rate. In a dedicated experiment both coefficients might influence the asymmetry. A measurement of Eq.~\eqref{eq:tildea} might also be sensitive to $\chi_r^{0l}$ and $\tilde{\chi}_i$, depending on the experimental setup.

\subsection{Conclusion}
To summarize, $\beta$ decay offers a unique way to study Lorentz violation in both the gauge and neutrino sector. The large variety of correlations allows for direct measurements of different components of $\chi$, while in the neutrino sector $\beta$ decay allows for the study of countershaded coefficients. 

In the gauge sector, strong bounds on the order of $10^{-6}$ - $10^{-8}$ exist from forbidden $\beta$-decay experiments. Unconstrained are the coefficients $\chi_i^{0l}$, which can be accessed in $\beta$ decay by considering the interaction of $\chi$ with two observables (Eq.~\eqref{eq:aschi}. Improving the existing bounds requires high statistics and precise knowledge of the systematic uncertainties. Beneficial for this would be to exploit the $\gamma_r^2$ enhancement of boosted $\beta$ decay or to consider electron capture. The real and imaginary part of $\chi$ can be constrained by measuring the asymmetries in Eq.~\eqref{eq:asfermi} and Eq.~\eqref{eq:asgt}, respectively. Such an effort could be combined with measurements of the Fierz-interference term.   

Further, we have discussed the possibilities to improve constraints on Lorentz violation in the countershaded neutrino sector. In that sector no dedicated experiment has been preformed so far, but using available data from tritium gives bounds of order $10^{-8}$ GeV. The parameters not constrained so far could be bound in $\beta$-decay correlation experiments. Lorentz violation gives a unique signal compared to other BSM physics when searched for in a dedicated experiment. 
Estimates for $0.1\%$ measurements of the coefficients $a$ and $B$ gives a constraint on Lorentz violation of $10^{-8}$ GeV, which shows the potential for these future experiments.

\section{Summary and discussion}\label{sec:discussion}
\noindent In this review we addressed the current status and role of nuclear and neutron $\beta$ decay in the search for physics beyond the SM. In these searches, the statistical precision is becoming increasingly important. However, systematic errors, despite improved detection methods, and higher-order corrections such as FSI, still appear to be the main limits. In the meantime, thanks to the evolution of EFT methods, constraints obtained in other fields weigh in, establishing bounds on the scalar and tensor contributions. This is illustrated in Fig.~\ref{fig:scalarlhcneutrino} and Fig.~\ref{fig:tensorlhcneutrino}, where measurements at LHC (Sec.~\ref{sec:LHC}) and limits from the neutrino mass (Sec.~\ref{sec:neutrino}) give constraints that outperformed the $\beta$-correlation measurements in the right-handed sector. This is quantified in Table~\ref{tab:LHCbeta}. 
	
The study of fundamental aspects of $\beta$ decay will be most fruitful in the study of left-handed scalar (Sec.~\ref{sec:scalar}) and tensor currents (Sec.~\ref{sec:nucten}), as these appear linearly in most observables via the Fierz-interference term. Fortunately, these interactions can be studied in parallel to precision studies of SM parameters (Sec.~\ref{sec:smpara}). For example, extracting the CKM matrix element $V_{ud}$ from superallowed Fermi transitions has, as a by-product, the most strict limit on left-handed scalar interactions. Lacking still is a similar limit on tensor contributions. An interesting option to obtain such a bound could come from measuring the detailed shape of the $\beta$ spectrum in Gamow-Teller transitions. Also the potential of mirror transitions, both for obtaining tensor limits and for obtaining a value for $V_{ud}$ independent of the superallowed Fermi transitions, has been recognized. In Table \ref{tab:pro} we indicate the precision required to impose new bounds on left- and right-handed scalar and tensor currents. Measuring  the  Fierz interference term  in $\beta$ decay remains  competitive in determining bounds on left-handed coupling constants. In contrast, Table V shows that, for right-handed couplings, the limits from LHC and the limits derived from the neutrino mass are by far superior to the best bounds derived from the $\beta\nu$-correlation $a$, and future experiments in $\beta$ decay are unlikely to reach this precision.
	
	Concerning the most fundamental measurement of T-violation, we discussed in Sec.~\ref{sec:time} the strong bounds on the triple-correlation coefficients $D$ and $R$ derived from the limits on permanent EDMs. These bounds are summarized in Table~\ref{tab:EDMbeta}. Not only are the bounds from EDMs several orders of magnitude stronger than those of $\beta$ decay, but the EDM limits also have a large potential to improve faster than those from $\beta$ decay. One reason is that EDMs can be measured in stable or long-lived particles, but also because of the widely perceived urgency for improved limits in this sector. 
	
A new twist to the discussion of symmetry violations in $\beta$ decay has been added, since $\beta$ decay also offers an interesting sensitivity to Lorentz violation in the weak interaction. In Sec.~\ref{sec:LIV}, we reviewed these limits for the first time. Because the discrete symmetries C, P, and T are each violated in the weak interaction, this interaction is a promising portal to search for new physics when considering CPT violation and thus Lorentz violation. The familiar $\beta$-decay correlations are now extended to include correlations between spin and momentum and a Lorentz-violating background tensor. Consequently, spin and momentum will appear to have preferred directions in absolute space, resulting in unique signals that can be distinguished from other BSM searches.

	In weak decays the LIV has been parametrized with the complex tensor $\chi$.  The bound on most components of this tensor are of the order of $10^{-6}$ to $10^{-8}$ (Sec.~\ref{sec:forb}). Fine tuning between the tensor components allow to weaken these bounds. Relatively simple new experiments can improve these bounds using very strong sources, also removing the possibility of fine tuning.  Obtaining sufficient high counting statistics is the main challenge. The searches for Lorentz violation can be expanded in a parallel effort with the more traditional searches. Alternatively, one can study $\beta$ decay in flight, exploiting the $\gamma^2_r$ enhancement. In this respect there may be as yet unexplored possibilities related to semileptonic decays in high-energy physics. Because this field of research is relatively unexplored, both experimentally and theoretically, the best approach may still emerge.

Improvements in theory and experimental techniques, as well as new radioactive-beam facilities, provide new possibilities to study fundamental aspects of $\beta$ decay, both in the search for exotic interactions and in the search for Lorentz violation. These studies should be done by considering also the other searches in high-energy physics and precision physics at low energies. Nuclear and neutron $\beta$ decay will remain an important topic on the research agenda.
	
\section*{Acknowledgments}
We are grateful to our colleagues  K. Bodek, J. Behr, W. Dekens, M. Gonz\'ales-Alonso, O. Naviliat-Cuncic, J. Noordmans, G. Onderwater, N. Severijns, J. de Vries, F. Wauters, and A. Young for fruitful discussions and updates on the experimental progress. We thank especially the organizers of the {\em 33$^{\,rd}$ Solvay Workshop on Beta Decay Weak Interaction Studies in the Era of the LHC} (Brussels, September 3-5, 2014).
This research was supported by the Dutch Stichting voor Fundamenteel Onderzoek der Materie (FOM) under Programmes 104 and 114. 	

\appendix
\section{Decay coefficients}\label{sec:app}
Our formalism can be linked to the original work of \textcite{Lee56,Jac57a}, where\footnote{Using our definition of $\gamma_5$ and neglecting pseudoscalar couplings.}
\begin{align}
\mathcal{L}_{\textrm{eff}} ={}& \bar{p}n\;\bar{e}(C_S  - C'_S\gamma_5)\nu_e \nonumber \\
{}& + \bar{p}\gamma^\mu n \;\bar{e}\gamma_\mu(C_V  - C'_V \gamma_5)\nu_e \nonumber \\
{}& + \bar{p}\gamma^\mu\gamma_5 n \;\bar{e}\gamma_\mu(C'_A  - C_A \gamma_5)\nu_e \nonumber \\
{}& + \frac{1}{2} \bar{p}\sigma^{\mu\nu} n \;\bar{e}\sigma_{\mu\nu}(C_T  - C'_T \gamma_5)\nu_e   \ .
\end{align}
This notation can be related to our couplings in Eq.~\eqref{eq:lageff} by using the normalized couplings
\begin{equation}
C_i = \frac{G_F}{\sqrt{2}}V_{ud}\; \bar{C}_i \ ,
\end{equation}
and 
\begin{align}\label{eq:Clist}
\bar{C}_V {}& = g_V(a_L + a_R )\nonumber\ , \\
\bar{C}'_V {}& = g_V(a_L - a_R )\nonumber\ , \\
\bar{C}_A {}& = -|g_A|(a'_L + a'_R )\nonumber\ , \\ 
\bar{C}'_A {}& = -|g_A|(a'_L - a'_R )\nonumber\ , \\
\bar{C}_S {}& = g_S(A_L + A_R )\nonumber\ , \\
\bar{C}'_S {}& = g_S(A_L - A_R )\nonumber\ , \\
\bar{C}_T {}& = 2g_T(\alpha_L + \alpha_R )\nonumber\ , \\
\bar{C}'_T {}& = 2g_T(\alpha_L - \alpha_R ) \ . 
\end{align} 
For simplicity we have defined
\begin{align}\label{eq:list}
a_L {}& \equiv a_{LL} + a_{LR} \nonumber\ , \\
a_R {}& \equiv a_{RR} + a_{RL} \nonumber\ , \\
a'_L {}& \equiv a_{LL} - a_{LR} \nonumber\ , \\
a'_R {}& \equiv a_{RR} - a_{RL} \nonumber\ , \\
A_L {}& \equiv A_{LR} + A_{LL} \nonumber\ , \\
A_R {}& \equiv A_{RL} + A_{RR}  \ .
\end{align} 
We write $\alpha_L$ and $\alpha_R$ as in Eq.~\eqref{eq:lageff}, because $\sigma_{\mu\nu}\gamma_5 = \frac{i}{2}\epsilon_{\mu\nu\alpha\beta}\sigma^{\alpha\beta}$. 
The coefficients $a_{\epsilon\delta}, A_{\epsilon\delta},$ and $\alpha_{\epsilon}$ are related to the $\epsilon$ coefficients in \textcite{Cir13, Nav13} by using 
\begin{equation}
\left\{a_{LL}, a_{LR}, a_{RL}, a_{RR}, A_{LL}+A_{LR}, A_{RR}+A_{RL}, \alpha_{L}, \alpha_{R}\right\} = 
\left\{1+\epsilon_L, \epsilon_R, \tilde{\epsilon}_L, \tilde{\epsilon}_R, \epsilon_S, \tilde{\epsilon}_S, 2\epsilon_T, 2\tilde{\epsilon}_T\right\}.
\end{equation}
A full list of correlation coefficients in allowed $\beta$ decay including Coulomb corrections is given in \textcite{Jac57a, Jac57}. Here we will give the most important decay coefficients in terms of couplings defined in Eq.~\eqref{eq:efflag}. We emphasize that only $b, B,$ and $N$ depend linearly on scalar and tensor couplings. We define $\lambda=|g_A|/g_V >0$ and neglect Coulomb interactions. 
The spin factors are
\begin{equation}\label{eq:lab}
\lambda_{J'J}= \begin{cases} 1, & J\rightarrow J'=J-1 \\ \frac{1}{J+1}, & J\rightarrow J'=J\\ \frac{-J}{J+1}, & J\rightarrow J'=J+1\end{cases},
\end{equation}
and
\begin{equation}\label{eq:lab2}
\Lambda_{J'J}= \begin{cases} 1, & J\rightarrow J'=J-1 \\ \frac{-(2J-1)}{J+1}, & J\rightarrow J'=J\\ \frac{J(2J-1)}{(J+1)(2J+3)}, & J\rightarrow J'=J+1\end{cases},
\end{equation}
where $J$ and $J'$ are the spin of the initial and final nucleus, respectively. In the following equations, $M_{F}$ and $M_{GT}$ are the Fermi and Gamow-Teller matrix elements, the upper (lower) sign refers to $\beta^-$ ($\beta^+$) decay, and $\gamma= \sqrt{1-\alpha^2 Z^2}$, with $Z$ the atomic number of the daughter nucleus and $\alpha$ the fine-structure constant.

Because of our normalization of the couplings in Eq.~\eqref{eq:lageff} we define
\begin{equation}
\bar{\xi} \equiv \frac{G_F^2 V_{ud}^2}{2} \xi \ ,
\end{equation}
with 
\begin{align}\label{eq:xibar}
\xi ={}& 2g_V^2 |M_F|^2\left\{|a_L|^2 +  |a_R|^2+\frac{g_S^2}{g_V^2}\left(|A_L|^2 +  |A_R|^2\right)\right\} \nonumber \\
{}& + 2 g_V^2 \lambda^2 |M_{GT}|^2 \left\{|a'_L|^2 + |a'_R|^2 + 4\frac{g_T^2}{g_A^2} \left(|\alpha_L|^2 +  |\alpha_R|^2\right)\right\} \ .
\end{align}

Neglecting Coulomb interactions, the decay coefficients are \cite{Jac57, Jac57a}
\begin{align}\label{eq:xia}
a\xi ={}& 2g_V^2 |M_F|^2 \left\{ |a_L|^2 +  |a_R|^2-\frac{g_S^2}{g_V^2} \left[|A_L|^2 +  |A_R|^2\right] \right\} \nonumber \\
 {}& + 2g_V^2 \lambda^2 \frac{|M_{GT}|^2}{3} \left\{ -|a'_L|^2 - |a'_R|^2+4\frac{g_T^2}{g_A^2} \left[|\alpha_L|^2 +  |\alpha_R|^2\right]\right\}  \ , \\
{}& \nonumber \\
b\xi  ={}& \pm 2 g_V^2\gamma \Big\{2|M_F|^2 \frac{g_S}{g_V} \Big[\textrm{Re}(A_L a^*_L) + \textrm{Re}(A_R a^*_R) \Big]   \nonumber \\
{}& - 4\frac{g_T}{|g_A|} \lambda^2 |M_{GT}|^2 \Big[\textrm{Re}(\alpha_L a^*_L)  + \textrm{Re}(\alpha_R a_R^*)\Big] \Big\} \ , \label{eq:bxi} \\
{}& \nonumber \\
c\xi ={}& 2g_V^2 \lambda^2 \Lambda_{J'J} |M_{GT}|^2 \left\{ -|a'_L|^2 - |a'_R|^2+4\frac{g_T^2}{g_A^2} \left[|\alpha_L|^2 +  |\alpha_R|^2\right]\right\}\label{eq:xic}  \ , \\
{}& \nonumber \\
A\xi = {}&\pm 2g_V^2 \lambda^2 |M_{GT}|^2 \lambda_{J'J}  \left\{ 4\frac{g_T^2}{g_A^2} \left[|\alpha_L|^2 - |\alpha_R|^2\right]  - \left[|a'_L|^2 - |a'_R|^2\right]\right\} \nonumber \\
 {}& + 2 g_V^2 \lambda \delta_{J'J}|M_F||M_{GT}|\sqrt{\frac{J}{J+1}}    \Big\{4\frac{g_T g_S}{|g_A| g_V} \Big[ \textrm{Re} (A_L \alpha^*_L) -  \textrm{Re}(A_R \alpha^*_R) \Big]  \nonumber \\
{}&  + 2 \Big[|a_{LL}|^2 - |a_{LR}|^2 - |a_{RR}|^2 + |a_{RL}|^2\Big] \Big\}  \ , \label{eq:Axi} 
\end{align}

\begin{align}
B\xi ={}& 2g_V^2 \lambda^2 |M_{GT}|^2 \lambda_{J'J}  \Big\{\frac{-4g_T}{|g_A|} \frac{m_e\gamma}{E_e} \Big[\textrm{Re}(\alpha_L a'^*_L)  - \textrm{Re}(\alpha_R a'^*_R)\Big]  \nonumber \\
{}&  \pm \frac{4g_T^2}{g_A^2} \left[|\alpha_L|^2 - |\alpha_R|^2\right] \pm \left[|a'_L|^2 - |a'_R|^2\right]\Big\} \nonumber \\
 {}& - 2g_V^2 \lambda \delta_{J'J}|M_F||M_{GT}|\sqrt{\frac{J}{J+1}}    \Big\{\frac{4g_T g_S}{g_V |g_A|} \Big[ \textrm{Re}(A_L \alpha^*_L) -  \textrm{Re}(A_R \alpha^*_R) \Big]  \nonumber \\
{}&  - 2  \Big[|a_{LL}|^2 - |a_{LR}|^2 - |a_{RR}|^2 + |a_{RL}|^2\Big]  \nonumber \\ 
{}& \pm \frac{m_e\gamma}{E_e} \Big(-\frac{2g_S}{g_V} \Big[\textrm{Re}(A_L a'^*_{L})- \textrm{Re}(A_R a'^*_R) \Big] + \frac{4g_T}{|g_A|} \Big[\textrm{Re}(a_L \alpha^*_L) - \textrm{Re}(a_R \alpha^*_R) \Big]\Big)\Big\} \ , \\
{}& \nonumber   \\
D\xi ={}& 2g_V^2 \lambda \delta_{J'J}|M_F||M_{GT}|\sqrt{\frac{J}{J+1}}    \Big\{\frac{4g_T g_S}{g_V |g_A|} \Big[ \textrm{Im}(A_L \alpha^*_L) +  \textrm{Im}(A_R \alpha^*_R) \Big]  \nonumber \\
{}&  + 2  \Big[ \textrm{Im}(a_L a'^{*}_L) +  \textrm{Im}(a_R a'^*_R) \Big] \Big\}  \ , \label{eq:Dap}\\
{}& \nonumber \\
R\xi ={}& \pm 2g_V^2 \lambda^2 |M_{GT}|^2 \lambda_{J'J} \frac{-4g_T}{|g_A|} \Big[\textrm{Im}(\alpha_L a'^*_L)  - \textrm{Im}(\alpha_R a'^*_R)\Big]  \nonumber \\
 {}& + 2g_V^2 \lambda \delta_{J'J}|M_F||M_{GT}|\sqrt{\frac{J}{J+1}}    \Big\{-2 \Big[ \textrm{Im}(A_L a'^*_L) -  \textrm{Im}(A_R a'^*_R) \Big]  \nonumber \\
{}& - \frac{4g_T}{|g_A|} \Big[\textrm{Im}(a_L \alpha^*_L) - \textrm{Im}(a_R \alpha^*_R) \Big]\Big\} \ , \label{eq:Rap}\\
\textrm{and}  \nonumber \\
N\xi ={}& 2g_V^2 \lambda^2 |M_{GT}|^2 \lambda_{J'J}  \Big\{\frac{m_e\gamma}{E_e} \Big[|a'_L|^2 + |a'_R|^2 + 4\frac{g_T^2}{g_A^2} \left[|\alpha_L|^2 +  |\alpha_R|^2\right]\Big]  \nonumber \\
{}& \mp \frac{4g_T^2}{g_A^2} \Big[\textrm{Re}(\alpha_L a^*_L)  + \textrm{Re}(\alpha_R a_R^*)\Big]\Big\} \nonumber \\
 {}& + 2g_V^2 \lambda \delta_{J'J}|M_F||M_{GT}|\sqrt{\frac{J}{J+1}} \Big\{-\frac{2g_S}{g_V} \Big[\textrm{Re}(A_L a'^*_{L})+ \textrm{Re}(A_R a'^*_R) \Big] \nonumber \\
{}&  + \frac{4g_T}{|g_A|} \Big[\textrm{Re}(a_L \alpha^*_L) + \textrm{Re}(a_R \alpha^*_R) \Big]  \pm \frac{\gamma m_e}{E_e}  \Big( \frac{4g_T g_S}{g_V |g_A|} \Big[ \textrm{Re}(A_L \alpha^*_L) +  \textrm{Re}(A_R \alpha^*_R) \Big] \nonumber \\
{}& - 2  \Big[|a_{LL}|^2 - |a_{LR}|^2 + |a_{RR}|^2 - |a_{RL}|^2\Big] \Big)\Big\}  \ .
\end{align}

The longitudinal electron polarization is \cite{Jac57, Jac57a}
\begin{equation}
P = \frac{G \frac{v_e}{c}}{1+b \left\langle \frac{m_e}{E_e}\right\rangle } \ ,
\end{equation}
with
\begin{align}
G\xi = {}&\pm 2|M_{F}|^2 g_V^2 \left\{\frac{g_S^2}{g_V^2} \left[|A_L|^2 - |A_R|^2\right]  - |a_L|^2 + |a_R|^2 \right\} \nonumber \\
{}& \pm 2|M_{GT}|^2 g_V^2 \lambda^2\Big\{\frac{4g_T^2}{g_A^2}\Big[|\alpha_L|^2 - |\alpha_R|^2\Big]- \Big[|a'_L|^2 - |a'_R|^2 \Big] \Big\} .
\end{align}

The neutron lifetime (Eq.~\eqref{eq:neutronlifetime}) depends on $V_{ud}$, which is extracted from the $0^+ \rightarrow 0^+$ superallowed Fermi decays. However, the extracted value of $V_{ud}$ might also depend on new physics. Taking into account this possibility,  
\begin{align}\label{eq:taucomplete}
\tau_n = K \frac{1-2 \tfrac{g_S}{g_V}A_L \gamma {\left\langle \tfrac{m_e}{E_e} \right\rangle}^{0^+\rightarrow 0^+} + \tfrac{g_S^2}{g_V^2}A_L^2 +\tfrac{g_S^2}{g_V^2}A_R^2}{1+\tfrac{g_S^2}{g_V^2}A_L^2+\tfrac{g_S^2}{g_V^2}A_R^2+3\lambda^2(1+4\tfrac{g_T^2}{g_A^2}\alpha_L^2+4\tfrac{g_T^2}{g_A^2}\alpha_R^2) + \gamma \left\langle \tfrac{m_e}{E_e}\right\rangle (2\tfrac{g_S}{g_V}A_L - 12\lambda^2 \tfrac{g_T}{|g_A|}\alpha_L)} \ ,
\end{align} 
where $\left\langle m_e/E_e \right\rangle^{0^+\rightarrow 0^+}$ is the inverse average energy of the superallowed decays, calculated by \textcite{Pat13}. The constant $K$ is \cite{Pat13}
\begin{equation}
K\equiv\frac{2\pi^3}{m_e^5 f_n(1+\Delta_{RC}) G_F^2 V_{ud}^2} = (1.9342\pm 0.002)\cdot 10^{-4}\ ,
\end{equation}
where $f_n=1.6887(2)$ is the statistical rate function \cite{Tow10} and $\Delta_{RC}$ are the SM electroweak corrections \cite{Cza04}.
 
The SM expressions can be obtained by setting $a_{LL}=1$ and neglecting all other couplings. Defining $\rho \equiv |g_A| M_{GT}/g_V M_F$, the remaining SM expressions are  
\begin{subequations}\label{eq:smexpr}
\begin{align}
a_{SM} {}& = \frac{1-\rho^2/3}{1+\rho^2} \ , \\
A_{SM} {}& = \frac{\mp \lambda_{J'J} \rho^2 + 2 \delta_{J'J}\sqrt{J/(J+1)}\rho}{1+\rho^2}\ ,  \\
B_{SM} {}& = \frac{\pm \lambda_{J'J} \rho^2 +2\delta_{J'J}\sqrt{J/(J+1)}\rho}{1+\rho^2}\ ,  \\
G_{SM} {}& = \mp 1 \ , 
\end{align}
\end{subequations}
while all other coefficients vanish. For neutron decay, $\rho= \sqrt{3}|g_A|$ and $J=J'=\tfrac{1}{2}$.

\subsection{Linear terms in $B$}
The $B$ coefficients contains terms linear in exotic couplings. Neglecting quadratic couplings, we can write
\begin{equation}
B\xi =   \pm\lambda_{J'J}\rho^2 + 2\rho  \delta_{J'J} \sqrt{J/(J+1)} +  \left\langle \frac{m_e\gamma}{E_e}\right\rangle  b_B\xi \ ,
\end{equation}
where 
\begin{align}
b_B\xi={}& -\rho^2 \lambda_{J'J} \frac{4g_T}{|g_A|} \textrm{Re}~\alpha_L \nonumber \\
{}&\mp  \delta_{J'J} \rho \sqrt{\frac{J}{J+1}}\Big[-\frac{2g_S}{g_V} \textrm{Re}~A_L   +\frac{4g_T}{|g_A|} \textrm{Re}~\alpha_L\Big] \ .
\end{align}
Most $B$ measurements measure
\begin{equation}
\tilde{B} = \frac{B_{SM}+b_B \gamma\left\langle  \frac{m_e}{E_e}\right\rangle } {1+b \left\langle \frac{m_e}{E_e}\right\rangle } \ . 
\end{equation}
For pure Gamow-Teller transitions, with $\rho \rightarrow \infty$, $\tilde{B}_{GT} =  \pm\lambda_{J'J} $ and the linear dependence cancels. For  neutron decay and assuming real couplings, $B_{SM}\simeq 1$ and
\begin{align}\label{eq:B}
\tilde{B}  {}& \simeq B_{SM}+\left\langle \frac{m_e}{E_e}\right\rangle (\gamma b_B - b B_{SM}) \nonumber \\
{}& \simeq \frac{2(\lambda +\lambda^2)}{1+3\lambda^2}+ \left\langle \frac{m_e \gamma}{E_e} \right\rangle  \frac{-\lambda -2\lambda^2+3\lambda^3}{(1+3\lambda^2)^2}\left[2\frac{g_S}{g_V} A_L + 4 \frac{g_T}{|g_A|}
\alpha_L \right] \nonumber \\
{}& \simeq 1 + \left[ 0.1 \frac{g_S}{g_V} A_L + 0.2 \frac{g_T}{|g_A|}\alpha_L\right]\left\langle \frac{m_e\gamma}{E_e}\right\rangle \ . 
\end{align}

For comparison, for neutron decay, the Fierz interference term is
\begin{align}
b_{\textrm{neutron}}{}& =\frac{2 \frac{g_S}{g_V}A_L-12\lambda^2 \frac{g_T}{|g_A|} \alpha_L}{1+3\lambda^2}\nonumber  \\
{}& \simeq 0.35 \frac{g_S}{g_V} A_L- 3.3  \frac{g_T}{|g_A|} \alpha_L \ .
\end{align}
For the measured $\tilde{A}$ coefficient in neutron decay, with $A_{\textrm{SM}}\simeq -0.11$, 
\begin{align}
\tilde{A}_{\textrm{neutron}} {}& = A_{\textrm{SM}} \mp A_{\textrm{SM}}  \frac{m_e\gamma}{E_e} (0.35 \frac{g_S}{g_V} A_L- 3.3 \frac{g_T}{|g_A|}\alpha_L) \ . 
\end{align}
So for neutron decay, $B$ actually has a reduced sensitivity to scalar and in particular tensor terms compared to for example $A$.

\section{Lorentz violation}\label{sec:appb}
The Lorentz-violating $\beta$ decay rate including Coulomb corrections and electron spin, to first order in $\chi^{\mu\nu}$, is \cite{Noo13a}
\begin{align}
dW  = {}&\frac{1}{(2\pi)^5}E_e p_e (E_0-E_e)^2  F(\pm Z,E_e)\bar{\xi}  dE_e d\Omega_e d\Omega_\nu \nonumber \\
{}& \times \; \left\{\left(1\mp\frac{\vec{p}_e\cdot\se}{E_e}\right)\left[\tfrac{1}{2}\left(1+B\frac{\vec{p}_\nu \cdot \langle \vec{J}\rangle }{J E_\nu}\right) + t +\frac{\vec{w}_1\cdot \vec{p}_\nu }{E_\nu} + \vec{w}_2\cdot \frac{\langle {\vec{J}}\rangle}{J} + T_1^{km}j^k j^m \right.\right.\nonumber \\
{}& \left.\left. + T_2^{kj} \frac{\langle J^k\rangle p_\nu^j}{J E_\nu} + \frac{S_1^{kmj}j^k j^m p_\nu^j}{E_\nu} \right]\right.\nonumber \\
{}& \left. +\left(\left(1\mp\frac{(E_e-\gamma m_e)(\vec{p}_e\cdot \vec{\sigma}_e)}{E_e^2-m_e^2}\right)\frac{p_e^l}{E_e}\mp\frac{\gamma m_e}{E_e}\hat{\sigma}_e^l \mp \frac{m_e}{E_e}\sqrt{1-\gamma^2}(\hat{p}_e \times \hat{\sigma}_e)^l\right) \right. \nonumber \\
{}& \left.\times\left[\tfrac{1}{2}A \frac{\langle J^k \rangle}{J} -\tfrac{3}{2}c'\frac{\vec{p}_\nu\cdot j}{E_\nu} j^l + \tfrac{1}{2}(a+c')\frac{p_\nu^l}{E_\nu} + w_3^l + \frac{T_3^{lj}p_\nu^j}{E_\nu} + T_4^{lk}\frac{\langle J^k \rangle}{J} \right. \right. \nonumber \\
{}& \left.\left. + S_2^{lmk} j^m j^k + \frac{S_3^{lmj}\langle J^m\rangle p_\nu^j}{JE_\nu} + \frac{ R^{lmkj} j^mj^k p_\nu^j}{E_\nu}\right]\right\} \ ,
\label{eq:appliv}
\end{align}
where $\gamma= \sqrt{1-\alpha^2 Z^2}$. The Lorentz-violating constants are\footnote{Note the sign error in $w_3^l$ in \textcite{Noo13a}.}
\begin{gather}
t = (a-\tfrac{1}{2}c')\chi_r^{00} \ , \notag \\
w_1^j = -x\chi_r^{0j} + \breve{g}(\chi_r^{j0}-\tilde{\chi}_i^j) \ , \qquad w_2^k = \breve{K}(\chi_r^{k0}-\chi_r^{0k})-\breve{L}\tilde{\chi}_i^k \ , \qquad w_3^l = -x\chi_r^{0l}+\breve{g}(\chi_r^{l0} + \tilde{\chi}_i^l) \ , \notag \\
T_1^{km} = \tfrac{3}{2}c'\chi_r^{km} \ , \qquad T_2^{kj} = \tfrac{1}{2}A\chi_r^{00}\delta^{jk} + \breve{L}(\chi_r^{jk}+\chi_i^{s0}\epsilon^{sjk})-\breve{K}(\chi_r^{kj}+\chi_i^{0s}\epsilon^{sjk}) \ , \notag \\
 T_3^{lj} = (x+\breve{g})\chi_r^{00}\delta^{lj}-(x\chi_i^{0s} + \breve{g}\chi_i^{s0})\epsilon^{sjl}-\breve{g}(\chi_r^{jl}+\chi_r^{lj}) \ , \notag \\ 
T_4^{lk} = \tfrac{1}{2}B\chi_r^{00}\delta^{lk}-\breve{L}(\chi_r^{lk} - \chi_i^{s0}\epsilon^{ksl})-\breve{K}(\chi_r^{kl}-\chi_i^{0s}\epsilon^{ksl}) \ , \notag \\
S_1^{kmj} = -\tfrac{3}{2}c'(\chi_r^{k0}\delta^{mj}-\chi_i^{ms}\epsilon^{sjk}) \ , \qquad S_2^{lmk} = -\tfrac{3}{2}c'(\chi_r^{m0}\delta^{kl}+\chi_i^{ms}\epsilon^{slk}) \ , \notag \\
S_3^{lmj} = \breve{L}\left(\chi_r^{l0}\delta^{jm}-\chi_i^{sl}\epsilon^{sjm}-\chi_r^{j0}\delta^{ml}+\tilde{\chi}_i^m \delta^{jl} - \chi_i^{sj}\epsilon^{lms}\right) \qquad\qquad  \notag \\ 
\qquad\qquad+ \breve{K}\left(\chi_i^{00}\epsilon^{ljm}-\chi_r^{0l}\delta^{jm}-\chi_r^{0j}\delta^{ml}+(\chi_r^{0m}+\chi_r^{m0})\delta^{jl}-\chi_i^{ms}\epsilon^{sjl}\right), \notag \\
R^{lmkj}= \tfrac{3}{2}c'\left(\chi_i^{m0}\epsilon^{lkj} - \chi_r^{mk}\delta^{lj} + \chi_r^{ml}\delta^{kj} + \chi_r^{mj}\delta^{kl}\right) \ ,
\label{eq:applivquantities}
\end{gather}
where $r$ and $i$ denote the real and imaginary part of $\chi^{\mu\nu}$, respectively, $\tilde{\chi}^l=\epsilon^{lmk}\chi^{mk}$, and $p^l$ denotes the electron momentum in the $l$-direction. 
$a, A, B,$ and $\bar{\xi}$ are the standard $\beta$-decay coefficients, given in Eq.~\eqref{eq:smexpr}, the other coefficients are 
\begin{gather}
\qquad x = \frac{1}{1+\rho^2} \ , \qquad y = \frac{-\rho}{1+\rho^2} \ ,  \notag \\
\qquad c' = (1-x)\bar{\Lambda}_{JJ'} \ , \notag \\
\breve{g} = \tfrac{1}{3}(1-x)(1+\tfrac{3}{2}\bar{\Lambda}_{JJ'}) \ , \qquad \breve{K}=-y \sqrt{\frac{J}{J+1}} \delta_{JJ'} \ , \qquad \breve{L} = \pm\tfrac{1}{2}\frac{\rho^2}{1+\rho^2}\lambda_{JJ'} \ ,
\label{defconstants}
\end{gather} 
where upper(lower) signs refer to $\beta^-(\beta^+)$ decay. 
The coefficient $\lambda_{J'J}$ is given in Eq.~\eqref{eq:lab} and 
$\bar{\Lambda}_{J'J} \equiv \Lambda_{J'J} \frac{\left\langle (\vec{J}\cdot\vec{j})^2\right\rangle -\tfrac{1}{3} J(J+1)}{J(2J-1)}$, with $\Lambda_{J'J}$ given in Eq.~\eqref{eq:lab2}.

\bibliography{refs}
\end{document}